%% file: arxiv.tex
\documentclass[11pt, reqno]{amsart}
\usepackage[utf8]{inputenc}
\usepackage[margin=1in]{geometry}
\usepackage{mathptmx}
\usepackage{scrextend}
\usepackage{url}
\usepackage{subfiles}
\usepackage{xcolor}
\usepackage{mathrsfs}
\usepackage{amsfonts}
\usepackage{bbm}

\usepackage{amsmath}
\usepackage{amsthm}
\usepackage{amssymb}
\usepackage{amsfonts}

\usepackage{enumitem}
\usepackage{setspace}
\usepackage{subcaption}

\newtheorem{theorem}{Theorem}[section]
\newtheorem{corollary}{Corollary}[theorem]
\newtheorem{lemma}[theorem]{Lemma}
\newtheorem{proposition}[theorem]{Proposition}
\theoremstyle{definition}
\newtheorem{definition}[theorem]{Definition}
\newtheorem{example}[theorem]{Example}

\newtheorem{remark}[theorem]{Remark}
\newtheorem{mech}[theorem]{Mechanism}

\newcommand{\todo}[1]{}
\renewcommand{\todo}[1]{{\color{red} {#1}}}

\newcommand{\F}{\mathcal{F}}

\newcommand{\D}{\mathcal{D}}

\newcommand{\DS}{\delta^{\F}} 
\newcommand{\dstar}{\delta}
\newcommand{\dprime}{\delta'}
\newcommand{\DSone}{\delta^{\F_1}}
\newcommand{\DStwo}{\delta^{\F_2}}
\newcommand{\DSthree}{\delta^{\F_3}}

\newcommand{\Supp}{\text{Supp}}

\newcommand{\prettypowerset}{\mathsf{Pow}}
\newcommand{\Mech}{A}
\newcommand{\DistMech}{\mathbb{A}}
\newcommand{\E}{\mathbb{E}}
\newcommand{\Mapping}{\mathcal{M}}
\newcommand{\OriginalDist}[3]{S^{#1,#2}_{#3}}
\newcommand{\CondDist}[3]{S^{C, #1,#2}_{#3}}
\newcommand{\UncondDist}[3]{S^{N, #1,#2}_{#3}}
\newcommand{\TildeCondDist}[3]{\tilde{S}^{C, #1,#2}_{#3}}
\newcommand{\TildeUncondDist}[3]{\tilde{S}^{N, #1,#2}_{#3}}
\newcommand{\DCondE}{d^{cond, \E}}
\newcommand{\DUncondE}{d^{uncond, \E}}
\newcommand{\DCondMMD}{d^{cond, MMD}}
\newcommand{\DUncondMMD}{d^{uncond, MMD}}
\newcommand{\notselected}{\bot}
\newcommand{\Map}[2]{M_{#1,#2}}
\newcommand{\MapNum}[2]{n_{#1,#2}}

\newcommand{\DSint}{\delta^{\mathsf{int}}}
\newcommand{\DSqual}{\delta^{\mathsf{quality}}}
\newcommand{\Clustering}[2]{(#1, #2)\text{-metric}}
\newcommand{\ClusteringBF}[2]{(#1, #2)\textbf{-metric}}

\newcommand{\Opipeline}{O_{pipeline}} 

\usepackage{fancyhdr}
\title{Individual Fairness in Pipelines}
\author{Cynthia Dwork$^*$ \and Christina Ilvento$^{**}$ \and Meena Jagadeesan$^{***}$}\thanks{\textit{Author names listed alphabetically.} \\
$^*$  Harvard John A Paulson School of Engineering and Applied Sciences, Radcliffe Institute for Advanced Study, and Microsoft Research. Research supported in part by  NSF grant CCF-1763665 and Microsoft Research. \\ 
$^{**}$ Harvard John A Paulson School of Engineering and Applied Sciences. This work was funded by the Sloan Foundation. \\
$^{***}$ Harvard University} 

\date{}
\begin{document}

% doublespace for editing
% \doublespacing

\maketitle

% Abstract: 0.5 pages

\subfile{sections/abstract}

% \inote{Test that notes are turned off.}
% \mnote{Test that notes are turned off.}
% \cnote{Test that notes are turned off.}
% \todo{Test that notes are turned off.}
% Introduction: 1 pages
\section{Introduction}
\subfile{sections/introduction}
\section{Model and Definitions}
\label{sec:modeldefinitions}
\subfile{sections/new_model}

% Conditions: 2 pages
\section{Conditions for Success}
\label{sec:conditionsforsuccess}
\subfile{sections/conditions}

% --------------- 7.75 pages ------------------------------
% Mechanisms: 1 pages (Meena: I think this should be 1.25 pages)
\section{Robust Mechanisms}
\label{sec:mechanisms}
\subfile{sections/mechanisms}
\section{Discussion and Future Work}
\label{sec:discussionfuturework}
\subfile{sections/discussion}
\section{Related Work}
\label{sec:related}
\subfile{sections/related}

% ------------- 10 pages ---------------------------

% Bibliography
% Appendix

\bibliographystyle{plain}
\bibliography{bibliography-forc.bib}
\appendix

\section{Extended Motivating Examples}
\label{appendix:extendedmotivation}
\subfile{sections/extended-motivating-examples.tex}
\clearpage % prevent weird table splitting
\section{Extended Details on Conditions}
\label{appendix:extendedconditionssuccess}
\subfile{sections/extended-conditionssuccess}

\section{Extended Details on Cohort Selection Mechanisms}
\label{appendix:existingmechs}
\subfile{sections/existingmechs.tex}

\section{Details for Section \ref{sec:mechanisms}}
\label{appendix:mechanismdetailed}
\subfile{sections/mechanisms-detailed.tex}

\section{Proofs for Section \ref{sec:modeldefinitions}}
\label{appendix:proofsmodel}
\subfile{sections/proofsmodel.tex}

\end{document}

%% file: sections/abstract.tex
\begin{abstract}
It is well understood that a system built from individually fair components may not itself be individually fair. In this work, we investigate individual fairness under pipeline composition. Pipelines differ from ordinary sequential or repeated composition in that individuals may drop out at any stage, 
and classification in subsequent stages may depend on the remaining ``cohort'' of individuals. As an example, a company might hire a team for a new project and at a later point promote the highest performer on the team. Unlike other 
repeated classification settings, where the degree of unfairness degrades gracefully over multiple fair steps, the degree of unfairness in pipelines can be arbitrary, even in a pipeline with just two stages.

Guided by a panoply of real-world examples, we provide a rigorous framework for evaluating different types of fairness guarantees for pipelines.  
We show that na\"{i}ve auditing is unable to uncover systematic unfairness and that, in order to ensure fairness, some form of dependence must exist between the design of algorithms at different stages in the pipeline. Finally, we provide constructions that permit flexibility at later stages, meaning that there is no need to lock in the entire pipeline at the time that the early stage is constructed.
\end{abstract}

%% file: sections/introduction.tex
As algorithms reach ever more deeply into our daily lives, there is increasing concern that they be {\em fair}.  
The study of the theory of algorithmic fairness was initiated by Dwork {\it et al.}~\cite{dwork2012fairness}, who introduced the solution concept of {\em individual fairness}.  Roughly speaking, individual fairness requires that similar individuals receive similar distributions on outcomes.  Dwork and Ilvento~\cite{DI2018} examined the behavior of individual fairness (and various group notions of fairness) under composition. % identifying two broad notions of composition that capture different real-life scenarios and showed ways in which fairness guarantees can fail to compose (and what can be done about this). 
They showed that although competitive composition, i.e. when two different tasks ``compete'' for individuals, can result in arbitrarily bad behavior under composition, fairness under simple repeated or sequential classifications (for the same task) degrades gracefully, similar to degradation of differential privacy loss under multiple computations.\footnote{Note that although fairness degrades gracefully in these scenarios, it does not rule out the existence of feedback loops which arbitrarily amplify unfairness, see e.g. \cite{HuC17,liu2018delayed}.}
%Nonetheless, fairness under simple sequential composition degrades gracefully, much in analogy to degredation of privacy under multiple computations. 
In this work we expand the investigation of individual fairness under sequential composition %(Dwork and Ilvento, ITCS 2019) 
to the case of {\em cohort pipelines}.  Cohort pipelines differ from ordinary sequential composition in that  %individuals may drop out of consideration at any stage.
each stage of the pipeline considers only the remaining cohort of individuals and may change its classification strategy conditioned on the set of individuals remaining.

% and not well characterized by the fairness properties of any single component in isolation.
Cohort pipelines are common: many data-driven systems consist of a sequence of cohort selection or filtering steps, followed by decision or scoring steps.  
%For example, a hiring pipeline might consist of an automated resume screening program followed by human resume review followed by interviews and a hiring decision. At each stage, remaining cohort is critical. For example, the interviewer's interview dates, etc. will be chosen to maximize the efficiency of the interview process for the set of candidates chosen in the resume review.
%\cnote{Want to say that because people can drop out later stages necessarily regard the universe as just the cohort.  This explains why we can have bad examples even in cases in which people don't drop out.}
A running exemplary scenario in this work will be a two-stage cohort pipeline: a company hires a team (cohort) of individuals to work on a  project and subsequently promotes the highest performer on the team to a leadership position. %\footnote{Necessarily some individuals will not be selected for the team, capturing that ``dropping out,'' for example, of the STEM pipeline, is not entirely voluntary.} 
Although the team selection may be fair in the sense that similarly qualified candidates have similar chances of being chosen for the team, the selection of the highest performer critically depends on the \textit{other members of the team}. %\inote{We need to add a sentence here justifying why the company should be allowed to adapt its policies and why comparison within the cohort should be OK.} Since the promotion stage has no access to the individuals not selected for the team, it must necessarily be {\em myopic}, in that its view of the universe of individuals is restricted to the cohort.
As we will see, being compared fairly to other members of the cohort in each stage doesn't imply fairness of the entire pipeline, as the competitive landscape can vary between similar individuals.  
%
%We refer to the team-formation stage as {\em cohort selection}~\cite{DI2018}.

Indeed, a fair cohort selection mechanism~\cite{DI2018} 
%policies for choosing the cohort that formally satisfy individual fairness, meaning that similarly qualified individuals have similar probabilities of being selected, 
can exploit the ``myopic'' nature of the promotion stage to skew overall fairness.  This can happen either through good intentions ({\it e.g.}, choosing teams so that members of a minority group always have a mentor on the team) or malice (\textit{e.g.}, ensuring that minority candidates are almost always paired with a more qualified majority candidate): in both these cases minorities suffer significantly reduced chances of promotion.%This can result in catastrophic unfairness under composition.
\footnote{See Appendix \ref{appendix:extendedmotivation} for additional examples.} 
Unlike  other repeated classification settings in which the degree of unfairness of multiple fair steps degrades gracefully, 
%As prior work \cite{DI2018} has shown, there is no guarantee that fair components will compose into a fair system. However, the case of cohort-based pipelines is of particular interest because unlike  other repeated classification settings in which the degree of unfairness of multiple fair steps is naturally bounded, 
the degree of unfairness in cohort pipelines can be arbitrary, even in a pipeline with just two stages. %Furthermore, we demonstrate that malicious exploitation of myopia of the later stages of the pipeline, under the guise of fairness, is not only feasible but straightforward.
Furthermore, we demonstrate that construction of  malicious pipelines under naïve auditing of fairness is straightforward and both computationally and practically feasible.

In this work we examine the subtle issues that arise in cohort-based pipelines, focusing on short pipelines consisting of a single cohort selection step followed by a scoring step. 
We formalize fairness desiderata capturing the issues unique to pipelines (not shared by ordinary sequential composition),
% and give a variety of motivating examples demonstrating how well-intentioned cohort selection and scoring policies can result in bad outcomes. %
and  give constructions for \textit{robust} cohort selection mechanisms that behave well under ({\it i.e.}, are robust to) pipeline composition with a variety of future scoring policies.  
%Thus, rather than prohibiting any competitive or comparative aspects of scoring functions or requiring monolithic pipeline designs, we recognize the reality of the need for decisionmaking pipelines that are constructed from components (potentially designed at different points in time by different stakeholders) that operate in the context of the cohort presented.
In particular, we demonstrate that it is possible to design cohort selection mechanisms that are robust to a rich family of subsequent scoring functions given a simple description of a {\em policy} governing the behavior of the family.\footnote{Formally, we can think of a policy as a description of a set of permitted scoring functions.}  This provides, for example, a means for enabling a company to choose an individually fair hiring procedure that will be robust to many possible compensation functions (all adhering to the policy) chosen at a later date. 
Guided by a panoply of real-world examples, this work provides a rigorous framework for evaluating and ensuring different types of fairness guarantees for pipelines.

We now summarize our contributions.
% \subsubsection*{Fair and robust cohort pipeline framework.}
First, we formalize what it means for the outcomes of a pipeline, which include both the outcome of the initial cohort selection step %$\Mech$
and the score %the scoring function  $f\in \mathcal{F}$
conditioned on being chosen, to be fair.\footnote{Bower et al. consider fairness in pipelines for a group-based definition of fairness, and primarily consider the accuracy of the final pipeline decision \cite{BowerKNSVV17}.} We then extend this fairness notion to describe how a cohort selection mechanism can be 
%and what it means for a cohort selection mechanism to be
\emph{robust} to a scoring policy, i.e. to compose fairly with any cohort scoring function chosen from a permissible set. 
%In particular, the cohort scoring function assigns a score to each individual in the cohort, and an individual's score may depend on the cohort "context".
Although the choice of scoring function may not depend on the cohort, the scores assigned to any individual may be highly dependent on their cohort ``context.''
 %As we will see in Section \ref{sec:modeldefinitions}, simply evaluating differences in expected score will not suffice to achieve important fairness desiderata.
% \subsubsection*{Sufficient conditions for robustness.}
Second, we determine how the scoring policy imposes conditions on the cohort selection mechanism. In particular, we show that there is a natural way to describe the set of cohort \emph{contexts} in which similar individuals %$(u,v)$ 
are treated similarly by all functions permitted by the policy, and we demonstrate that assigning similar individuals to similar \textit{distributions} over cohort contexts is sufficient (and sometimes necessary) to ensure pipeline robustness. %in the family in terms of a {\em mapping} between the set of cohorts containing $u$ and the set of cohorts containing $v$. 
% a pair of individuals. 
%We show that simple policies governing scoring functions can be expressed concisely as mappings.
%in which cohort contexts the family of functions treats a pair of individuals similarly by
%We then derive a set of conditions requiring that similar individuals be mapped to similar distributions over cohorts, using the mapping to specify which cohort contexts can be considered similar. 
%Second, we must understand how the family of scoring functions $\mathcal{F}$ imposes conditions on the way $\Mech$ chooses cohorts and determine a way of describing in which contexts $\mathcal{F}$ treats individuals similarly.
% \subsubsection*{Robust constructions for rich families of scoring functions.}
Third, we provide constructions for cohort selection mechanisms which are both robust to a rich set of practical scoring policies and permit flexibility in selection of the original cohort. %In particular, we demonstrate mechanisms that are robust to the family of scoring functions which have the property that their behavior doesn't change ``too much'' if a single individual in the cohort is replaced by another. %$\Mech$ which are robust for rich set of $\mathcal{F}$.

%% file: sections/new_model.tex
%In this section, we introduce the necessary background for individual fairness and the cohort selection problem, give motivating examples to illustrate the nuances of cohort-based pipelines and convert the intuition behind these examples into formal definitions for robustness in cohort-based pipelines.

\subsection{Preliminaries}
We base our model on individual fairness, as proposed in \cite{dwork2012fairness}. The intuition behind individual fairness is that ``similar individuals should be treated similarly.''
What constitutes similarity for a particular classification task is provided by a metric which captures society's best understanding of who is similar to whom.
%where similarity is encoded in a task specific metric.
%In this work, we will assume the metric is given. 
Below we formally define individual fairness as in \cite{dwork2012fairness} with a natural Lipschitz relaxation.
%Recall the definition of $\alpha$-Individual Fairness given in the Introduction:
\begin{definition}[$\alpha$-Individual Fairness  \cite{dwork2012fairness}]
\label{def:individualfairness} 
Given a universe of individuals $U$, and a metric $\D: U \times U \rightarrow [0,1]$ for a classification task with outcome set $O$, and a distance metric $d:\Delta(O) \times \Delta(O) \rightarrow [0,1]$ over distributions over outcomes, a randomized classifier $C: U \rightarrow \Delta(O)$  %$\left(C(u)\right)_{u \in U}$ 
is
$\alpha-$\textit{individually fair}
if and only if for all $u,v \in U$, $d(C(u),C(v)) \le \alpha \D(u,v)$.
\end{definition}
We use the phrase ``similar individuals are treated similarly'' as a shorthand for the individual fairness Lipschitz condition. 
%\noindent Individual fairness is a strong definition compared to others proposed in the literature, and although it is based on fairness constraints for individuals it is also a powerful tool for uncovering systemic discrimination.\footnote{Critically, it does not have the composition inconsistency problems of group or statistical definitions of fairness. See \cite{DI2018} for more details.}
Individual fairness was originally proposed in the context of independent classification, \textit{\textit{i.e.}} each individual is classified exactly once, independently of all others. However, in many practical settings individuals cannot be classified independently, particularly when there are a limited number of positive classifications available (\textit{e.g.} a university which can only accept a limited number of students each year, an advertiser with a limited budget).
Dwork and Ilvento formalized this problem as the ``cohort selection problem,'' in which a set of exactly $n$ individuals must be selected such that the probabilities of selection conform to individual fairness constraints \cite{DI2018}.
\begin{definition}[Cohort Selection Problem \cite{DI2018}]\label{def:basiccohort}
Given a universe $U$ of individuals, an integer $n$, and a task with metric $\D$, select a cohort $C \subseteq U$ of exactly $n$ individuals such that
$|\Pr[u \in C] - \Pr[v \in C]| \leq \D(u,v)$.
%the probability of selection is 1-Lipschitz with respect to $\D$, where the probability of selection is taken over all randomness in the system. We call a mechanism that satisfies these criteria an
We call such a mechanism an \textit{individually fair cohort selection mechanism}.
\end{definition}

\input{sections/terminologytable.tex}

Our work extends the investigation into fair composition by considering composition within a \textit{pipeline} of cohort selection and scoring steps.
We focus on the case of a two-step pipeline, and we assume for simplicity that the metric for the cohort selection and scoring functions are the same.%\footnote{We discuss generalization to longer pipelines in Appendix~\ref{appendix:generalizingmultiplecohortselectionsteps}.} 

\begin{definition}[Two-stage Cohort pipeline]\label{def:cohortpipeline}
Given a universe of individuals $U$, a two-stage cohort pipeline consists of: a set of permissible cohorts $\mathcal{C} \subseteq \prettypowerset (U)\backslash \emptyset$ (where $\prettypowerset(U)$ indicates the power set of $U$), a single (randomized) cohort selection mechanism $\Mech$ which outputs a single cohort $C \subseteq \mathcal{C}$, a set of scoring functions $\mathcal{F}: \mathcal{C}\times U \rightarrow [0,1]$, and a scoring function $f \in \mathcal{F}$.
The two-stage cohort pipeline procedure is $\Mech \circ f$.
\end{definition}

We now briefly introduce supporting terminology (summarized in Table \ref{tab:terminology}).
For $C \in \mathcal{C}$, let $\DistMech(C)$ denote the probability that $\Mech$ outputs  $C$, where the probability is over the randomness in the cohort selection mechanism~$\Mech$ operating on the universe~$U$.
%Notice that the probability that $\Mech$ selects a particular individual $u\in U$ maybe split between different cohorts. 
We denote the set of cohorts containing $u$ as $\mathcal{C}_u$, and the probability that $\Mech$ selects $u$ can be expressed $p(u)=\sum_{C \in \mathcal{C}_u}\DistMech(C)$. %(These terms are summarized in Table \ref{tab:terminology} for reference.)
As initial constraints on $\Mech$ and $\mathcal{F}$, we assume that $\Mech$ is an individually fair cohort selection mechanism and that each $f \in \mathcal{F}$ is individually fair within the cohort it observes, \textit{i.e.}, it is \textit{intra-cohort individually fair}: %(Note that $f$ may be designed at a later point in time by different individuals than the designers of $\Mech$, so it's not reasonable for $f$ to maintain the full context of $\Mech$ and $U$.)
%More formally, we define \textit{intra-cohort individual fairness:}
\begin{definition}[Intra-cohort individual fairness] Given a cohort $C$, a scoring function
 $f:\mathcal{C}\times U \rightarrow [0,1]$ is \textit{intra-cohort individually fair} if for all $C \in \mathcal{C}$, $\D(u,v) \ge |f(C, u) - f(C,v)|$ for all $u, v \in C$. %\in \mathcal{C}$.
\end{definition}
\noindent Although intra-cohort fairness constrains $f$ to be individually fair \textit{within} a particular cohort, $f(C_1, u)$ can differ arbitrarily from $f(C_2, u)$ if $C_1 \neq C_2$.  
%Consider a cohort $C$ and an individual $u \in C$. 
For ease of exposition we sometimes refer to $C$ as the ``cohort context'' or simply the ``context" of $u$ for $u \in C$.
%For ease of exposition we will often refer to an individual $u$ appearing in a ``cohort context'' to discuss how the cohort apart from $u$ impacts the treatment of $u$ or how swapping $u$ for a different individual $v$ changes outcomes. We use $\Context{C}\cup\{ u\}$ and $\Context{C} \cup \{v\}$ to indicate a context  $\Context{C} \in \{C \backslash \{x\} \mid C \in \mathcal{C}, x \in C \}$, where it is assumed $u,v \notin \Context{C}$.
%\cnote{I commented out the definition of the $\Context{C}$ notation. I stopped editing here.}
\begin{remark}[Intra-cohort individual fairness is insufficient.]\label{remark:nofreepipelines}
A pipeline consisting of an individually fair cohort selection mechanism and intra-cohort individually fair scoring function may result in arbitrarily unfair treatment. For example, suppose $\mathcal{X} = \{X_1,X_2,\ldots\}$ is a partition of $U$, and $\Mech$ chooses a cohort $X_i$ uniformly at random. Suppose $f$ assigns score $1$ to all members of the cohort corresponding to $X_*$, and otherwise assigns score 0. $\Mech$ is not only individually fair, it selects each element with an equal probability; $f$ is not only intra-cohort individually fair, it treats all members of a given cohort equally; yet the pipeline can result in arbitrarily large differences in scores for similar individuals. Furthermore, this observation holds for \textit{any} partition including adversarially chosen partitions. Although this abstract example suffices to prove the point, we include an extensive set of realistic pipeline examples, analogous to the ``Catalog of Evils'' of \cite{dwork2012fairness}, in Appendix \ref{appendix:extendedmotivation}. We also include a practical method for \textit{malicious} pipeline construction in Appendix \ref{appendix:existingmechs}.
\end{remark}

An important part of the pipeline definition is the contextual behavior of $f$, \textit{i.e.}, the behavior of the second stage of the pipeline may depend on the selected cohort $C$. % before it chooses which $f \in \mathcal{F}$ to apply. %Indeed, the second stage of the pipeline may know \textit{nothing} about $A$ other than its choice of $C$. 
%Roughly speaking, the problems with pipelines stem from the fact that later stages in the pipeline behave fairly within the context of the cohort they observe, but the initial cohort selection mechanism \textit{doesn't guarantee that similar individuals have similar probability of appearing with the same cohort complement.} One's first thought might be to design $f$ in lockstep with $\Mech$, or to require $f$ to adapt to an arbitrary choice of $\Mech$. Unfortunately,
%In practice, $f$ may be designed at a later point in time by an entirely different set of individuals who have no context for the original choice of $\Mech$.
% The key property of cohort pipelines which allows for this bad behavior is the fact that later steps operate only in the context of the cohorts presented to them, without the context of how they were selected. 
The simplistic solution to this problem is to design and evaluate the whole pipeline for fairness as a single unit, \textit{i.e.} requiring that similar individuals have similar distributions over $\Delta(\Opipeline)$. Although such evaluation would catch unfairness, it (1) doesn't provide explicit guidance for designing any given component, (2) may miss certain pipeline-specific fairness issues (see Examples \ref{example:conditional} and \ref{example:certainty}), and (3) ``locks'' the pipeline into a single monolithic strategy, which is highly impractical. For example, employers frequently need to change compensation policies due to changing market conditions. However, changing compensation policies due to disliking a particular member of a cohort, \textit{e.g.} switching to equal bonuses for all team members if the company does not like the individual who would have received the largest bonus, is not permitted in our model.
Indeed, later stages in the pipeline may be completely ignorant of the existence of prior stages, \textit{e.g.} a manager deciding on employee compensation may be unaware of automated resume screening.

This motivates our design goal of \textit{robustness}: designing the cohort selection mechanism $\Mech$ which composes well with \textit{every} function in $\mathcal{F}$, rather than expecting 
%We take the view that it is more practical to design the initial cohort selection steps to be robust to composition with a rich family of scoring functions than to expect
the scoring function to properly analyze and respond to the choices made in the original cohort selection mechanism design. As a result, the only communication necessary between the steps is the description of $\mathcal{F}$. %, \textit{i.e.}, the policy scoring functions are expected to follow in the future.
%For example, the designers of a hiring procedure might not know exactly how compensation will be decided at a later date, but they know their company has fundamental policies regarding compensation that won't change. 
%may know that their company's fundamental operating policies require that similarly qualified individuals receive similar compensation within a particular cohort. % Thus, our goal is to formalize what it means to design a hiring procedure that composes fairly with a given compensation \textit{policy} dictating the set of permissible scoring functions $\mathcal{F}$. % (that is, a collection of intra-cohort individually fair score functions $\mathcal{F}$ that, for example, may consist of compensation strategies of a particular type). %More specifically, we aim to design a cohort-selection mechanism $\Mech$ that is robust to composition with all $f \in \mathcal{F}$.
With this in mind, a deceptively(!) simple extension of Definition~\ref{def:individualfairness} gives our fairness desideratum for pipelines.  %Let $\Mech$ be a cohort selection mechanism that, given a universe of individuals, selects a cohort.  Let $\mathcal{F}$ be a family of scoring functions, each mapping pairs $(C,u)$, where $C$ is a cohort and $u$ is a member of the cohort, to a score in $[0,1]$.  

\begin{definition}[$\alpha$-Individual Fairness and Robustness for Pipelines (informal)]
Consider the pipeline consisting of $(\mathcal{C},\Mech,\mathcal{F})$, with outcome space $\Opipeline$. For $f \in \F$, the pipeline instantiated with $f$ satisfies $\alpha-$individual fairness with respect to the similarity metric $\D$ and a distance measure $d: \Delta(\Opipeline) \times \Delta(\Opipeline) \rightarrow [0,1]$ %for a single $f \in \mathcal{F}$ if 
if $\forall u,v \in U$,
$d([f \circ \Mech](u), [f \circ \Mech](v)) \le \alpha \D(u,v)$.

If the pipeline satisfies $\alpha-$individual fairness with respect to \textit{every} $f \in \mathcal{F}$, \textit{i.e.}, if 
$\forall f \in \mathcal{F}$ and
$\forall u,v \in U$,
$d([f \circ \Mech](u), [f \circ \Mech](v)) \le \alpha \D(u,v)$, 
we say that $\Mech$ is $\alpha-${\em robust} to $\mathcal{F}$ with respect to $d,\D$.
\end{definition}

We model the contextual nature of the problem by allowing the behavior of each $f\in \F$ to depend on the cohort, rather than allowing $f$ to be chosen adaptively in response to the selected cohort. This modeling choice still allows us to capture the contextual nature of scoring policies, while keeping our abstractions clean.\footnote{See Appendix \ref{appendix:extendedmotivation} for explicit examples of modeling adaptation to changing market conditions.}

\subsection{Fairness of pipelines}
Lurking in this informal definition are two  subtle choices critical to pipeline fairness: (1) how should distributions over $\Opipeline$ be interpreted, and (2) what distance measure $d$ is appropriate for measuring differences in distributions over $ \Opipeline$. 
In the remainder of this section, we consider these two questions and frame the notion of robustness parametrized by the two axes: distribution and distance measure over distributions.

%We devote the remainder of this section to formalizing our fairness requirements in considering what it means for similar individuals have similar distributions for pipeline outcomes.

%In particular, we must formalize what it means for similar individuals to have similar distributions over pipeline outcomes, as there are multiple components of the pipeline outcome, \textit{i.e.} ``selected'' or not and score conditioned on selection.
%There are two axes we must consider: (1) what is the appropriate distribution over outcomes, \textit{i.e.}, conditioned on being selected or not, and (2) how should we measure distances between distributions over outcomes.

\subsubsection{Choosing the interpretation of the distribution}
%First, we consider over which distributions to evaluate fairness.
%It's critical 
To account for the fact that individuals not selected by $\Mech$ never receive a score from $f$ the relevant outcome space is the union of possible scores and ``not selected,'' \textit{i.e.} $\Opipeline: = [0, 1]\cup \{\notselected\}$. Thus conditioning on whether an individual was selected or not changes the interpretation of the distribution over the outcome space and, more importantly, changes the \textit{perception} of fairness.

\begin{example}[Perception of conditional probability]\label{example:conditional}
Suppose Alice ($a$) and Bob ($b$) are similar but not equal job candidates, \textit{i.e.} $\D(a,b) \in (0,0.1]$. Consider an individually fair cohort selection mechanism, $\Mech$ which either selects a cohort containing one of Alice or Bob or neither and satisfies $p(a) = p(b) = p^*$. Consider the fairness constraint on the scoring function $f$ for the unconditional distribution over $\Opipeline$: 
$|p(a)f(a) - p(b)f(b))| \leq \D(a,b)$, which simplifies to $p^*|f(a) - f(b))| \leq \D(a,b)$.
(Note: as Alice and Bob never appeared together in a cohort, there is no intra-cohort fairness condition.) The constraint on the difference in treatment by $f$ is essentially diluted by a factor of $p^*$.  

Enforcing fairness on the unconditional distribution essentially allows the company to hand out job offers of the following form: ``Congratulations you are being offered a position at Acme Corp., you can expect a promotion after one year with probability $x\%$.'' Alice and Bob may \textit{receive} offers will equal probability, but the values of $x$ printed on the offer may be wildly different, and as such they will perceive the value of the job offer differently. 
\end{example}
% \begin{example}[Perception of conditional probability]\label{example:conditional}
% Suppose Alice and Bob are two similarly (but not identically) qualified job candidates. Alice has probability $p(a)$ of being hired and Bob has probability $p(b)$ of being hired. One year after being hired, Alice and Bob are each awarded a bonus such that the difference in their expected unconditioned bonuses over the randomness of the entire pipeline is equal to  $\D(a, b)$. However, conditioned on Alice (or Bob) having been hired, the difference in their bonuses could be much larger than $\D(a,b)$, as $|p(a)f(a)-p(b)f(b)| =\D(a,b)$ does not imply that $f(a)-f(b) \leq \D(a,b)$. From the perspective of Alice, Bob, and their teammates who have limited insight into the original hiring procedure and who likely consider probability of being hired one year prior to be irrelevant to the question of compensation in the present,
% % \mnote{I commented out a line here in an effort to trim a bit (see Tex file)}
% this creates the perception that Alice and Bob are being treated much more differently than their true difference in qualification should allow. %because the probability of either being hired is opaque and irrelevant to the question of compensation at a later date.
% By using the unconditioned distribution we allow the probability of being hired to ``dilute'' the constraint on the treatments conditioned on being hired.
% \end{example}
The choice of  conditional or unconditional distribution boils down to what perception of fairness is  important. In the case of bonuses or promotions  awarded long after hiring, the conditional perception may be particularly important. However, on shorter time frames or if the only consequential outcome is the final score, the unconditional distribution may be more appropriate (\textit{e.g.} resume screening immediately followed by interviews).\footnote{Although in this work we consider pipelines with a single relevant metric, the conditional versus unconditional question is critically important when metrics differ between stages of the pipeline. For example, the metric for selecting qualified members of a team may be different than the metric for choosing an individual from the team to be promoted to a management role, as the two stages in the pipeline require different skillsets.}
% save\footnote{We will later show that the constraints imposed by robustness w.r.t the unconditional distribution are weaker than those imposed by the conditional distribution.}
We consider two approaches which capture these different perspectives:
the \textbf{unconditional distribution} $\UncondDist{\Mech}{f}{u}$, treats the $\notselected$ outcome as a score of $0$
and the \textbf{conditional distribution} $\CondDist{\Mech}{f}{u}$
 conditions on $u$ being selected in the cohort. More formally:

%\cnote{Unconditional vs non-conditional: what is the relationship between these?  Purely from a notational point of view, we have $S^{N,A,f}$ for the non-conditional distribution but we write $d^{uncond,*}$ for the distributions that depend on these $S^{N,A,f}$.  Can we get to words that both begin with N or both begin with U?.  Along the same lines, perhaps we could use a mnemonic of the form N/C (or U/C) instead of Notion 1 and Notion 2, eg, Notion-U and Notion-C.}
\begin{definition}[Conditional and unconditional distributions]
\label{def:condandnoncond}
Let $\OriginalDist{\Mech}{f}{u} \in \Delta(O_{\text{pipeline}})$ be the distribution over outcomes arising from the pipeline, \textit{i.e.} $f \circ \Mech$. $\OriginalDist{\Mech}{f}{u}$ places a probability of $1 - p(u)$ on $\notselected$, and  for $s \in [0,1]$, $\OriginalDist{\Mech}{f}{u}$ places a probability of $\sum_{C \in \mathcal{C}} \Pr[f(C,u)=s]A(C)$ on $s$.
  \begin{itemize}[leftmargin=*]
    \item  The \textbf{unconditional distribution} $\UncondDist{\Mech}{f}{u}$ is identical to $\OriginalDist{\Mech}{f}{u}$ with the exception that it treats the $\notselected$ outcome as if it had score $0$. That is, for $0 < s \le 1$, $\UncondDist{\Mech}{f}{u}$ places a probability of $\sum_{C \in \mathcal{C} }\Pr[ f(C, u) = s] \DistMech(C)$ on $s$; at $s = 0$, $\UncondDist{\Mech}{f}{u}$ has a probability of $1-p(u) + \sum_{C \in \mathcal{C} } \Pr[ f(C, u) = 0]\DistMech(C)$.
    \item The  \textbf{conditional distribution} $\CondDist{\Mech}{f}{u}$ has probability $\frac{\sum_{C \in \mathcal{C} }\Pr[ f(C, u) = s]\DistMech(c)}{p(u)}$ for each score $s \in [0,1]$, \textit{i.e.}, it is $\OriginalDist{\Mech}{f}{u}$ conditioned on the positive outcome of $\Mech(C)$.\footnote{\label{pu:consideration} This definition is not defined if $p(u) = 0$, since it does not make sense to consider a ``conditional distribution'' if $u$ is never selected to be in the cohort (and thus never receives a score). In defining robustness of a cohort selection mechanism, we should thus  restrict to considering $u \in U$ where $p(u) > 0$ (and individual fairness of the cohort selection mechanism on its own would provide fairness guarantees over the probabilities $p(u)$). For simplicity, we do not explicitly mention this modification.}
    % \cnote{Add a footnote for the $p(u)=0$ case.}
  \end{itemize}
\end{definition}
\noindent Each of these approaches can be viewed as a method for converting a distribution $\OriginalDist{\Mech}{f}{u}$ over $\Opipeline$ to distributions $\CondDist{\Mech}{f}{u}$ and $\UncondDist{\Mech}{f}{u}$ over $[0,1]$. %Why is this helpful? It is much more intuitive from both a fairness and a technical perspective to define distance measures over distributions over $[0,1]$, which we will consider in the next section.  

\subsubsection{Distance measures over distributions}
%Now that we have enumerated our choices of distributions over $[0,1]$,
%we consider which distance measures over these distributions to use. 
The natural approach for measuring distances between distributions would be to use expectation: that is, $\DUncondE(\OriginalDist{\Mech}{f}{u}, \OriginalDist{\Mech}{f}{v}) := |\mathbb{E}[\UncondDist{\Mech}{f}{u}] - \mathbb{E}[\UncondDist{\Mech}{f}{v}]|$ and \\ $\DCondE(\OriginalDist{\Mech}{f}{u}, \OriginalDist{\Mech}{f}{v}):=|\mathbb{E}[\CondDist{\Mech}{f}{u}] - \mathbb{E}[[\CondDist{\Mech}{f}{v}]|$. %(Note: we adopt the convention of specifying distance measures over distributions in terms of individuals, with the specific distribution (either conditional or unconditional) and the distance measure over distributions noted in the superscript for ease of notation.)
Difference in expectation generally captures the unfairness in the examples discussed thus far. However, a subtle issue can arise from the \textit{certainty} of outcomes, which requires greater insight into the distribution of scores.

\begin{example}[Certainty of outcomes]\label{example:certainty}
 Consider two equally qualified job candidates, Charlie and Danielle. As these two candidates are equally qualified, they should clearly be offered jobs and promotions with equal probability. Recall the company's pleasant form letter for job offers from Example \ref{example:conditional}, ``Congratulations you are being offered a position at Acme Corp., you can expect a promotion after one year with probability $x\%$.''
 Danielle receives an offer with $x=70\%$ (with probability $p^*$), but Charlie receives either an offer with $x=100\%$ (with probability $0.7p^*$) or an offer with $x=0\%$ (with probability $0.3p^*$). Although both are offered jobs with equal probability and their expectations of promotion are equal, Charlie's offers have \textit{certainty} of promotion (or no promotion) whereas Danielle's promotion fate is uncertain. 
\end{example}

% \begin{example}[Certainty of outcomes]\label{example:certainty}
%   Consider two individuals, Charlie and Danielle, who are equally qualified. Whenever Charlie is hired, he is either hired onto a team where he is guaranteed promotion, or a team where he is guaranteed not to be promoted.
%   Danielle, on the other hand, is always hired into teams where promotion is uncertain. Even if Charlie and Danielle have equal probability of promotion over the randomness of hiring and promotion, Charlie has higher certainty about his outcome than Danielle does. This might inform Charlie's decisions to stay at the company or leave to try to find better opportunities for advancement. For example, Charlie may be assigned to either work on the company's flagship product or on code maintenance, whereas Danielle is assigned to work on a risky project with uncertain outcome.
% \end{example}
As Example \ref{example:certainty} illustrates,
 expected score does not entirely capture problems related to the \textit{distribution} of scores rather than the average score. Although total-variation distance is a natural choice for evaluating such distributional differences, it is too strong for this setting. %, since we do not want perturbations by a small score (\textit{i.e.} by $\D(u,v)$) to result in a large metric distance.
 For example, if Charlie receives a score of $0.7$ with probability $1$ (over randomness of the entire pipeline), while Danielle receives a score of $0.7 - \epsilon$ with probability $0.5$ and a score of $0.7 + \epsilon$ with probability $0.5$, then the total variation distance would be $1$, though these outcomes are intuitively very similar. We therefore introduce the notion of \textit{mass-moving distance} over probability measures. Mass-moving distance combines total variation distance with earthmover distance to reflect that similar individuals should receive similar distributions over close (rather than identical) sets of scores.
 %in order to require that similar people receive similar distributions over $[0,1]$ but still take into account that similar scores (\textit{e.g.} $0.75 + \epsilon$) in the support are still intuitively treated similarly. 
 %using the fact that there is a natural metric over the outcome space $[0,1]$ induced by $\ell_1$ distance (\textit{i.e.} difference).
%  \cnote{Revise definition so that all supports are finite.}
 \begin{definition}[Mass-moving distance]
 \label{def:mmd}
 Let $\gamma_1$ and $\gamma_2$ be probability mass functions over finite sets $\Omega_1 \subseteq [0,1]$ and $\Omega_2 \subseteq [0,1]$, respectively. Let $V \subseteq [0,1]$ be the set of real values $v \in [0,1]$ such that there exist probability mass functions $\tilde{\gamma}_1$ and $\tilde{\gamma}_2$ over $[0,1]$ with finite supports $\tilde{\Omega}_1$ and $\tilde{\Omega}_2$, respectively, where:
 \begin{enumerate}[leftmargin=*]
     \item \textit{Nothing moves far and mass is conserved.} \label{def:mmd:1} For $i = 1,2$, there is a function $Z_i: [0,1] \rightarrow \Delta(\tilde{\Omega}_i)$ such that: 
     \begin{enumerate}
         \item \textit{Nothing moves far.} For all $x \in [0,1]$ and $y \in \text{Supp}(Z_i(x))$, it holds that $|x - y| \le 0.5 v$.
         \item \textit{Mass is conserved.} For all $y \in \tilde{\Omega}_i$, it holds that $\tilde{\gamma}_i(y) = 
    \sum_{x \in \Omega_i} z^x_i(y) \gamma_i(x)$, where $z^x_i$ is the probability mass function of the distribution $Z_i(x)$. 
     \end{enumerate}
    \item \textit{Total variation distance is small.} \label{def:mmd:2} It holds that $0.5 v \ge TV(\tilde{\gamma}_1, \tilde{\gamma}_2) :=  \frac{1}{2} \sum_{w \in \tilde{\Omega}_1 \cup \tilde{\Omega}_2} |\tilde{\gamma_1}(w) - \tilde{\gamma}_2(w)|$. 
 \end{enumerate}
  Then we let $MMD(\gamma_1, \gamma_2) = \inf(V)$.

 \end{definition}
\noindent 
A simple way to think about  mass-moving distance is to break the definition down into two steps: (1) transforming the original distributions over scores into distributions over a single shared set of \textit{adjusted scores} and (2) moving mass between the distributions over adjusted scores. 

Since there is a natural association between probability distributions over $[0,1]$ and probability mass functions over $[0,1]$, Definition \ref{def:mmd} also gives a notion of distance between probability distributions.\footnote{We slightly abuse notation and use $MMD(\mathcal{X}_1, \mathcal{X}_2)$ for probability distributions $\mathcal{X}_1$ and $\mathcal{X}_2$, to denote $MMD(\gamma_1, \gamma_2)$ where $\gamma_1$ is the probability mass function associated to $\mathcal{X}_1$ and $\gamma_2$ is the probability mass function associated to $\mathcal{X}_2$.} 
%Notice that Condition (1) provides a total variation style constraint, and Condition (2) provides the flexibility to ``perturb mass'' by a small distance.
In the example of Charlie and Danielle receiving scores of $0.7$ or $0.7\pm \varepsilon$ described above, the mass-moving distance is at most $2 \epsilon$ since $\tilde{\gamma}_1$ and $\tilde{\gamma}_2$ can both be taken to be the probability measure that places the full mass of $1$ on $0.7$. 

Using mass-moving distance, we specify two additional  complementary distance measures: \\ $\DCondMMD(\OriginalDist{\Mech}{f}{u}, \OriginalDist{\Mech}{f}{v}) := MMD(\CondDist{\Mech}{f}{u}, \CondDist{\Mech}{f}{v})$ and $\DUncondMMD(\OriginalDist{\Mech}{f}{u},\OriginalDist{\Mech}{f}{v}) := MMD(\UncondDist{\Mech}{f}{u}, \UncondDist{\Mech}{f}{v})$.

\subsection{\emph{Robustly} fair pipelines}

% With this in mind, a deceptively(!) simple extension of Definition~\ref{def:individualfairness} gives our fairness desideratum for pipelines.  Let $\Mech$ be a cohort selection mechanism that, given a universe of individuals, selects a cohort.  Let $\mathcal{F}$ be a family of scoring functions, each mapping pairs $(C,u)$, where $C$ is a cohort and $u$ is a member of the cohort, to a score in $[0,1]$.  

% \begin{definition}[$\alpha$-Individual Fairness for Pipelines (informal)]
% The pipeline consisting of $(\Mech,\mathcal{F})$, with outcome space $\Opipeline$ satisfies individual fairness with respect to the similarity metric $\D$ and a distance $d: \Delta(\Opipeline) \times \Delta(\Opipeline) \rightarrow [0,1]$ if $\forall f \in \mathcal{F}$ and $\forall u,v \in U$,
% $d([f \circ \Mech](u), [f \circ \Mech](v)) \le \alpha \D(u,v)$.

% In this case we also say that $\Mech$ is {\em robust} to $\mathcal{F}$ with respect to $d,\D$.
% \end{definition}

%Interestingly, by more tightly controlling the class $\mathcal{F}$ of functions we can permit the mechanism $\Mech$ to be more creative, for example constructing cohorts with a balanced set of interests and skills.  The intuition is that $\Mech$ does not have to ``hedge its bets" so carefully if there is less freedom at later stages of the pipeline.}

%\subsection{Formalizing robustly fair pipelines}
Recall our informal notion that a cohort selection mechanism $\Mech$ is robust to a family of scoring functions $\mathcal{F}$ if the composition of $\Mech$ and any $f \in \mathcal{F}$ is individually fair.
We can now formalize robustness
as either \textbf{conditional} or \textbf{unconditional} with respect to either \textbf{expected score} or \textbf{mass moving distance} over score distributions. By evaluating the properties of each combination of distribution and distance measure, we can capture a range of subtle fairness desiderata in pipelines.\footnote{Note that these choices for $d$ are not the only possible choices, and the framework can be extended to different choices of distribution and distance measure to address other fairness concerns.}
% For convenience, we explicitly define four distance measures over pipeline outomces capturing the possible combinations of these parameters:
% \begin{enumerate}
%     \item $\DCondE(u,v) = |\E[\CondDist{\Mech}{f}{u}] - \E[\CondDist{\Mech}{f}{v}]|$
%     \item $\DUncondE(u,v) = |\E[\UncondDist{\Mech}{f}{u}] - \E[\UncondDist{\Mech}{f}{v}]|$
%     \item $\DCondMMD(u,v) = MMD(\CondDist{\Mech}{f}{u}, \CondDist{\Mech}{f}{v}])$
%     \item $\DUncondMMD(u,v) = MMD(\UncondDist{\Mech}{f}{u}, \UncondDist{\Mech}{f}{v}])$
% \end{enumerate}
% \noindent which we now use to formally specify our notion of robustness.
% %We now formalize this notion of robustness with two key properties purposely parameterized to express a range of important settings: (1) the distribution over outcomes and (2) the distance measure over distributions.
% %As we will see below, the choice distribution over outcomes, in particular whether to condition on an individual being selected by $\Mech$ is critical for understanding the \textit{perception} of fairness, and the distance measure over distributions is important in cases where \textit{certainty} impacts our understanding of fairness.
\begin{definition}[Robust pipeline fairness]
Given a universe $U$, a metric $\D$, %and a pipeline outcome space $\Opipeline$,
let $\Mech$ be an individually fair cohort-selection mechanism and let $\mathcal{F}$ be a collection of intra-cohort individually fair scoring functions $\mathcal{C} \times U \rightarrow [0,1]$. 
%let $\UncondDist{\Mech}{f}{u}$ (resp. $\UncondDist{\Mech}{f}{u}$)
Choose $d \in \{\DCondE, \DUncondE,$ $ \DCondMMD, \DUncondMMD\}$, a distance measure over $\OriginalDist{\Mech}{f}{u}$.
We say $\Mech$ is $\alpha$-\textbf{robust w.r.t $\mathcal{F}$} for $d$
 if $d(\OriginalDist{\Mech}{f}{u},\OriginalDist{\Mech}{f}{v}) \leq \alpha\D(u,v)$ for all $u,v \in U$ and for all $f \in \mathcal{F}$. 
\end{definition}

Throughout the rest of this work, we will examine robustness properties in terms of particular settings of $d$.
%A natural question to ask is how the choices of $d$ relate to each other. 
As one might expect, mass moving distance over score distributions is a stronger condition than expected score, %, as we show in Appendix \ref{appendix:proofsmodel}.\footnote{See Proposition \ref{prop:mmdimpliesexpectedscore}. 
% Specifically, we show that if $\gamma_1$ and $\gamma_2$ are probability measures corresponding to distributions $\mathcal{H}_1$ and $\mathcal{H}_2$ and have a mass moving distance of $d$, then $|\mathbb{E}[\mathcal{H}_1] - \mathbb{E}[\mathcal{H}_2]| \le 3d$.}
 and %Moreover, we show in  that the
 conditional robustness implies unconditional robustness up to a Lipschitz relaxation.\footnote{See Propositions \ref{prop:mmdimpliesexpectedscore} and   \ref{prop:implicationexpectedscore}. Interestingly, we show in Theorem \ref{thm:necessarydefexpectedscore} that for some classes of score functions, guaranteeing individual fairness w.r.t mass-mover distance fairness is ``equivalent'' to guaranteeing individual fairness w.r.t expected score.}

%% file: sections/terminologytable.tex
% \cnote{Add $\delta^{\cal F}$} \inote{I'm adding a separate table for mapping terminology See Table \ref{tab:mappingterminology}. We can combine these, but I thought it would be easier to wait to introduce the other terminology in section 3.}
% \cnote{Latex artifact.  this was supposed to appear with the table.  I meant: add $\delta^{\cal F}$ to the tabler.}
\begin{table}[]
        \centering
        
\begin{tabular}{|p{0.2\textwidth}|p{0.8\textwidth}|}
\hline 
 Term & Definition \\
\hline 
 $U $ & The universe of individuals \\
 \hline 
 $\D: U \times U \rightarrow [0,1] $ & The individual fairness metric \\
\hline 
 $\mathcal{C} \subseteq \prettypowerset(U)\backslash \emptyset$ & The set of permissible cohorts \\
\hline 
 $\displaystyle \mathcal{F} $ & The family of permitted scoring functions.  \\
\hline 
 $\displaystyle f:\mathcal{C} \times U \rightarrow [0,1]$ & A scoring function.  $f(C,x)$ is {\em undefined} whenever $x \notin C$, and throughout this work, whenever we write $f(C,x)$, where $x$ is any element in $U$, it is the case that $x \in C$. \\
\hline 
 $\Mech : U \rightarrow \mathcal{C}$ & An individually fair cohort selection mechanism.\\
\hline 
 $\DistMech(C)\in [0,1]$ & The probability that $\Mech$ outputs the cohort $C$.\\
\hline 
 $\mathcal{C}_u \subseteq \mathcal{C}$& The subset of permissible cohorts containing $u$.\\
 \hline 
 $p(u) \in [0,1]$& The probability $\Mech$ outputs a cohort containing $U$.\\
\hline 

%   & Conditional distribution over pipeline outcomes \\
% \hline 
%   & Unconditional distribution over pipeline outcomes \\
% \hline 
%   & Mass-moving distance measure over pipeline outcomes \\
% \hline 
%   & Expected Score distance measure over pipeline outcomes \\
%  \hline
\end{tabular}
        \caption{Terminology }\label{tab:terminology}
        \end{table}

%% file: sections/conditions.tex
In this section, we describe conditions on $\Mech$ that will result in our desired robustness properties with respect to a class of scoring functions $\mathcal{F}$. 
We first consider the description of $\F$ available to $\Mech$, \textit{i.e.} the policy.
The simplest method of specifying the policy by describing all $f \in \F$ prohibits adding $f$ with similar or identical fairness properties to $\F$ at a later point and is highly unrealistic (and potentially intractable). 
%is both highly unrealistic and potentially intractable. It also prohibits adding $f$ with similar or identical fairness properties to $\F$ at a later point. 
In practice, we expect policies to govern how differently $f$ can treat individuals within different contexts, rather than enumerating the permitted functions. %, \textit{i.e.}, that the policy restricts the difference in $f(C_1,u)$ and $f(C_2,v)$.
To that end, we propose policies in the form of 
%has two drawbacks: (1) if $\F$ is large, it may be intractable to communicate the full specification, and (2) it disallows adding $f^*$ to $\F$ at a later date which has ``morally equivalent'' behavior to other $f \in \F$. As we will see below, 
%A natural option would be to give $\Mech$ access to $\F$ in order to carry out composition planning. However, this is undesirable for a number of reasons. First, if the class of scoring functions in $\F$ is large, it could be intractable to communicate the full specification of every function in $f \in \F$.
%Second, if  an additional scoring function is added to $\F$ that is very ``similar'' to the some of the existing scoring functions in $\F$, this would require additional communication that is intuitively unnecessary. Third, to achieve robustness, 
%$\Mech$ does not need access to the specific values of functions $f \in \F$, but rather a description of %, only to the differences of
%how differently $f$ behaves on different (cohort, individual) pairs. 
%For these reasons, we propose that $\Mech$ instead be provided with a summary of $\F$ in the form of
a \textit{distance function over (cohort, individual) pairs}, $\DS: (\mathcal{C} \times U) \times (\mathcal{C} \times U) \rightarrow [0,1]$. %\footnote{Like with score functions, we say that $\DS((C_1,u), (C_2,v))$ is undefined if $u \not\in C_1$ or $v\not\in C_2$.} 
This distance function specifies the maximum difference in score between two (cohort, individual) pairs $\DS((C_1, u), (C_2, v)) := \sup_{f \in \F} |f(C_1, u) - f(C_2, v)|$.
$\DS$ captures the salient fairness behavior of the family of scoring functions, while being succinct in comparison to maintaining a list of all supported $f$ directly. In fact, as we will show in Lemma \ref{lemma:mappingreduction}, a partial description or an overestimate of $\DS$ will also suffice. %As a result, $\DS$ serves as the shared information between the designers of the cohort selection mechanism and the designers of the compensation scheme.
% $\DS$ heavily influences the strength of conditions on $A$, and so composition planning for the cohort selection mechanism needs to be tailored. 
To illustrate our policy descriptions, consider the following two families:
\begin{enumerate}[leftmargin=*]
    \item $\F_1$ ignores the cohort context entirely, and treats each $u\in U$ the same regardless of the cohort, \textit{i.e.} $\F_1=\left\{f \mid \exists g: U \rightarrow [0,1] \text{ s.t. } f(C, u) = g(u) \text{ for all } (C,u) \in \mathcal{C} \times U \right\}$.
    \item $\F_2$ treats $u$ and $v$ similarly within the same context, but has no constraint on treatment in different contexts, \textit{i.e.} 
 $\F_2 = \left\{f \mid f((C\backslash \{u\})\cup \{v\},v) - f(C,u)| \leq \D(u,v) \text{ for all }u,v \in U \text{ and }\forall C \in \mathcal{C} \text{ s.t. }u \in C, v\notin C\right\}.$
 
\end{enumerate}
Recall that intra-cohort individual fairness requires that the scoring functions in both families must treat $u$ and $v$ similarly if they appear in the same cohort, \textit{i.e.} $\D(u,v)\geq |f(C,u)-f(C,v)|$.

For the family $\F_1$, we observe that $\DSone ((C_1,u),(C_2,v)) =  \D(u,v)$, and, intuitively, the designers of $\Mech$ will not need to consider the behavior of $\F$ in their design of $\Mech$. %require composition planning beyond achieving the basic individual fairness guarantees of individually fair cohort selection.
On the other hand, for $\F_2$, we observe that $\DStwo((C,u),(C,v)) = \D(u,v)$ for any cohort $C$, but $\DStwo((C,u),(C',v))$ can be much greater than $\D(u,v)$ for $C' \neq C$. For this reason, composition planning for $\Mech$ is non-trivial. As one would expect, $\DS$ heavily influences the strength of conditions on $A$. %In the next subsection, we formalize and analyze the conditions imposed on $\Mech$ by a general distance function $\DS$. 

%A natural question to ask is: does $\Mech$ require knowledge of the exact behavior of $\DS$ in order to achieve robustness with respect to $\F$? We show that $\Mech$ can carry out composition planning with a summary of $\F$ which overestimates $\DS$: that is, a distance function $\dprime: (\mathcal{C} \times U) \times (\mathcal{C} \times U) \rightarrow [0,1]$ (not necessarily a metric) that provides an upper bound for $\DS$ on all pairs of cohort contexts. In particular, if the behavior of differences of scores is only known on a subset of $\mathcal{C} \times U$, this partial knowledge can easily be completed to a distance function on $\mathcal{C} \times U$ which overestimates $\DS$, by taking the unknown distances to be the maximal value of $1$. In Lemma \ref{lemma:mappingreduction}, we show in the next subsection that providing $\Mech$ with a distance function that overestimates $\DS$ will still guarantee robustness with respect to $\F$.

\subsection{$A$'s Task: Designing Mechanisms Compatible with $\DS$}\label{subsec:procedure}
We now describe how to design $A$ to guarantee robustness with respect to $\F$, given (possibly overestimates of) the distance function $\DS$ over (cohort, individual) pairs describing $\F$. %Speaking intuitively, given the stability in treatment between two individuals in the same cohort (guaranteed by the intra-cohort individual fairness of $f$), provided similar individuals have similar probabilities of appearing in each cohort, robustness properties should follow. In this context, 
The conditions on $\Mech$ will roughly consist of  making sure that $\Mech$ assigns \textit{similar individuals to similar distributions over cohort contexts}, where similarity of (cohort, individual) pairs is defined with respect to $\DS$. 
% \inote{Please see the below text about $\DS$ as a policy.} \mnote{Added something explicit about this being the only communication above and reorganized a bit.} 

%However, working with $\DS$ directly for this step can still be cumbersome and unintuitive. %\footnote{After defining mappings, in Section \ref{sec:mechanisms}, we give an example of a function family $\F_3$, where thinking about mappings can be much more intuitive than $d^{\mathcal{F}_3}$.} 
%Instead, we translate $\DS$ into a ``mapping'', which is a collection of simpler objects.
Although $\DS$ is a succinct description of a policy, it is more intuitive when designing with composition in mind to translate $\DS$ into a set of ``mappings'' specifying which (cohort, individual) pairs will be treated similarly by $f \in \F$. That is, for each pair $u,v\in U$, we can describe $\DS$ as a partitioning $\mathcal{P}_{u,v}$ of $(\mathcal{C}_u \times u )\cup (\mathcal{C}_v \times v) $ such that each partition or ``cluster'' has small diameter with respect $\DS$, \textit{i.e.} within a cluster $\DS((C_1,u),(C_2,v)) \leq \D(u,v)$. 
%Let  $\mathcal{C}_u$ (resp. $\mathcal{C}_v$) be the set of cohorts containing $u$ (resp. $v$). 
%Roughly speaking, for each pair of individuals $u$ and $v$, we consider the set of (cohort, individual) pairs involving $u$ and $v$, and construct a partition of this set into clusters with small diameter with respect to $\DS$ (that is, where the diameter is at most $\D(u,v)$).
The collection of partitions over all pairs of individuals then defines the mapping. % More formally, for a distance function $\dstar$ over cohort contexts: 

\begin{definition}[Mapping based on $\dstar$]
\label{def:mapping}
%$\forall x \in U$, let $\mathcal{C}_x\subseteq \mathcal{C}$ denote the subset of permissible cohorts containing $x \in U$. 
For each pair of distinct individuals $u$ and $v$, consider the subset $\mathcal{P}_{u,v} := (\mathcal{C}_u \times \left\{u\right\}) \cup (\mathcal{C}_v \times \left\{v\right\})$ of (cohort, individual) pairs. Consider a partition of $\mathcal{P}_{u,v}$ into clusters that respects $\dstar$, \textit{i.e.} that satisfies the following condition: if $(C_1, x), (C_2, y)$ are in the same cluster\footnote{Note that $x,y \in \{u,v\}$. Recall that $(C_1,u)$ and $(C_2,u)$ may appear in the same cluster, and thus it is possible that $x=y$.}, then $\dstar((C_1, x), (C_2, y)) \le \D(u,v)$.  Let $n_{u,v}$ (and $n_{v,u}$) be the number of clusters of the partition. We call a collection of such partitions for each pair $u, v \neq U$ a \textbf{mapping} of $\mathcal{C}$ that \textbf{respects $\dstar$}.
\end{definition}
Mappings interact well with distance functions $\dprime$ that overestimate $\DS$, as larger distances between (cohort, individual) pairs imposes more strict conditions on cluster membership. Lemma \ref{lemma:mappingreduction} states that a mapping that respects $\dprime$ will also respect $\DS$, although the resulting conditions on the mapping might be more restrictive. % than if $\Mech$ instead used $\DS$.
\begin{lemma}
\label{lemma:mappingreduction}
Let $\dprime: (\mathcal{C} \times U) \times (\mathcal{C} \times U) \rightarrow [0,1]$ be a distance function. Suppose that $\dprime$ has the property that for all pairs of cohort contexts $(C_1, x), (C_2, y) \in \mathcal{C} \times U$, it holds that $\dprime((C_1, x), (C_2, y)) \ge \DS((C_1, x), (C_2, y))$. If a mapping respects $\dprime$, then the mapping also respects $\DS$. 
\end{lemma}
\begin{proof}
Consider any pair of individuals $u$ and $v$, and consider any mapping that respects $\dprime$. In the partition corresponding to $u$ and $v$, if $(C_1, x)$ and $(C_2, y)$ are in the same cluster, then it holds that $\DS((C_1, x), (C_2,y)) \le \dprime((C_1, x), (C_2,y)) \le \D(u,v)$. Thus, the mapping respects $\DS$, as desired.
\end{proof}
%\begin{remark}
We now briefly introduce supporting terminology for policies and mappings (summarized in Table \ref{tab:mappingterminology}).
To succinctly refer to the clusters in a mapping, we define label functions
$\Map{u}{v}: \mathcal{C}_u\rightarrow \mathbb{N}$ and $\Map{v}{u}: \mathcal{C}_v\rightarrow \mathbb{N}$ such that $\Map{u}{v}(C)$ is the label of the cluster containing $(C,u)$ and $\Map{v}{u}(C)$ is the label of the cluster containing $(C,v)$. 
We use $n_{u,v}$ (or $n_{v,u}$) to denote the number of clusters in a mapping.
We also refer to the set of functions $(\Map{u}{v})_{u \neq v \in U}$, which entirely specify the partitions, as a mapping.
% In order to refer to the clusters in a mapping, we label the clusters in each partition $1, \ldots, n_{u,v}$. We then define the label functions $\Map{u}{v}: \mathcal{C}_u\rightarrow \mathbb{N}$ and $\Map{u}{v}: \mathcal{C}_v\rightarrow \mathbb{N}$ such that $\Map{u}{v}(C)$ is the label of the cluster containing $(C,u)$ and $\Map{v}{u}(C)$ is the label of the cluster containing $(C,v)$.  
% We also refer to the set of functions $(\Map{u}{v})_{u \neq v \in U}$, which entirely specify the partitions, as a mapping.
%\end{remark}
%\begin{remark}
Valid mappings for $\dstar$ are not necessarily unique, as there may be more than one way to partition $\mathcal{P}_{u,v}$ into clusters with diameter bounded by $\D(u,v)$. We let $\Mapping_{\dstar}$ be the set of mappings that respect $\dstar$. 
%\end{remark}
\input{sections/mappingterminologytable}

Given a mapping of $\DS$ (or of an overestimate $\dprime$), we can now interpret ``distributions over cohorts'' induced by $A$ as ``distributions over clusters'' induced by $A$. %As before, let $\DistMech$ be the probability distribution over $\mathcal{C}$ induced by $\Mech$.
Formally, we convert the distributions over cohorts into measures over $[\MapNum{u}{v}]$ for each pair $(u,v) \in U \times U$. 
As a result, ``similar distributions over cohorts'' will turn out to mean ``similar measures over $[\MapNum{u}{v}]$.'' % \inote{Consider cutting this sentence}

%We first define the two measures on $[\MapNum{u}{v}]$:

\begin{definition}
\label{def:measures}
Let $(\Map{u}{v})_{u \neq v \in U}$ be a mapping of $\mathcal{C}$. For $u, v \in U$, we define measures $q^1_{u,v}$ and $q^2_{u,v}$ over the sample space $[\MapNum{u}{v}]$ as follows: 
\begin{enumerate}[leftmargin=*]
    \item The \textbf{unconditional measure over cohorts $q^1_{u,v}$} on the sample space $[\MapNum{u}{v}]$ for each $(u,v)$ ordered pair is defined as follows. For $ i \in[\MapNum{u}{v}]$, we let $q^1_{u,v}(i) = \sum_{C \in \mathcal{C}_u \mid \Map{u}{v}(C) = i} \DistMech(C)$.\footnote{This is not necessarily a probability measure, since the total sum on the sample space is $p(u) \le 1$, but it is finite.}\label{def:measures:1}
    \item The \textbf{conditional measure over cohorts $q^2_{u,v}$} on the sample space $[\MapNum{u}{v}]$ for each $(u,v)$ ordered pair is defined as follows. For $ i \in [\MapNum{u}{v}]$, we let $q^2_{u,v}(i) = \frac{\sum_{C \in \mathcal{C}_u \mid \Map{u}{v}(C) = i} \DistMech(C)}{p(u)}$.\footnote{Observe that this is in fact a probability measure since $p(u) = \sum_{C \in \mathcal{C}_u} \DistMech(C) = \sum_{i=1}^{M_{u,v}} \sum_{C \in \mathcal{C}_u \mid \Map{u}{v}(C) = i} \DistMech(C)$.}\footnote{Like in Definition \ref{def:condandnoncond}, this definition is not defined if $p(u) = 0$, since it does not make sense to consider a ``conditional distribution'' if $u$ is never selected to be in the cohort (and thus never receives a score). We should thus  restrict to considering $u \in U$ where $p(u) > 0$ (and individual fairness of the cohort selection mechanism on its own would provide fairness guarantees over the probabilities $p(u)$). For simplicity, in this extended abstract, we do not explicitly mention this modification.}\label{def:measures:2}
\end{enumerate}
\end{definition}

We now specify sufficient conditions for robustness in terms of distances between these measures over $[n_{u,v}]$. The conditions require that for each pair $u, v \in U$, %and each cluster $1 \le i \le n_{u,v}$ in the mapping,
$\Mech$ assigns similar probabilities to cohorts containing $u$ and cohorts containing $v$ within each cluster.
%$\Mech$ places similar total probability on cohorts corresponding to $u$ in cluster $i$ and on cohorts corresponding to $v$ in cluster $i$.

%Roughly speaking, we allow $\Mech$ to distribute probability mass in any way it wants between cohorts mapped to the same cluster, but we restrict how the probability mass can be distributed between different clusters. Our conditions will essentially require that for each pair of individuals $u, v \in U$, and each cluster $1 \le i \le n_{u,v}$ in the mapping, $A$ places similar total probability on cohorts corresponding to $u$ in cluster $i$ and on cohorts corresponding to $v$ in cluster $i$. %That is, $A$ places similar total probability on $\left\{C \mid (C, u) \in \text{ cluster i } \right\}$ and on $\left\{C \mid (C, v) \in \text{ cluster i } \right\}$.  

 %\noindent Now, we specify conditions on $\Mech$ using the total-variation distances of these new measures over $[n_{u,v}]$. Similar to our robustness conditions, we specify requirements for both the conditional and unconditional robustness variants.\footnote{We also show in Appendix \ref{appendix:extendedconditionssuccess} that Notion 2 is a stronger condition than Notion 1. See Proposition \ref{prop:implicationnotions}.}
\begin{definition}[$\alpha$-Notions 1 and 2]
\label{def:conditions}
Let $(\Map{u}{v})_{u \neq v \in U}$ be a mapping of $\mathcal{C}$. For $u, v \in U$, let $q^1_{u,v}$ and $q^2_{u,v}$ be defined as in Definition \ref{def:measures}. We define $\alpha$-Notions 1 and 2 as follows:
\begin{enumerate}[leftmargin=*]
    \item For $\alpha \ge 0.5$, we say that $A$ satisfies $\alpha$-\textbf{Notion 1} if for all $u, v \in U$ such that $\D(u,v ) < 1$, $TV(q^1_{u,v}, q^1_{v,u}) \le (\alpha - 0.5) \D(u,v)$.  (The $0.5$ arising in Notion 1 comes from having to ``smooth out'' $q^1_{u,v}$ to a probability measure in a later step.)  \label{def:conditions:notion1}
    \item For $\alpha \ge 0$, we say that $A$ satisfies $\alpha$-\textbf{Notion 2} if for all $u, v \in U$ such that $\D(u,v ) < 1$, $TV(q^2_{u,v}, q^2_{v,u}) \le \alpha \D(u,v)$. \label{def:conditions:notion2}
\end{enumerate}
\end{definition}

Our main result is that these conditions guarantee pipeline robustness for composition with any $f\in \mathcal{F}$ with respect to mass-moving distance (and thus expected score).\footnote{See Corollary \ref{cor:robustnessexpectedscore} for a formal statement of the relationship between MMD and expected score.}
%That is, so long as $\Mech$ selects a mapping that respects $\DS$ and satisfies the conditions in Definition \ref{def:conditions} for this mapping,  the composition of $\Mech$ with any scoring function in $\F$ will be individually fair with respect to the relevant distribution and score measure. %As previously discussed, since mass-moving distance is a stronger condition than expected score, this automatically gives robustness for expected scores.\footnote{See Corollary \ref{cor:robustnessexpectedscore} for a formal statement.} 
Theorem \ref{thm:manytoonemapping} states that so long as $\Mech$ satisfies Notion 1 (resp. 2) for the mappings associated with $\mathcal{F}$, then $\Mech$ will be robust with respect to $\mathcal{F}$. % \inote{Consider cutting this sentence, somewhat redundant.}
\begin{theorem}[Robustness to Post-Processing]
\label{thm:manytoonemapping}
Let $\F$ be a class of scoring functions, let $\alpha \ge 0.5$ be a constant. Suppose that $\left(\Map{u}{v}\right)_{u \neq v \in U}$ is in $\Mapping_{\frac{1}{2 \alpha}\DS}$. If $\Mech$ is individually fair and satisfies $\alpha$-Notion 1 (resp. $\alpha$-Notion 2) for $\left(\Map{u}{v}\right)_{u \neq v \in U}$, then $\Mech$ is $2\alpha$-robust w.r.t. $\F$ for $\DUncondMMD$ (resp. $\DCondMMD$). 
\end{theorem}
\noindent
The proof of Theorem \ref{thm:manytoonemapping} appears in Appendix \ref{appendix:proofsconditionssuccess}.

Furthermore, these conditions are necessary both for mass-moving distance and the weaker condition of expected score for certain rich classes of scoring functions. % and is even necessary for the weaker condition of expected score for certain classes of scoring functions.
%A natural question is whether Theorem \ref{thm:manytoonemapping} is ``tight'', \textit{i.e.} whether there are weaker conditions that would also guarantee robustness. We show that our definition is necessary for mass-moving distance for certain rich classes of scoring functions and is even necessary for the weaker condition of expected score for certain classes of scoring functions.
\begin{theorem}[Informal]
\label{thm:necessary}
Let $d$ be any metric in $\left\{\DUncondMMD, \DCondMMD, \DCondE, \DUncondE\right\}$. %Under certain conditions on $\F$,
Loosely speaking, given $\F$ described by mappings such that inter-cluster distances are much larger than intra-cluster distances, 
the requirements on $\Mech$ in Theorem \ref{thm:manytoonemapping} are \textbf{necessary} for achieving robustness w.r.t. $d$. 
\end{theorem} %\inote{I think we want intuition for what these classes look like.}
\noindent We formalize Theorem \ref{thm:necessary} in Appendix \ref{appendix:extendedconditionssuccess}.\footnote{See Theorem \ref{thm:necessarydefmmd} and Theorem \ref{thm:necessarydefexpectedscore}.}
% The intuition for why our conditions are necessary is that in there are cases in which there is only one possible collection of mappings for $\DS$, and in such cases cohort selection mechanisms which deviate from the conditions must necessarily assign similar individuals to different sets of cohorts. It is then straightforward to construct a scoring function which takes advantage of the different cohort assignments for similar individuals, violating individual fairness. \mnote{moved to Appendix}

%% file: sections/mappingterminologytable.tex
\begin{table}[]
        \centering
        
\begin{tabular}{|p{0.2\textwidth}|p{0.8\textwidth}|}
\hline 
 Term & Definition \\
\hline 
$\DS:(\mathcal{C} \times U) \times (\mathcal{C} \times U) \rightarrow [0,1]$. & distance function specifying the maximum difference in treatment between (cohort,individual) pairs by any $f\in \F$. $\DS((C_1,u), (C_2,v))$ is undefined if $u \not\in C_1$ or $v\not\in C_2$.\\
\hline 
% $\mathcal{P}_{u,v}$ & \\
% \hline
$ \Map{u}{v}: \mathcal{C}_u\rightarrow \mathbb{N}$ & a mapping of the cohorts containing $u$ to clusters containing $(C,u)$. \\
\hline 

$n_{u,v}$ & The number of clusters in a mapping \\
\hline 
$\Mapping_{\dstar}$ & the set of all mappings which respect $\dstar$.\\
\hline 
%   & Conditional distribution over pipeline outcomes \\
% \hline 
%   & Unconditional distribution over pipeline outcomes \\
% \hline 
%   & Mass-moving distance measure over pipeline outcomes \\
% \hline 
%   & Expected Score distance measure over pipeline outcomes \\
%  \hline
\end{tabular}
        \caption{Policy and mapping terminology }\label{tab:mappingterminology}
        \end{table}

%% file: sections/mechanisms.tex
Although the conditions specified in the previous section are quite strict, and indeed some pathological scoring function families admit no robust solutions, we can nonetheless construct robust cohort selection mechanisms for rich classes of scoring policies.\footnote{See Appendix \ref{appendix:pathologicalexample} for an example of $\F$ which admits no robust $A$.} 
We exhibit mechanisms robust to two broad classes of policies:
\begin{enumerate}[leftmargin=*]
    \item \textbf{Individual interchangeability:} replacing a single individual in the cohort does not change treatment of the cohort too much, \textit{i.e.} policies like $\DStwo$.
    \item \textbf{Quality-based treatment:} cohorts with similar quality ``profiles'' are treated similarly. That is, the scoring function only considers the set of qualifications represented within a cohort and is agnostic to the specific individual(s) exhibiting a given qualification.
\end{enumerate}
These policies cover a wide range of realistic scenarios and allow for significant flexibility and adaptability in the choice of $f$. In this section, we demonstrate that these policies also admit a variety of efficient and \textit{expressive} constructions for $\Mech$, \textit{i.e.} $\Mech$ that may assign a wide range of probabilities $p(u)$ to individuals.

\begin{remark}
As previously noted, robustness is trivial for the class of scoring functions which ignore the cohort context ($\F_1$). We formalize this observation in the following proposition:
\begin{proposition}
\label{prop:easiestmech}
Consider the mapping that, for each pair of individuals $u$ and $v$, places all of the cohort contexts in $(\mathcal{C}_u \times \left\{u\right\}) \cup (\mathcal{C}_v \times \left\{v\right\})$ into the same cluster. If $\Mech$ is individually fair, then $\Mech$ satisfies $0.5$-Notion 1 and $0.5$-Notion 2 w.r.t. this mapping. 
\end{proposition}
\end{remark}

\subsection{Individual interchangeability}\label{subsec:F2}
% $\F_2$ is formally defined as the following subset of all intra-cohort individually fair scoring functions $\F_I$:

To describe the interchangeability policy, we specify a distance function $\DSint: (\mathcal{C} \times U) \times (\mathcal{C} \times U) \rightarrow [0,1]$ that requires that ``swapping'' any individual in a cohort does not result in significantly different treatment. More formally: %overestimates $\DStwo$, but still preserves enough of the fairness structure to construct the desired mapping. This distance function $\dprime$ is defined as follows: 
\begin{definition}[Individual interchangeability policy]
\[ 
\DSint((C, u), (C',v)) = 
\begin{cases}
\D(u,v) & \text{ if } C = C' \\
\D(u,v) & \text{ if } C' = (C \setminus \left\{u\right\}) \cup \left\{v\right\}. \\
1 & \text{ otherwise}.
\end{cases}
\]
\end{definition}

\noindent 
$\DSint$ can be viewed as an overestimate of $\DStwo$, or as a partial specification of the distance function on a subset of $(\mathcal{C} \times U) \times (\mathcal{C} \times U)$, trivially completed to $1$ on other pairs of cohort context pairs. %As expected, $\dprime$ does not take advantage of the true differences between scores for some pairs of cohort contexts. %In particular, given a cohort $C$ that contains $u$ and $w$ but does not contain $v$, we know that $|f((C\backslash \{u\}) \cup \{v\},w) - f(C,w)| \le |f((C\backslash \{u\}) \cup \{v\},w) - f((C\backslash \{u\}) \cup \{v\},v)| + |f((C\backslash \{u\}) \cup \{v\},v) - f(C,u)| + |f(C,u) - f(C,w)| \le \D(u,v) + \D(u, w) + \D(w, v)$, which might be much less than $1$, even though $\dprime((C\backslash \{u\} \cup \{v\},w), (C,w)) := 1$. 
%
% To formalize the policy for individual interchangeability, recall the example function family $\F_2$:  
% $\F_2 = \left\{f \mid f((C\backslash \{u\})\cup \{v\},v) - f(C,u)| \leq \D(u,v) \text{ for all }u,v \in U \text{ and }\forall C \in \mathcal{C} \text{ s.t. }u \in C, v\notin C\right\}.$
% Recall 
%\[\F_2= \left\{f \in \F_I \text{ s.t. } |f((C\backslash \{u\}) \cup \{v\},v) - f(C,u)| \leq \D(u,v) \text{ for all } u,v \in U \text{ and }\forall C \in \mathcal{C} \text{ s.t. }u \in C, v\notin C\right\}, \] where $\F_I$ is the set of all intra-cohort individually fair scoring functions.
%$\F_2$ enjoys the following ``interchangeability'' property.
%Nonetheless, $\dprime$ preserves key aspects of the policy, as demonstrated with the following mapping.
$\DSint$ is naturally translated into a simple mapping: 
for any pair of individuals $u$ and $v$, the partition corresponding to $u$ and $v$ in the mapping consists of clusters of size $2$ consisting of ``corresponding'' (cohort, individual) pairs. This follows from observing that if an individual $u$ receives some score $f(C, u)$ in a cohort $C$, if $u$ were replaced by $v \notin C$, then $v$ would receive a score in $[f(C,u) - \D(u,v), f(C, u) + \D(u,v)]$.  More formally:
\begin{definition}[Swapping Mapping]\label{def:swapping}
Let $\mathcal{C}$ be the set of all subsets of $U$ with exactly $k$ individuals. The \textbf{swapping mapping} is defined as follows. For each pair of individuals $u, v \in C$: 
\begin{enumerate}
    \item For $C \in \mathcal{C}$ such that $u, v \in C$, the partition includes the cluster $\left\{(C,u), (C,v)\right\}$.
    \item For $C \in \mathcal{C}$ such that $u \in C$ and $v \not\in C$, the partition includes the cluster $\left\{(C,u), ((C \setminus \left\{u\right\}) \cup \left\{v\right\},v)\right\}$.
\end{enumerate}
\end{definition}
\noindent It is straightforward to verify that the swapping mapping respects $\DSint$. %(and thus respects $\DStwo$ by Lemma \ref{lemma:mappingreduction}). %Moreover, the swapping mapping clearly illustrates the ``interchangeability'' property through the cluster structure.

% From a practical perspective, the swapping mapping imposes the constraint that swapping a single member of a cohort cannot result in arbitrarily large changes to the scores assigned to the cohort members. Returning to our motivating setting of employment, notice that such policies for compensation or promotion are not all that restrictive. Indeed, simply asking the question, ``How would my decision change if I could change any one member of the team?'' would be sufficient in most cases to check whether a scoring function conforms to the swapping mapping.

For the swapping mapping, there is a simple condition under which cohort selection mechanisms satisfy unconditional robustness (Notion 1): monotonicity. 
\begin{definition}[Monotonic cohort selection]
% Suppose that $\mathcal{C}$ is the set of cohorts of size $k$. A cohort selection mechanism $\Mech$ is \textbf{monotonic} if the following conditions are satisfied, for all pairs of individuals $u$ and $v$. WLOG, assume that $p(u) \le p(v)$. For any $C' \subseteq U$ such that $|C'| = k-1$ and $u, v \not\in C'$, it must hold that: $\DistMech(C' \cup \left\{u \right\}) \le \DistMech(C' \cup \left\{v\right\})$.
Suppose that $\mathcal{C}$ is the set of cohorts of size $k$. A cohort selection mechanism $\Mech$ is \textbf{monotonic} if for all pairs of individuals $u,v \in U$, for any $C' \subseteq U$ such that $|C'| = k-1$ and $u, v \not\in C'$, if $p(u) \leq p(v)$ then $\DistMech(C' \cup \left\{u \right\}) \le \DistMech(C' \cup \left\{v\right\})$. %\inote{@Meena are we really using the size $k$ condition in the proof other than to require that we can always swap? Reading through now, this seems overly restrictive.}
\end{definition}
The intuition for the link between the monotonicity property and the swapping mapping is that the probability masses on a cohort containing $u$ and a cohort containing $v$ that are paired in the swapping mapping are directionally aligned and cannot diverge by more than $\D(u,v)$. 
\begin{lemma}
\label{lemma:monotone}
Suppose that $\mathcal{C}$ is the set of cohorts of size $k$. If $\Mech$ is monotonic, then $\Mech$ satisfies $0.5$-Notion 1 for the swapping mapping.
\end{lemma}
\noindent 
Both PermuteThenClassify and WeightedSampling, cohort selection mechanisms proposed in \cite{DI2018}, are monotonic, efficient and have a high degree of expressivity.\footnote{See Appendix \ref{appendix:existingmechs} for detailed descriptions of these mechanisms and formal proofs of the monotonicity property.} 

However, monotonicity alone is not sufficient to guarantee conditional robustness (Notion 2) for the swapping mapping (see Appendix \ref{appendix:existingmechs}).
%It turns out that it is more difficult to satisfy Notion 2 for the swapping mapping, and monotonicity alone is not sufficient, as we show in Appendix \ref{appendix:existingmechs}. Nonetheless,
Borrowing intuition from PermuteThenClassify, we give a novel, efficient, individually fair cohort selection mechanism that achieves conditional robustness (Notion 2) for the swapping mapping:
\begin{mech}[Conditioning Mechanism]
\label{mech:modifiedPTC}
Given a target cohort size $k$, a universe $U$ and a distance metric $\D$, initialize an empty set $S$. 
For each individual $u \in U$:
\begin{enumerate}
    \item Assign a  weight $w(u)$ such that $|w(u) - w(v)| \le \D(u,v)$, \textit{i.e.}, the weights are individually fair.
    \item Draw from $\mathbbm{1}_u \sim \text{Bern}(w(u))$, (\textit{i.e.} flip a biased coin with weight $w(u)$). If $\mathbbm{1}_u$, add $u$ to $S$. % to provisionally assign $u$ to the cohort.
\end{enumerate} 

\noindent If $|S|\geq k$, return a uniformly random subset of $S$ of size $k$.\footnote{One might imagine a mechanism that conditions on exactly $k$ individuals being chosen, but this mechanism can be arbitrarily far from individually fair. Consider $k-1$ individuals with weight $1$ and $|U| - k- 1$ individuals with weight $0.9$. Conditioning exactly $k$ individuals would cause $|p(u) - p(v)|$ to diverge arbitrarily for $w(u)=.9$ and $w(v)=1$. %to be blow up significantly if $u$ is one of the individuals with probability $1$ and $v$ is one of the individuals with weight $0.9$.
} Otherwise, repeat the mechanism.

%We condition on \textit{at least} $k$ individuals overall being selected, and uniformly at random choose $k$ of these individuals.

%Given a weight function $w: U \rightarrow [0,1]$, for each $u \in U$, independently draw from $\mathbbm{1}_u \sim \text{Bern}(w(u))$. Denote the set of individuals with $\mathbbm{1}_u=1$ as $S$. %If $\mathbbm{1}_u = 1$ for $\ge k$ individuals in $U$, then let the cohort be $k$ of these individuals drawn uniformly at random. Otherwise, start over and repeat.
%If $|S|\geq k$, choose a cohort of $k$ individuals uniformly at random from $S$. Otherwise, repeat.
\end{mech}

\noindent We show that under mild conditions, the Conditioning Mechanism is satisfies Notion 2, concludes in a small number of rounds, and allows for a high degree of expressivity. (See Lemma \ref{lemma:modifiedPTC} in Appendix \ref{appendix:mechanismdetailed} for a formal statement and proof details.)
% \begin{lemma}[Informal]
% \label{lemma:informalmodifiedPTC}
% Let $\mathcal{C}$ be the set of all subsets of $U$ of size $k$, and let $C_1$,  $T$, and $C_2$ be appropriately chosen constants independent of $|U|$ and $k$. Consider the Conditioning Mechanism with $1/C_1$-individually fair weight function $w: U \rightarrow [0,1]$, such that $\sum_{x \in U} w(x) =  3k/2$. Then, the Conditioning Mechanism is individually fair, satisfies $C_2$-Notion 2, concludes in expectation within $T$ rounds, and enjoys a high degree of expressivity. 
% \end{lemma}
% \noindent We present the detailed version of Lemma \ref{lemma:informalmodifiedPTC} in Appendix  \ref{appendix:mechanismdetailed}. The intuition for the $C_2$-Notion 2 property of the Conditioning Mechanism is that this mechanism behaves ``close to'' independently deciding if each individual is in the cohort, though some mild correlations must be introduced to ensure there are exactly $k$ individuals in the cohort. 

\subsection{Quality-based treatment.% Mechanisms compatible with $\F_3$
}\label{subsec:F3}
One downside of the monotonic mechanisms proposed for $\DSint$ is that they require that any cohort with a single individual swapped is considered with nearly the same probability as the original cohort. In practice, this is problematic when $\Mech$ needs to ensure that each cohort has a certain structure. For example, when hiring a team of software engineers, designers and product managers, the proportion of each type of team member is important, and arbitrary swaps are not desirable from the perspective of team structure. 
By restricting to scoring functions that only consider the quality profile of a cohort, \textit{i.e.} how many individuals from each quality group are represented in a cohort, $\Mech$ can construct highly \textit{structured} cohorts, so long as the structure of the cohort is valid with respect to the fairness metric $\D$.

We now consider robust mechanisms for policies predicated on additional structure within the metric over $U$. In particular, we assume the existence of 
%We define $\F_3$ for metrics endowed with an additional object:
a partition of the universe $U$ into one or more ``quality groups'' $q_1, \ldots, q_n$. 
These quality groups satisfy the property that the distances within a quality group are smaller than distances between quality groups. {\em How much} smaller is determined by a parameter~$\beta$. More formally, %we define this notion for a metric on a general set $S$ as follows: 
\begin{definition}
\label{def:alphaclustered}
Let $\beta \le 1$ be a constant and $n \ge 1$ be an integer. Consider a partitioning of a $U$ into subsets $q_1, \ldots, q_n$, \textit{i.e.} ``quality groups'', and let $\D^*$ be a metric on $U$. Now, we define metrics $D$ on $\left\{1, \ldots, n\right\}$ and $\D^i$ for $1 \le i \le n$ on $q_i$ as follows: we let $D(i,j) = \inf_{u \in q_i, v \in q_j} \D^*(u,v)$ and $\D^i$ be the restriction of $\D^*$ to $q_i$. We call the metric $\D^*$ endowed with quality groups $q_1, \ldots, q_n$ \textbf{$\beta$-quality-clustered} if for all $1 \le i \le n$, we have that 
\[\max_{u,v \in q_i} \D^i(u,v) \le \beta \min_{j \neq i} D(i,j).\] 
\end{definition}
\noindent Notice that any metric $\D^*$ is trivially $1$-clustered with respect to the trivial quality group $q_1 = U$. The benefit of endowing $\D^*$ with a greater number of quality groups is to exploit additional structure of the metric, when any exists. 

For simplicity in the specification of the relevant policy and family of scoring functions we introduce  
%We now define a family of scoring functions $\F_3$ that are determined by the quality groups. We first define 
a \textbf{quality profile function} $P$ to count the number of individuals in each quality group in a cohort: that is, $P: 2^U \rightarrow \left\{(x_1, \ldots, x_n) \mid x_i \in \mathbb{Z}^{\ge 0}\right\}$, and the $i$th coordinate of $P(C)$ is $|C \cap q_i|$. 
Loosely speaking, the quality-based treatment policy requires that the only information about a cohort utilized by the scoring functions is its quality profile.
We now formally define $\F_3$ and an associated policy $\DSqual:$
\begin{definition}\label{def:F3}
Let $\beta \le 1$ be a constant. Suppose that $\D$ is endowed with quality groups $q_1, \ldots, q_n$ and $\D$ is $\beta$-quality-clustered. We define $\F_3$ to be the set of intra-cohort individually fair score functions $f: \mathcal{C} \times U \rightarrow [0,1]$ satisfying the following conditions:
\begin{enumerate}
    \item For $C, C' \in \mathcal{C}$ satisfying $P(C) = P(C')$, if $u$ and $v$ that are in the same quality group, then $f(C,u) = f(C',v)$. 
    \item For integers $1 \le i \neq j \le n$, $C, C' \in \mathcal{C}$ satisfying $P(C) = P(C')$, and any individuals $u \in q_j$ and $v \in q_j$, it holds that $|f(C,u) - f(C',v)| \le D(i,j)$. 
\end{enumerate}
\end{definition}
\noindent When each quality group is homogeneous in terms of individual ``quality'', this corresponds to score functions that are determined purely by ``quality''.\footnote{In this case, $\F_3$ includes Equal Treatment, Promotion, Stack Rank, and Fixed Bonus (discussed in Appendix \ref{appendix:extendedmotivation}) when scores are based on the ``quality'' of ``performance'' of individuals.}
As in Section \ref{subsec:F2}, we specify a distance function $\DSqual: (\mathcal{C} \times U) \times (\mathcal{C} \times U) \rightarrow [0,1]$ that overestimates $\DSthree$, but still preserves enough of the fairness structure to construct the desired mapping. %This distance function $\dprime$ is defined as follows:
\begin{definition}[Quality-based treatment policy]
Given a universe $U$, a set of permissible cohorts $\mathcal{C}$ and distance metrics and quality groups as in Definition \ref{def:F3},
\begin{enumerate}[leftmargin=*]
    \item For $C, C' \in \mathcal{C}$ satisfying $P(C) = P(C')$, if $u \in q_j$ and $v \in q_j$, then $\DSqual((C,u), (C',v)) = 0$. 
    \item For integers $1 \le i \neq j \le n$, $C, C' \in \mathcal{C}$ satisfying $P(C) = P(C')$, and any individuals $u \in q_j$ and $v \in q_j$, we set $\DSqual((C,u), (C',v)) = D(i,j)$. 
\end{enumerate}
\end{definition}

The core intuition is that a nice mapping exists when $\mathcal{C}$ is ``symmetric with respect to individuals in each quality group.''
It is helpful here to consider a bipartite graph $G=(A,B,E)$, where $A$ has one vertex for each subset of the universe~$U$, $B$ has one vertex for each possible profile of a subset of $U$, and there is an edge $(a,b) \in E$ precisely when $b$ is the profile of $a$, that is $b = P(a)$.

Fix any $\mathcal{C}$, and consider the subgraph $G'=(A',B',E')$ of $G$ induced by the vertices in $A$ corresponding to members of $\mathcal{C}$, the edges incident on these vertices, and the subset of $B$ induced by these edges.   We say that $\mathcal{C}$ is \textbf{quality-symmetric} if for all $b' \in B'$ it is the case that $E'$ contains all the edges in $E$ (in the original graph) incident on $b'$.

%We call $\mathcal{C}$ \textbf{quality-symmetric} if $\mathcal{C}$ is exactly the preimage of some set $\left\{(x_1, \ldots, x_n) \mid x_i \in \mathbb{Z}^{\ge 0}\right\}$ under the quality profile function $P$. 
 That is, $\mathcal{C}$ contains all cohorts obtained by swapping out individuals from the same quality group. If $\mathcal{C}$ is quality-symmetric, then consider the following mapping. 
\begin{definition}[Quality-Based Mapping]
Let $\beta \le 1$ be a constant. Suppose that $\D$ is endowed with quality groups $q_1, \ldots, q_n$ and $\D$ is $\beta$-quality-clustered. Suppose $\mathcal{C}$ is quality-symmetric. The \textbf{quality-based mapping} is defined as follows. For each pair of individuals $u, v \in C$, let $\mathcal{P}_{u,v} = (\mathcal{C}_u \times \left\{u \right\}) \cup (\mathcal{C}_v \times \left\{v\right\})$. 
%Cluster the cohorts $C\in\mathcal{C}_u \cup \mathcal{C}_v$ according to the profiles:
For each $(x_1, \ldots, x_n) \in P(\mathcal{C}_u \cup \mathcal{C}_v)$, the partitioning of $\mathcal{P}_{u,v}$ contains a cluster of the form $\left\{(C,x) \in \mathcal{P}_{u,v} \mid P(C) = (x_1, \ldots, x_n)\right\}$. 
\end{definition}
\noindent We verify that the quality-based mapping indeed respects $\DSqual$ (and thus respects $\DSthree$ by Lemma \ref{lemma:mappingreduction}). If $u$ and $v$ are in the same quality group, then the diameter of each cluster under $\DSqual$ is $0$, which is trivially upper bounded by $\D(u,v)$. On the other hand, if $u$ and $v$ are in different quality groups $q_i$ and $q_j$ respectively, then the diameter of each cluster is no more than $D(i,j) \le \D(u,v)$. Thus, the properties of a mapping are satisfied by the quality-based mapping. 

In this scenario, the quality-based \textit{mapping} captures the intuition for the fairness structure of $\F_3$ much better than $\DSqual$. The mapping groups together all cohorts with the same quality profile (\textit{i.e.} the same number of individuals in each quality group), capturing the intuition that the only information that a score function in $\F_3$ utilizes about a cohort is the quality profile.

As the score function behavior does not depend on the specific individuals in a quality group, $\Mech$ should have significant freedom to choose individuals within each quality group while still satisfying robustness w.r.t. $\F_3$. We will show that once the number of members of each quality group in the cohort is decided, utilizing any individually fair cohort selection mechanism within each quality group will satisfy our conditions. Moreover, our mechanisms have some flexibility in deciding the quality profile as well.  

\begin{mech}[Quality Compositional Mechanisms]\label{mech:qualitycomp}

Let $\beta \le 1$ be a constant, and suppose that $\D$ endowed with quality groups $q_1, \ldots, q_n$ is $\beta$-quality-clustered. Suppose also that $\mathcal{C}$ is quality-symmetric. For each $1 \le i \le n$ and each $1 \le x_i \le |q_i|$, let $\Mech_{i, x_i}$ be a $\D^i$-individually fair mechanism selecting $x_i$ individuals in $q_i$. We define the \textbf{quality compositional mechanism} for $\left\{\Mech_{i, x_i}\right\}$ as follows. Let $\mathcal{X}$ be any distribution over $n$-tuples of nonnegative integers $(x_1, \ldots, x_n) \in P(\mathcal{C})$. 

\begin{enumerate}
    \item Draw $(x_1, \ldots, x_n) \sim \mathcal{X}$. 
    \item Independently run $\Mech_{i,x_i}$ for each $1 \le i \le n$, and return the union of the outputs of all of these mechanisms. 
\end{enumerate}
\end{mech}

In the next lemma, we show that when a quality composition mechanism only selects cohorts whose quality projection vectors $(x_1, \ldots, x_n)$ are ``close'' to an inter-quality group distance multiple of $(|q_1|, \ldots, |q_n|)$, Notion 1 is achieved. (This requirement essentially says that the relative proportion of selected individuals in each quality group needs to be approximately reflective of the relative proportion of individuals in each quality group in the universe, scaled by the difference between the quality groups in the original metric. This type of requirement turns out be necessary for basic individual fairness guarantees, by the constrained cohort impossibility result in \cite{DI2018}.) Moreover, under stronger conditions, we show that Notion 2 is also achieved. 
\begin{lemma}
\label{lemma:distquality}
Let $\beta \le 0.5$ be a constant, and suppose that $\D$ endowed with quality groups $q_1, \ldots, q_n$ is $\beta$-quality-clustered. Suppose also that $\mathcal{C}$ is quality-symmetric, and let $\mathcal{X}$ be any distribution over $(x_1, \ldots, x_n) \in P(\mathcal{C})$ such that $|\frac{x_i}{|q_i|} - \frac{x_j}{|q_j|}| \le (1 - 2 \beta) D(i,j)$. If $\Mech$ is a quality compositional mechanism, then:

\begin{enumerate}
    \item $\Mech$ is always individually fair.
    \item $\Mech$ always satisfies $0.5$-Notion 1.
    \item $\Mech$ satisfies $0.5$-Notion 2 for $\D$ and $\DS$ if \textbf{either} of the following conditions hold:
    \begin{enumerate}
        \item (One set) $|\text{Supp}(\mathcal{X})| = 1$ (\textit{i.e.} one ``canonical'' $(x_1, \ldots, x_n)$), or
        \item (0-1 metric) $D(i,j) = 1$ for $1 \le i \neq j \le n$ and $\D^i(u,v) = 0$ for $1 \le i \le n$.
    \end{enumerate}
\end{enumerate}
\end{lemma}

The quality compositional mechanisms provide a greater degree of structure in cohort selection than the monotone mechanisms giving in Section \ref{subsec:F2}. The Conditioning Mechanism and similar monotone mechanisms are forced to select individuals essentially independently, with the only dependence stemming from the cohort size constraint. However, structured cohorts are necessary in a number of practical applications, as previously noted. Although $\DSqual$ imposes more constraints on the permitted $\F$ than $\DSint$, the basis for these constraints is likely to be tolerated well in legitimate use cases in which structure is important. %For example, suppose that a company wanted to always select a balanced teams of highly experienced, moderately experienced, and inexperienced team members. Under the assumption that the scoring function (\textit{e.g.} bonus determination) solely depends on experience levels, the quality compositional mechanism would enable the company to achieve this balance by guaranteeing $x_{high}$ highly experienced individuals, $x_{medium}$ moderately experienced individuals, and $x_{low}$ inexperienced individuals, as long as this roughly reflects the proportions of experience in the universe. 

Moreover, the company has flexibility in selecting individuals within each experience group, as any individually fair mechanism can be utilized. This offers significantly more flexibility than selecting members in each quality group uniformly at random. Such flexibility is particularly crucial, for example, if a company further wants to ensure that tech company teams have a mixture of software engineers and product managers. The individually fair mechanisms within each quality group can help achieve this balance through selecting balanced subsets of engineers and product managers. In essence, the quality compositional mechanisms allow flexibility in cohort selection while still satisfying robustness for $\F_3$, due to restrictions on the behavior of scoring functions in $\F_3$.

%% file: sections/discussion.tex
We have presented a framework for evaluating the robustness of cohort selection as part of a pipeline.
We've demonstrated that naive auditing strategies concerning average cohort quality or score are unable to uncover significant fairness problems. We've also shown that many reasonable policies for cohort selection and subsequent scoring can conflict with each other resulting in very poor fairness outcomes. Furthermore, we've demonstrated that a malicious pipeline designer can easily use composition problems to disguise bad behavior.
Despite these hurdles, we've shown that it is possible to construct pipelines that are fair. In particular we've shown that constructing cohort selection mechanisms that are robust to composition with a family of scoring functions is possible. By framing the problem in terms of robustness, we address the concern that placing requirements on future designs is nearly unenforceable, whereas designing the current stage to be robust to a large class of potential future policies can give much better practical guarantees.
Finally, we've shown robust cohort selection mechanisms that compose well with reasonable scoring function families.

In the process of exploring robustness and fairness in pipelines, we uncovered a number of interesting questions for future work. \textbf{Policy complexity}: we have considered a set of concise and practical policies in this work, but the trade-off between policy complexity and the expressiveness of cohort selection has not been fully characterized. \textbf{Fair Matching}: choosing a cohort is very similar to the problem of assigning an individual to an existing cohort. However, in the traditional matching literature, significant emphasis is placed on individuals' and teams' preferences over placements, rather than external fairness criteria. Is it possible to simultaneously achieve a good matching, in the sense of satisfying preferences or stability, and individual fairness? \textbf{Quantifying the tradeoff}:
There are significant differences in the difficulty for constructing mechanisms which satisfy the conditional, versus unconditional, notion of robustness. Is it possible to more directly quantify the tradeoff in mechanism expressivity between these two settings?
\textbf{Different metrics}:
Handling different metrics in the pipeline: we considered just one metric throughout the entire pipeline, but using different metrics for different stages of the pipeline may be valid. For example, in the case of promoting an individual contributor to a management position, the metric for ``manager'' may be different.
\textbf{Ranking instead of scoring}: although ranking with hard cutoffs does not satisfy individual fairness, it is frequently used in practice. Can the model we have outlined with respect to scoring be translated to ranking, \textit{e.g.}, incorporating the results of \cite{evidencerankings2019}?

%% file: sections/related.tex
There is a wide variety of work concerning fairness in machine learning \cite{gillen2018online,jung2019eliciting,rothblum2018probably,kim2018fairness,chouldechova2017fair,DBLP:journals/corr/KleinbergMR16,zemel2013learning,madras2018learning,hardt2016equality,kilbertus2017avoiding,kearns2017gerrymandering,hebert2017calibration,HuC17,liu2018delayed,lambrecht2016algorithmic,datta2015automated,ritov2017conditional,dwork2012fairness}. Individual fairness was introduced by Dwork \textit{et al.} \cite{dwork2012fairness}. Dwork and Ilvento studied composition of combination of individually fair and group fair classifiers~\cite{DI2018}. %, the direct predecessor of this work, ``Fairness Under Composition'' \cite{DI2018} appeared at ITCS 2019 \inote{Cynthia, what is the elegant way to include the ITCS reference?}\mnote{both this paper and Cynthia's original paper appeared at ITCS right?}. 
%Ilvento~\cite{I2019} has shown how to learn a metric on individuals using few queries to an oracle endowed with this information.
Two other recent lines of work have also considered composition problems and fair systems. First, several works have studied the problem of feedback loops, in which decisions that previous time steps, such as where to send law enforcement officers, influence outcomes at later time steps potentially unrelated to the original decision \cite{HuC17, FNSV17, lum2016predict}. 
Bower \textit{et al.} study fairness in a pipeline of decisions under a group-based notion of fairness \cite{BowerKNSVV17}. They primarily consider the combination of multiple non-adaptive sequential decisions, evaluating fairness at the end of the pipeline. 
Second, several works have considered competitive scenarios, such as advertising, in which many (potentially fair or unfair) classifiers compete for individuals \cite{celis2019toward, CIJ2019}. Although not explicitly addressing composition, recent work considering fairness in rankings, \textit{e.g.} \cite{evidencerankings2019}, also address fairness in a setting in which outcomes, in this case rankings, naturally depend on the outcomes of others.

%\inote{I think we discussed before that this is a bit confusing -- the original paper didn't actually address composition and assumed a one-shot setting. It also doesn't hold that two fair classifiers composed in sequence are fair, which is what this implies. }
%\mnote{I'm wondering if we should bring up : ``In \cite{dwork2012fairness}, the authors consider sequential classifier composition. Let $M: U \rightarrow \Delta(A)$ be a classifier, and let the metric over the outcomes be total variation distance or $\ell_{\infty}$ distance. Suppose that the post-processing function is $f: A \rightarrow B$. It follows immediately that if $M$ is individually fair, $f \circ M$ is also individually fair. While these composition results capture some practical settings in the context of classifiers, this framework is quite limited in the context of cohort selection, since it does not allow $f$ to consider the interaction between individuals in a cohort in the post-processing function.'' Although this post-processing result is limited, in a sense, our result is a direct generalization of this framework.}

%% file: sections/extended-motivating-examples.tex
To complement the motivating examples included throughout the text, we include a ``Catalog of Evils'' relevant to pipelines.

% now discuss a more comprehensive set of examples with formal specifications of the scoring functions $\mathcal{F}$ and cohort selection mechanisms ($\Mech$) that give rise to the problematic behavior. We consider both unintentional oversight and malicious motivations for these examples to illustrate the ease with which bad actors can obscure their inentions with the veneer of individual fairness, and unwitting designers can undermine their own good intentions.
% We consider a single domain, employment, to keep our examples concise, but similar concerns obviously arise in other areas (the reader familiar with the literature on gerrymandering may see striking similarities to concerns raised in that domain).
% In many of these examples, we compare treatment of different potential cohorts under a given scoring function to illustrate unfairness. In our stylized model, however, only a single cohort is selected and subsequently scored. The reader can view these examples as comparing the treatment of potential or hypothetical cohorts (only one of which is realized) or as a comparison of multiple cohorts chosen and scored via the same policies, assuming a sufficiently large pool to form multiple cohorts.
% This allows us to observe how the score function behaves on multiple teams and compare treatment.

In each example, we consider a universe $U$ comprised of individuals belonging to two groups, a majority group $S$ and a minority group  $T$, such that the majority group is $k$ times as large as the minority group (i.e. $k |T| = |S|$). For the particular employment task in question, there is a known metric  $\mathcal{D}$ which specifies who is similar to whom for the purposes of this task. For simplicity, we assume that $\mathcal{D}$ is one-dimensional, i.e. each individual $u$ has a qualification $q_{u}\in [0,1]$, and $\D(u,v):=|q_u-q_v|$.
We assume that $S$ and $T$ have an equal distribution of talents: more specifically, for every qualification level $q$, there are exactly $k$ times as many individuals with qualification $q$ in $S$ as there are in $T$.
We assume that there is a nontrivial range of qualifications in $[0,1]$, and we will generally assume that the company prefers to hire the most highly qualified candidates, but in order to fill the number of positions open cannot hire only maximally qualified candidates. We use $Q_H$ to refer to the subset of individuals who are highly qualified.

Our examples are based on a set of facially neutral company compensation policies. We now give precise descriptions of these policies in the form of a scoring function, and indicate where the scoring policies must be adjusted to give intra-cohort individual fairness. (As we will see later, even adjusting the policies to be intra-cohort individually fair won't be enough to prevent bad behavior under composition.)
\begin{enumerate}[leftmargin=*]
    \item \textbf{Fixed Bonus Pool}: A fixed pool of bonus money $B$ is assigned to each team and is split between the members of each team, with the highest achieving members receiving larger portions of the pool.
    More formally, given a cohort of individuals $C=\{x_{1},\ldots,x_{c}\}$ of size $c$ with qualifications $\{q_{x_1},\ldots, q_{x_c}\}$, the scoring function  $f_{B}$ assigns a bonus share $b_{i}$ to each individual $x_{i}$ such that $\sum_{u\in C}b_u=1$, optimized to ensure that individuals with higher qualification receive larger bonuses.

    In particular, $f_B$ can either be a simple proportional mechanism, e.g. $f_B(u) \propto q_u$, or it can be optimized for specific goals, e.g. maximizing the difference in compensation between the most and least qualified individuals, creating an even spread of compensations, etc. For example, the company could choose $f_B$ using the following optimization to choose the largest ``weighted spread'' to maximize the objective of increasing the difference in compensation based on difference in qualification: 
    \begin{align*} &argmax_{\{b_u \in [0,1]\}} \{\sum_{u,v \in C}(b_u - b_v)(q_u-q_v)\} \\
    &\text{subject to }\\
    &|b_u - b_v|\leq |q_u - q_v| \text{ for all }u,v \in C\\
    &\sum_{u \in C} b_u=1
    \end{align*}

    This optimization will tend to choose bonus shares that maximize the differences in bonuses between individuals with significantly different qualifications within the cohort. 
    Notice that the scoring function has no way of knowing what other cohorts may or may not appear and with what probabilities, and so it only optimizes within the particular cohort $C$.
    %However, in practice it's easy to see that this optimization will result in unfair allocation of bonuses in the case of examples x,y,z. In particular, consider the case of packing. Individuals assigned to cohorts with more talented individuals will tend to receive smaller bonuses than equally talented individuals in cohorts with fewer talented individuals.

    \item \textbf{Stack Rank}: The bottom 10\% of each team may be fired or put on ``performance plans''.
    Formally, 
   \[f(C,u):= \begin{cases}
       1 \text{ if } \frac{|\{v \mid q_u > q_v\}|}{|C|} \leq 0.1,\\
       0 \text{ otherwise}
       \end{cases}
   \]

   However, this strict cut off violates intra-cohort individual fairness, as two nearly equally qualified individuals might find themselves on opposite sides of the cutoff.
   Alternatively, we can construct a scoring function which closely approximates the desired policy but still satisfies intra-cohort individual fairness,  by optimizing subject to the intra-cohort fairness constraints. For example, taking $\mathbb{O}_u$ to be the indicator that $u$ is in the bottom 10\% of the cohort, one could use the following optimization to maximize the probability that only the bottom 10\% are placed on performance plans
   \begin{align*} &argmax_{f} \sum_{u \in C}f(C,u)\mathbb{O}_u + (1-f(C,u))(1-\mathbb{O}_u) \\
   &\text{subject to }\\
   &|f(C,u) - f(C,v)|\leq |q_u - q_v| \text{ for all }u,v \in C
   \end{align*}
Alternatively, if exactly 10\% of the cohort should be put on performance plans, Permute-Then-Classify can be applied or an additional constraint on the expected number of employees placed on performance plans could be added to the optimization above in order to satisfy \textit{intra-cohort} individual fairness.

   %\inote{reuse text later?}
   %As previously noted, even though this policy may be individually fair in the context of a particular cohort, individuals assigned to cohorts with many other talented individuals will be more likely to put on performance plans than their equally qualified counterparts assigned to cohorts with fewer other talented individuals.\footnote{Notice that in this case the average qualification of the teams may be equivalent, but the rankings may still be different.}
    \item \textbf{Equal Treatment}: Each team's bonus is determined by average performance of the team (assumed to be proportional to average quality) and awarded equally to each member. Formally, the scoring function $f$ first chooses the total bonus amount $B_C\propto B\sum_{u \in C}q_u$, and then assigns $b_u=\frac{B_C}{|C|} $ for all $u \in C$. Intra-cohort individual fairness for $f$ is trivial, as every individual is treated equally. %\inote{@Cynthia -- I left this as average, because the bonus is chosen proportionally. Can you clarify the concern?}%But as we observed before, individuals on teams with higher average quality will tend to receive higher bonuses than equally talented individuals on teams with lower average quality.
    \item \textbf{Promotion}: Choose the single most qualified person on the team to promote, based on performance. As in the case of stack ranking, strictly implementing this policy will violate intra-cohort individual fairness, as nearly equal individuals may be treated very differently. As above we can satisfy \textit{intra-cohort} individual fairness by posing the relevant optimization question, and Permute-then-Classify (see Appendix \ref{appendix:existingmechs}) can be used to select exactly one individual for promotion.
\end{enumerate}

We now show that these compensation policies can cause significant unfairness for $T$ when combined with simple hiring protocols.
In each case, we state the set of cohorts the company intends to select from, and we assume that the company uses a method similar to the one described in Appendix \ref{appendix:ws} to derive a fair set of weights to use to sample a single cohort in an individually fair way.\footnote{We omit the details of the method and the particulars of the conditions on the set of cohorts specified as they are easy to fulfill in these settings. In particular, each set of cohorts we specify can clearly be used to form a partition of $U$, fulfilling the requirements of Theorem \ref{theorem:craftedcohorts}. }
First, we consider the ``packing'' hiring protocol. %\inote{@Cynthia -- all of the examples in this ``catalog'' are focused on hiring/employment. I think it's good to reinforce that language to emphasize that these are realistic protocols for hiring. I don't feel super strongly about it though.}
\begin{example}[Packing]\label{example:packing}
Suppose that in the past, the company had a particular problem retaining employees from the minority group $T$ and in order to address this problem, the company ensures that individuals with high potential from  $T$ are always hired together into the same team for mutual support. On the other hand, talented members of $S$ are spread out between the other teams, to make sure that there is at least one highly talented individual on each team.
Formally, the company specifies the set of cohorts $\mathcal{C}_{packing}= \{C \in \mathcal{C} \mid (|C \cap T \cap Q_H|>1 \wedge |C \cap S \cap Q_H|=0) \oplus (|C \cap T \cap Q_H|=0 \wedge |C \cap S \cap Q_H|=1)\}$, where $Q_H$ is the set of highly qualified candidates, and samples a single cohort from the set such that individual fairness is satisfied.
\end{example}

\paragraph*{``Packing'' results in lower compensation for $T$ for Fixed Bonus Pool, Stack Rank, and Promotion compensation policies.} ``Packing'' causes talented members of $T$ to be on teams of higher average quality than those with talented members of $S$. As a result, members of $T$ will receive lower bonuses and promoted less often than members of $S$. Thus, this seemingly beneficial practice can backfire when composed with certain compensation policies.

\medskip
One may imagine that utilizing a ``splitting'' strategy, where qualified members of $T$ are separated from other qualified members to increase their chance of ``standing out'' on teams, would solve this issue.
\begin{example}[Splitting]\label{example:splitting}
The company chooses teams where highly qualified members of $T$ are always the only highly qualified member of their team, giving them the opportunity to stand out and be recognized for their talent. More formally, the company chooses from the set of cohorts  $\mathcal{C}_{splitting}= \{C \in \mathcal{C} \mid (|C \cap T \cap Q_H|=1 \wedge |C \cap S \cap Q_H|=0) \oplus (|C \cap T \cap Q_H|=0 \wedge |C \cap S \cap Q_H|\geq 1)\}$. In each cohort containing a highly qualified member of $T$, there are no other highly qualified individuals (from either $T$ or $S$).
%Pack qualified $S$ into one team, and place qualified members of $T$ each on their own team in order to give minority members a higher chance of ``standing out'' on their teams.
\end{example}
\noindent Though this policy no longer leads to lower compensation for $T$ for Stack Rank, Fixed Bonus Pool, and Promotion, ``Splitting'' results in lower compensation for $T$ for Equal Treatment, %. This results from ``Splitting'' causing 
because the practice causes talented members of $T$ to be on teams of lower average quality than talented members of $S$. As a result, with Equal Treatment, qualified $S$ will receive greater compensation than qualified $T$.
%Similar to Example \ref{example:steering}, 
Splitting can also occur when members of $T$ are primarily hired via outreach. For example, suppose that a company has been trying to form a team to work on a difficult or low prestige task (e.g. Fortran code maintenance). All of the talented candidates in $S$ pass on the job offer because they are confident they can do better, so HR reaches out more aggressively to candidates in $T$. These candidates may be more willing to take the job because they are less confident about their other options. Thus, even without an explicit policy in place to choose minority candidates to be the singular most qualified member on a less qualified team, these situations can still arise from the interactions between the hiring procedure and the job market.

\begin{remark}
  The motivation for both of these policies could be malicious, and determining whether the stated goals or justifications were legitimate aims of the policy would be difficult.
\end{remark}

One may imagine that these issues could be addressed by ensuring that  %enforcing that 
qualified members of $T$ and qualified members of $S$ appearing on teams with similar average quality. However, a malicious company can still cause members of $T$ to receive lower compensation.
\begin{example}[Adversarial ranking]\label{example:ranking}
Suppose that the company did not want any member of the $T$ to be chosen for promotion or wished to depress their compensation relative to the members of $S$. The company decides to choose teams such that, for each team, there is a correspondence between the members of $T$ and $S$ included in the team, such that the members of $S$ are almost always more talented than their counterparts in $T$. (Given the equal distribution of talents of $T$ and $S$, there may be an excess member of $T$ that is allowed to be the most qualified, but this is a singular case.) More formally, the company chooses from $\mathcal{C}_{adv.ranking}= \{C \in \mathcal{C} \mid \exists G: C \cap T \rightarrow C \cap S $ s.t. $\forall u \in C \cap T$, $q_u < q_{G(u)}\}$.
%Spread out qualified $S$ and qualified $T$ between teams. However, place an exceptionally qualified member of $S$ on most of the teams with talented, but exceptionally talented, $T$, and offset this with an extremely unqualified member of $S$. Avoid placing qualified $S$ and exceptionally qualified $S$ on the same team.
\end{example}
\noindent ``Adversarial Ranking'' is particularly catastrophic for $T$ for Promotion or Stack Ranking if the hard cutoff (not intra-cohort individually fair) versions are used. %, which likely happens frequently in practice. 
Although ensuring intra-cohort individual fairness helps, members of $T$ will always be seeing depressed levels of promotion, higher levels of firing, and lower levels of compensation except in the case of Equal Treatment.
Thus ``Adversarial Ranking'' keenly illustrates that average team quality is not sufficient to ensure that individuals are truly being treated fairly in cohort-based pipelines. We stress that Adversarial Ranking can also be efficiently achieved using the procedure in Theorem \ref{theorem:craftedcohorts}.

\subsection{Sample Cohorts}

To illustrate these issues, we include
Figures \ref{fig:bonus_comparison} and \ref{fig:promotion_comparison} to compare the example scoring functions for a pair of cohorts, demonstrating the issues outlined above.

\noindent \begin{figure}[h]
    \centering
    \begin{subfigure}{6cm}
        \begin{tabular}{|l|p{1cm}|p{1cm}|p{1cm}|}
        \hline
         & Quali-fication & Fixed Pool Bonus & Equal Bonus \\
        \hline \hline
        \textbf{Cohort 1} & & & \\
        \hline
        Alice & $0.8$ & 35	&60 \\
        \hline
        Bob & 0.7	&25&	60 \\
        \hline
        Charlie & 0.5	&5	&60\\
        \hline
        Dan & 0.2	&0&	60 \\
        \hline
        Eve & 0.8	&35	&60 \\
        \hline \hline
        
        \textbf{Cohort 2} & & & \\
        \hline
        Frank &0.8&	57	&40 \\
        \hline
        George &0.6&	36	&40 \\
        \hline
        Harriet & 0.1	&0&	40\\
        \hline
        Ivan & 0.2	&0	&40 \\
        \hline
        Julia & 0.3	&7	&40 \\
        \hline
        \end{tabular}
        \caption{Bonus score function comparisons for two  cohorts, each containing five individuals of varying qualifications.
        Cohort 1 has an average qualification of 0.6, and Cohort 2 has an average qualification of 0.4.
        In the fixed pool bonus, a total pool of 100 is split between the members of the cohorts. The same optimization is used for both cohorts, that is according the maximum possible bonus to the most qualified individual(s). Notice that in Cohort 1, Alice and Eve have to share the top bonus (35 each), but in Cohort 2, Frank doesn't have to split the top bonus (57).
        Notice also that George and Julia receive higher bonuses than Bob and Charlie, even though they are (much) less qualified.
        On the other hand, in the equal bonus setting Frank receives a lower bonus than both Alice and Eve, even though he's equally qualified. } \label{fig:bonus_comparison}

    \end{subfigure}
    \quad \quad
    \begin{subfigure}{7.1cm}
    \begin{tabular}{|l|p{1cm}|p{1cm}|p{1cm}|p{1cm}|}
        \hline
         & Quali-fication & Pro-motion & Stack Rank (IF) & Stack Rank (exact, not IF) \\
        \hline \hline
        \textbf{Cohort 1} & & & \\
        \hline
        Alice & 0.8	&35\%	&0&	0 \\
        \hline
        Bob & 0.7&	25\%	&10\%	&0 \\
        \hline
        Charlie & 0.5	&5\%	&30\%	&0\\
        \hline
        Dan & 0.2&	0	&60\%	&1 \\
        \hline
        Eve & 0.8	&35\%	&0&	0 \\
        \hline \hline
        
        \textbf{Cohort 2} & & & \\
        \hline
        Frank &0.8&	57\%	&0&	0 \\
        \hline
        George &0.6&36\%&	0	&0\\
        \hline
        Harriet & 0.1	&0	&43\%	&1\\
        \hline
        Ivan & 0.2	&0	&33\%	&0\\
        \hline
        Julia & 0.3	&7\%	&24\%	&0 \\
        \hline
        \end{tabular}
        \caption{Promotion score function comparison of the cohorts from Figure \ref{fig:bonus_comparison}.
        The promotion policy attempts to maximize the probability of promotion for the most qualified individuals, subject to the individual fairness constraints and that the expected number of promotions is 1. In this case, essentially the same observations apply as in the fixed pool bonus setting.
        In the case of Stack rank, both cohorts are optimized to maximize the probability of placing the least qualified person on a performance plan. Notice that Dan is much more likely to be placed on a performance plan than the equally qualified Ivan, due to the larger number of less qualified individuals in Cohort 2. Although it might seem that the exact stack rank policy, rather than the individually fair version, would be less likely to have this problem, in fact in this case Dan is still treated differently than Ivan.} \label{fig:promotion_comparison}

    \end{subfigure}
    
\end{figure}

%% file: sections/extended-conditionssuccess.tex
In this section, we provide proofs and some additional results mentioned in Section \ref{sec:conditionsforsuccess}. 
In proving results, we consider the mass-moving distance between probability distributions over scores. That is, for every pair of individuals $u$ and $v$ and every score function $f \in \F$, we consider the mass-moving distance between  $\UncondDist{\Mech}{f}{u}$ and $\UncondDist{\Mech}{f}{u}$ (resp. $\CondDist{\Mech}{f}{u}$ and $\CondDist{\Mech}{f}{v}$).

A simple way to think about our notion of mass-moving distance is to break the definition down into two steps: (1) transforming the original distributions over scores into distributions over a single shared set of \textbf{adjusted scores} and (2) moving mass between the distributions over adjusted scores. %As we discussed previously, simply using total variation distance is too strict in the sense that similar scores (\textit{e.g.}, $\pm \varepsilon$) are treated the same as very different scores (\textit{e.g.}, $\pm 0.5$).
By introducing the transformation in step (1), we take the two distributions over scores (which may have disjoint supports) and transform them into distributions over a single support of adjusted scores so that similar scores are mapped to similar adjusted scores.
% In this context, we need to define probability measures over adjusted scores that are close to each other with respect to total variation distance. We will call these probability measures over adjusted scores $\TildeUncondDist{\Mech}{f}{u}$ and $\TildeUncondDist{\Mech}{f}{v}$ (resp. $\TildeCondDist{\Mech}{f}{u}$ and $\TildeCondDist{\Mech}{f}{v}$). 

% We focus on defining $\TildeUncondDist{\Mech}{f}{u}$ and $\TildeCondDist{\Mech}{f}{u}$, since $\TildeUncondDist{\Mech}{f}{v}$ and $\TildeCondDist{\Mech}{f}{v}$ are defined analogously.
% The high-level idea is as follows.  
% We base $\TildeUncondDist{\Mech}{f}{u}$ on $q^1_{u,v}$ (resp. $\TildeCondDist{\Mech}{f}{u}$ on $q^2_{u,v}$), where $q^1_{u,v}$ is the distribution over clusters induced by $\Mech$. 

The next consideration is how we can choose adjusted scores and write distributions over adjusted scores in a way that takes advantage of what we know about the mapping and similar treatment of similar cohort contexts by $f$. To do this, we write the distributions over adjusted scores ($\TildeUncondDist{\Mech}{f}{u}$ and $\TildeUncondDist{\Mech}{f}{v}$ (resp. $\TildeCondDist{\Mech}{f}{u}$ and $\TildeCondDist{\Mech}{f}{v}$)) in terms of the distributions over clusters induced by $A$ ($q^1_{u,v}$ (resp. $q^2_{u,v}$)). 

Why does this help? Given a cluster, we can propose an adjusted score based on its extreme behavior under $f$, \textit{i.e.} the highest and lowest possible scores in the cluster. More formally, given a mapping,
we define a function $Q: \left\{1, \ldots, \MapNum{u}{v}\right\} \rightarrow [0,1]$ to transform the cluster labels that form the sample space of $q^1_{u,v}$ (resp. $q^2_{u,v}$) into adjusted scores in $[0,1]$ that form the sample space of $\TildeUncondDist{\Mech}{f}{u}$ (resp. $\TildeCondDist{\Mech}{f}{u}$). $Q$ will map each cluster in the partition corresponding to $u$ and $v$ to an adjusted score given by an ``average'' score in the cluster. Let $S(i) = \left\{(C, u) \mid C \in \Map{u}{v}^{-1}(i) \right\} \cup  \left\{(C, v) \mid C \in \Map{v}{u}^{-1}(i) \right\}$ be the (cohort, individual) pairs appearing in the cluster $i$. Let $a_i$ and $b_i$ be the minimum and maximum scores in $\left\{f(C,x) \mid (C,x) \in S(i)\right\}$. Now, we let $Q(i) = \frac{a_i + b_i}{2}$, which can be viewed as an ``average'' score in the cluster. Briefly, this choice of definition for $Q$ will guarantee that intra-cluster differences in treatment are bounded by $\mathcal{D}(u,v)$, and thus mapping scores within a cluster to this ``average'' will conform to the requirement that the transformation to adjusted scores doesn't move any score ``too far.'' (See Definition \ref{def:mmd}.\ref{def:mmd:1}.)

We can now write the distributions $\TildeCondDist{\Mech}{f}{u}$ and $\TildeUncondDist{\Mech}{f}{u}$ in terms of clusters. For each $j \in [0,1]$, notice that $Q^{-1}(j)$ gives the set of clusters which correspond to that score (if such a cluster exists). $q^{*}_{u,v}(Q^{-1}(j))$ then yields the probability assigned to each cluster corresponding to the score $j$ by $A$. More formally, we
%Using $Q$ to map clusters to scores, we define $\TildeCondDist{\Mech}{f}{u}$ and $\TildeUncondDist{\Mech}{f}{u}$ from $q^2_{u,v}$ and $q^1_{u,v}$ by mapping clusters. For every $j \in Q([n_{u,v}])$, we use the fact that $Q^{-1}(j)$ is a set of cluster labels. We 
define $\TildeCondDist{\Mech}{f}{u}(j)$ as follows:
\[\TildeCondDist{\Mech}{f}{u}(j)  = 
\begin{cases}
q^2_{u,v}(Q^{-1}(j)) & \text{ if } j \in Q([n_{u,v}])\\
0 & \text{ if } j \not\in Q([n_{u,v}]).
\end{cases}
\]
We define $\TildeUncondDist{\Mech}{f}{u}(j)$ similarly, with the slight modification that we place an additional $1 - p(u)$ mass at $0$ to account for the fact that not being selected in a cohort corresponds to a score of $0$. 
\[\TildeUncondDist{\Mech}{f}{u}(j)  = 
\begin{cases}
q^1_{u,v}(Q^{-1}(j)) & \text{ if } j \in Q([n_{u,v}]), \text{ if } j \neq 0\\
q^1_{u,v}(Q^{-1}(0)) + 1 - p(u) & \text{ if }  j = 0 \\
0 & \text{ otherwise}.
\end{cases}
\]
We analogously define these quantities for $v$.

\subsection{Proofs for Section \ref{sec:conditionsforsuccess}}
\label{appendix:proofsconditionssuccess}

Using the distributions over adjusted scores, we can now prove Theorem \ref{thm:manytoonemapping}.

\begin{theorem}[Restatement of Theorem \ref{thm:manytoonemapping}]
\label{thm:restatementmanytoonemapping}
Let $\F$ be a class of scoring functions, let $\alpha \ge 0.5$ be a constant. Suppose that $\left(\Map{u}{v}\right)_{u \neq v \in U}$ is in $\Mapping_{\frac{1}{2\alpha}\DS}$. If $\Mech$ is individually fair and satisfies $\alpha$-Notion 1 (resp. $\alpha$-Notion 2) for $\left(\Map{u}{v}\right)_{u \neq v \in U}$, then we have that $\Mech$ is $2\alpha$-robust w.r.t. $\F$ for $\DUncondMMD$ (resp. $\DCondMMD$). 
\end{theorem}
\begin{proof}[Proof of Theorem \ref{thm:restatementmanytoonemapping}]

Pick a pair of individuals $u \neq v \in U$. Pick any $\alpha \ge 0.5$. Assuming that $\Mech$ satisfies $\alpha$-Notion 1 (\textit{i.e.} Definition \ref{def:conditions}.1), we construct measures in the mass-moving definition that achieve a distance of no more than $\alpha \D(u,v)$. We use $\TildeUncondDist{\Mech}{f}{u}$ and $\TildeUncondDist{\Mech}{f}{v}$ (resp. $\TildeCondDist{\Mech}{f}{u}$ and $\TildeCondDist{\Mech}{f}{v}$) as defined above as our finite support measures in the mass-moving distance definition.

The proof consists of three steps: First, we take as given that $\TildeUncondDist{\Mech}{f}{u}$ and $\TildeUncondDist{\Mech}{f}{v}$ (resp. $\TildeCondDist{\Mech}{f}{u}$ and $\TildeCondDist{\Mech}{f}{v}$) satisfy Definition \ref{def:mmd}.\ref{def:mmd:1} (\textit{i.e.}, that there exists some $Z$ satisfying the first condition) and we show that \ref{def:mmd}.\ref{def:mmd:2} is satisfied. This follows from a straightforward computation of total variation distance. 
Next, we exhibit the appropriate $Z$ for \ref{def:mmd}.\ref{def:mmd:1} by linking the adjusted scores given by $Q(i)$ to the original scores and showing: (A) no score moves too far when adjusted and (B) mass is conserved. Both arguments follow from the construction of the function $Z$.

First, we consider Condition \ref{def:mmd}.\ref{def:mmd:2} (\textit{i.e.} ``total variation distance is small'') for the distributions over adjusted scores. We use the fact that 
\[TV(\TildeCondDist{\Mech}{f}{u}, \TildeCondDist{\Mech}{f}{v}) = \sum_{j \in \text{Supp}(\TildeCondDist{\Mech}{f}{u})} \left|\sum_{i \in Q^{-1}(j)} (q^2_{u,v}(i) - q^2_{v,u}(i))\right| \le \sum_{i=1}^{n_{u,v}} |q^2_{u,v}(i) - q^2_{v,u}(i)| = TV(q^2_{u,v},q^2_{v,u}) \le \alpha \D(u,v),\] where the last step follows from $\alpha$-Notion 2. A similar argument shows that: 
\[TV(\TildeUncondDist{\Mech}{f}{u}, \TildeUncondDist{\Mech}{f}{v}) \le TV(q^1_{u,v}, q^1_{u,v})) + 0.5 |(1 - p(u)) - (1- p(v))| \le \alpha \D(u,v),\] where the last step follows from  $\alpha$-Notion 1 and individual fairness of $A$.

Now, we show Condition \ref{def:mmd}.\ref{def:mmd:1} (\textit{i.e.} ``nothing moves far and mass is conserved'') for the conversion of distributions over scores to distributions over adjusted scores. 
We handle the unconditional case (and a very similar argument works for the conditional case). We show condition 1 for $\TildeCondDist{\Mech}{f}{u}$, since an analogous argument shows condition 1 for $\TildeCondDist{\Mech}{f}{v}$. We use the following approach to move from the distribution over scores $\CondDist{\Mech}{f}{u}$ to the distribution over adjusted scores $\TildeCondDist{\Mech}{f}{u}$. First, we couple the distributions over scores and over adjusted scores into a carefully chosen joint distribution $\mathcal{X}_u \in \Delta([0,1] \times \Supp(\TildeUncondDist{\Mech}{f}{u}))$, where the $x$-coordinate can be thought of as the score and $y$-coordinate can be thought of as the adjusted score. Then, we implicitly specify the function $Z: [0,1] \rightarrow \Delta(\text{Supp}(\TildeCondDist{\Mech}{f}{u}))$ using the joint distribution $\mathcal{X}_u$.

% label the condition 1 and condition 2; condition 2.9.1 and 2.9.2 to clarify this

We define $\mathcal{X}_u$ as follows. We link the $x$-coordinate (score) and $y$-coordinate (adjusted score) through cohorts: that is, for each cohort $C$, we place probability of $\DistMech(C)$ on the (score, adjusted score) ordered pair given by $(f(C,u), Q(M_{u,v}(C)))$. More formally, for $x \in [0, 1]$ and $y \in \Supp(\TildeCondDist{\Mech}{f}{u}))$ such that $(x,y) \neq (0,0)$, we define:
\[\mathcal{X}_u((x,y)) = 
\begin{cases}
\sum_{C \in \mathcal{C}_u, f(C, u) = x, Q(\Map{u}{v}(C)) = y} \DistMech(C) & \text{ if } (x,y) \neq (0,0) \\
1 - p(u) + \sum_{C \in \mathcal{C}_u, f(C, u) = 0, Q(\Map{u}{v}(C)) = 0} \DistMech(C) & \text{ if } (x,y) = (0,0) \\
0 & \text{ otherwise}.
\end{cases}
\]
It is straightforward to observe that the marginal distribution of the $x$-coordinates of $\mathcal{X}_u$ is $\CondDist{\Mech}{f}{u}$, and the marginal distribution of the $y$-coordinates of $\mathcal{X}_u$ is $\TildeCondDist{\Mech}{f}{u}$.

Now, we are ready to define the function $Z: [0,1] \rightarrow \Delta(\text{Supp}(\TildeCondDist{\Mech}{f}{u}))$. For each $x \in [0,1]$, we define $Z(x)$, which is a probability measure on $\text{Supp}(\TildeCondDist{\Mech}{f}{u})$, as follows. First, we define the distribution $\mathcal{X}_u^x \in \Delta(\left\{x\right\} \times \text{Supp}(\tilde{S}_u^{N, A, f}))$ to be $\mathcal{X}_u$ conditioned on the $x$-coordinate being $x$. Then, $Z(x)$ is given by the marginal distribution of the $y$-coordinates of $\mathcal{X}_u^x$.

First, we show that the first sub-condition (\textit{i.e.} that ``nothing moves far'') is satisfied. It suffices to show that for all $C \in \mathcal{C}$, it holds that $|f(C, u) - Q(\Map{u}{v}(C))| \le \alpha \D(u,v)$. Suppose that $(C, u)$ is in cluster $i$. We know that $\frac{1}{2 \alpha} |f(C, u) - Q(\Map{u}{v}(C))| \le \frac{1}{2 \alpha}  0.5 (b_i - a_i) \le 0.5\D(u,v)$, since scores in cluster $i$ have small diameter by the conditions required for a mapping respecting $\frac{1}{2 \alpha} \DS$. This means that $|f(C, u) - Q(\Map{u}{v}(C))| \le (0.5) (2\alpha) \D(u,v)$, 

Now, we show that the second sub-condition on $\text{Supp}(\TildeCondDist{\Mech}{f}{u})$ (\textit{i.e.} that ``mass is conserved'') is satisfied. Let $\gamma$ be the probability mass function associated with $\mathcal{X}_u$. Moreover, for each $x \in [0,1]$, $Z(x)$ is a probability distribution over $\text{Supp}(\TildeCondDist{\Mech}{f}{u})$, and we let $z^x$ be its probability mass function. For each $y^* \in \text{Supp}(\TildeCondDist{\Mech}{f}{u})$, we wish to show that:
\[\TildeCondDist{\Mech}{f}{u}(y^*) = \sum_{x \in \text{Supp}(\CondDist{\Mech}{f}{u})} z^x(y^*) \CondDist{\Mech}{f}{u}(x).\]  Using the fact that the marginal distribution of $\mathcal{X}_u$ on the $x$-coordinates is $\CondDist{\Mech}{f}{u}$, along with the fact that $Z(x)$ is the distribution on the $y$-coordinates conditional on the $x$-coordinate being $x$, we can deduce that $\CondDist{\Mech}{f}{u}(x) z^x(y^*) = \gamma((x,y^*))$. Thus, we have that 
\[\sum_{x \in \text{Supp}{\CondDist{\Mech}{f}{u}}} z^x(y^*) \CondDist{\Mech}{f}{u}(x) = \sum_{(x, y^*) \in \text{Supp}{\mathcal{X}_u}} \gamma((x,y^*)).\] This is the probability mass at $y^*$ of the marginal distribution of the $y$-coordinates of $\mathcal{X}_u$. Using that the marginal distribution of the $y$-coordinates of $\mathcal{X}_u$ is $\TildeCondDist{\Mech}{f}{u}$, we know that $\sum_{(x, y^*) \in \text{Supp}{\mathcal{X}_u}} \gamma'((x,y^*)) = \TildeCondDist{\Mech}{f}{u}(y^*)$ as desired.
\end{proof}

\subsection{Additional results}

First, we state a corollary of Theorem \ref{thm:manytoonemapping} that gives conditions for robustness w.r.t expected score, using the fact that the weaker notion of expected score robustness follows from the stronger notion of mass-moving distance robustness. 
\begin{corollary}[Robustness to Post-Processing w.r.t Expected Score]
\label{cor:robustnessexpectedscore}
Let $\F$ be a class of scoring functions, and let $\alpha \ge 0.5$ be a constant. Suppose that $\left(\Map{u}{v}\right)_{u \neq v \in U}$ is in $\mathcal{M}_{\frac{1}{\alpha} \DS}$. If $\Mech$ satisfies $\alpha$-Notion 1 (resp. $\alpha$-Notion 2) for $\left(\Map{u}{v}\right)_{u \neq v \in U}$, then we have that $\Mech$ is $6\alpha$-robust w.r.t. $\mathcal{F}$ for $\DUncondE$ (resp. $\DCondE$).
\end{corollary}
\begin{proof}[Proof of Corollary \ref{cor:robustnessexpectedscore}]
This is implied by Theorem \ref{thm:manytoonemapping} and the using the relationship between mass-moving distance and expected score in Proposition \ref{prop:mmdimpliesexpectedscore}.
\end{proof}

Now, we show that Notion 1 is a ``weaker'' notion than Notion 2, which aligns with our result in Proposition \ref{prop:implicationexpectedscore} that unconditional fairness guarantees are ``weaker'' than conditional fairness  guarantees. More specifically, we show in Proposition \ref{prop:implicationnotions} shows that for a given mapping and individually fair \Mech, Notion 2 is stronger than Notion 1 up to Lipschitz factors.
\begin{proposition}
\label{prop:implicationnotions}
Let $\alpha \ge 0.5$ be a constant, and suppose that  $\left(\Map{u}{v}\right)_{u \neq v \in U}$ is a mapping and $\Mech$ is an individually fair cohort selection mechanism. If $A$ satisfies $\alpha$-Notion 2, then $A$ satisfies $\alpha$-Notion 1.
\end{proposition}
\begin{proof}[Proof of Proposition \ref{prop:implicationnotions}]
Notion 2 (\textit{i.e.} Definition \ref{def:conditions}.2) and individual fairness tell us that:
\[\frac{1}{2} \sum_{i=1}^{\MapNum{u}{v}} |q^2_{v,u}(i) - q^2_{u,v}(i)| \le (\alpha - 0.5)  \D(u,v),\]
\[|p(u) - p(v)| \le \D(u,v).\] We want to show that:
\[\frac{1}{2} \sum_{i=1}^{\MapNum{u}{v}} |q^1_{v,u}(i) - q^1_{u,v}(i)| \le \alpha \D(u,v). \]
We can write the first condition as:
\begin{align*}
  \frac{1}{2} \sum_{i=1}^{\MapNum{u}{v}} |\frac{q^1_{v,u}(i)}{p(v)} - \frac{q^1_{u,v}(i)}{p(u)}| &\le (\alpha - 0.5) \D(u,v)  \\
  \frac{1}{2} \sum_{i=1}^{\MapNum{u}{v}} |\frac{p(u) q^1_{v,u}(i)}{p(u) p(v)} - \frac{p(v) q^1_{u,v}(i)}{p(u) p(v)}| &\le \alpha \D(u,v) \\
  \frac{1}{2} \sum_{i=1}^{\MapNum{u}{v}} |p(u) q^1_{v,u}(i) - p(v) q^1_{u,v}(i)| &\le p(u) p(v) (\alpha - 0.5) \D(u,v) \\
   \frac{1}{2} \sum_{i=1}^{\MapNum{u}{v}} |p(u) q^1_{v,u}(i) - p(u) q^1_{u,v}(i) + p(u) q^1_{u,v}(i) - p(v) q^1_{u,v}(i)| &\le p(u) p(v) (\alpha - 0.5) \D(u,v).
\end{align*}
Now, we use the fact that $|A| - |B| = |A| - |-B| \le |A + B|$. We see that this means that
\[
  \frac{1}{2} \sum_{i=1}^{\MapNum{u}{v}} |p(u) q^1_{v,u}(i) - p(u) q^1_{u,v}(i)| - \frac{1}{2} \sum_{i=1}^{\MapNum{u}{v}} |p(u) q^1_{u,v}(i) - p(v) q^1_{u,v}(i)| \le p(u) p(v) (\alpha - 0.5)  \D(u,v).\]
 This means that:
 \begin{align*}
  \frac{1}{2} p(u) \sum_{i=1}^{\MapNum{u}{v}} |q^1_{v,u}(i) - q^1_{u,v}(i)| &\le p(u) p(v) (\alpha - 0.5)  \D(u,v) + \frac{1}{2} |p(u) - p(v)| \sum_{i=1}^{\MapNum{u}{v}}q^1_{u,v}(i) \\
  \frac{1}{2} \sum_{i=1}^{\MapNum{u}{v}} |q^1_{v,u}(i) - q^1_{u,v}(i)| &\le \frac{p(u) p(v) (\alpha - 0.5)  \D(u,v) + \frac{1}{2} |p(u) - p(v)| p(u)}{p(u)} \\
  &\le (p(v) (\alpha - 0.5) + \frac{1}{2}) \D(u,v) \\
  &\le \alpha \D(u,v).
\end{align*}
\end{proof}

We now show that satisfying $\alpha$-Notion 1 (or $\alpha$-Notion 2) is required for pipeline fairness when the metric $\DS$ is of a certain form. That is, we consider metrics $\dstar$ on (cohort, individual) pairs with the following structure. For each pair of individuals $u$ and $v$, consider the metric $\dstar_{u,v}$ defined by $\dstar$ restricted to the set $\mathcal{P}_{u,v} = (\mathcal{C}_u \times \left\{u \right\}) \times (\mathcal{C}_v \times \left\{v \right\})$. We focus on the case in which $\mathcal{P}_{u,v}$ has a partitioning into clusters s.t. $\dstar_{u,v}$ is large across clusters and small within clusters. (In fact, the condition that we place on each cluster bears some resemblance to the standard requirements of an $(\alpha, \beta)$-cluster of a graph \cite{mishra2007}, though our condition is adapted to metric spaces.) We formally define ``$(\alpha, \beta)$ metrics'' as follows:
\begin{definition}[$(\alpha,\beta)$-Metrics]
Let $d$ be a metric over some finite set $S$, and let $\beta > \alpha \ge 0$ be constants. Suppose that there exists a partition of $S$ into clusters that satisfies the following conditions: if $s_1, s_2 \in S$ are in the same cluster, then $d(s_1, s_2) \le \alpha$; if $s_1, s_2 \in S$ are in different clusters, then $d(s_1, s_2) \ge \beta$. Then, we say that $d$ is an $\ClusteringBF{\alpha}{\beta}$, and we call the partition into clusters the \textbf{induced partition}.\footnote{It is straightforward to verify that if such a partition exists, then it is unique.} 
\end{definition}
\noindent 
Suppose that a metric $\dstar$ over (cohort, individual) pairs has the property that  for all pairs of individuals $u$ and $v$, $\dstar_{u,v}$ is a $\Clustering{\D(u,v)}{1}$. The collection of induced partitions for each $\dstar_{u,v}$ gives a mapping. We call this mapping a \textbf{coarsest mapping} for $\dstar$, because for every pair $u$ and $v$, it is not possible for the partition to merge two clusters and still respect $\dstar$ (as per the requirements of Definition \ref{def:mapping}). Moreover, it is straightforward to verify that that this mapping is the \textit{unique} coarsest mapping, using the fact that $\dstar_{u,v}$ is a $\Clustering{\D(u,v)}{1}$.

Now, suppose that $\DS$ is such that for all pairs of individuals $u$ and $v$, $\frac{1}{\alpha}\DS_{u,v}$ is a $\Clustering{\D(u,v)}{\frac{1}{\alpha}}$. For metrics of this form, we show that satisfying $\alpha$-Notion 1 (resp. $\alpha$-Notion 2) is necessary for pipeline fairness: the intuition for necessity is that the conversion of $\frac{1}{\alpha} \DS_{u,v}$ into a coarsest mapping is not lossy from a fairness perspective. That is, if $\Mech$ does not satisfy $\alpha$-Notion 1 (resp. $\alpha$-Notion 2), we can construct a scoring function violating pipeline fairness: this scoring function can take advantage of similar individuals having different distributions of cohorts across clusters. 

Our proofs will rely on a standard lemma about extensions of functions on metric spaces and we present a proof for the sake of being self-contained. 
\begin{lemma}
\label{lemma:functioncompletion}
Let $\alpha > 0$ be a constant, and let $d$ be a pseudo-metric over $U$ and let $U' \subseteq U$. If $f: U' \rightarrow [0,1]$ is $\alpha$-Lipschitz w.r.t. $d|_{U'}$, then there exists a function $g: U \rightarrow [0,1]$ such that $g |_{U'} = f$ and that is $\alpha$-Lipschitz w.r.t $d$. 
\end{lemma}
\begin{proof}
Let $g': U \rightarrow \mathbb{R}$ be defined by $g'(x) = \inf_{u \in U'} \left\{f(u) + \alpha d(x,u)\right\}$. Observe that $g'|_{U'} = f$. Moreover, $\forall x \in U$, observe that for every $\epsilon > 0$, there exists $u^* \in U'$ such that: $g'(x) \ge f(u^*) + \alpha d(x,u^*) - \epsilon$. Moreover, we have by definition that $g'(y) \le  f(u^*) + \alpha d(y,u^*)$. Thus:
 $g'(y) - g'(x) \le f(u^*) + \alpha d(y,u^*) - (f(u^*) + \alpha d(x, u^*)) + \epsilon \le \alpha (d(y, u^* - d(x, u^*)) + \epsilon  \le \alpha d(x,y) + \epsilon$. Thus, $g'$ is $\alpha$-Lipschitz w.r.t. $d$. Now, we let $g(x) = \min(1, g'(x))$. We see that $g|_{U'} = g'|_{U'} = f$ and $g$ is $\alpha$-Lipschitz w.rt. $d$ since $|g(y) - g(x)| \le |g'(y) - g'(x)|$.   
\end{proof}

We first prove that the conditions in Theorem \ref{thm:manytoonemapping} are necessary for mass-moving distance,
using the structure of $\Clustering{\D(u,v)}{\frac{1}{\alpha}}$s.
\begin{theorem}[Necessity for mass-moving-distance]
\label{thm:necessarydefmmd}
Let $\F$ be a family of scoring functions, and let $\alpha \ge 1$ be a constant. Suppose that $\DS$ has the property that  for all pairs of individuals $u$ and $v$, $\frac{1}{\alpha} d^{\F}_{u,v}$ is a $\Clustering{\D(u,v)}{\frac{1}{\alpha}}$. Let $\left(\Map{u}{v}\right)_{u \neq v \in U}$ be the coarsest mapping for $\frac{1}{\alpha} \DS$. Suppose that $\D(u,v) < \frac{1}{\alpha \MapNum{u}{v}}$. Suppose that $\Mech$ does not satisfy $\alpha$-Notion 1 (resp. $\alpha$-Notion 2) is not satisfied for  $\left(\Map{u}{v}\right)_{u \neq v \in U}$. Moreover, suppose that $|p(u) - p(v)| = \D(u,v)$. Then, $\Mech$ is not $\alpha$-robust w.r.t. $\mathcal{F}$ for $\DUncondMMD$ (resp. $\DCondMMD$).
\end{theorem}
\begin{proof}[Proof of Theorem \ref{thm:necessarydefmmd}]
Suppose that $\Mech$ does not satisfy $\alpha$-Notion 1 (resp. $\alpha$-Notion 2) for $\left(\Map{u}{v}\right)_{u \neq v \in U}$. Then, there exists some pair of individuals $u$ and $v$ such that $TV(q_{u,v}^1, q_{v,u}^1) > (\alpha - 0.5) \D(u,v)$ (resp. $TV(q_{u,v}^2, q_{v,u}^2) > \alpha \D(u,v)$). We construct a scoring function $g$ where the mass-moving distance between $\UncondDist{\Mech}{g}{u}$ and $\UncondDist{\Mech}{g}{v}$ (resp. $\CondDist{\Mech}{g}{u}$ and $\CondDist{\Mech}{g}{v}$) is larger than $\alpha \D(u,v)$. 

Fix $\epsilon > 0$ sufficiently small. As before, let $\mathcal{P}_{u,v} = (\mathcal{C}_u \times \left\{u \right\}) \cup (\mathcal{C}_v \times \left\{v \right\})$ be the set of all (cohort, individual) pairs involving $u$ or $v$. We define a scoring function $f: \mathcal{P}_{u,v} \rightarrow [0,1]$. As before, for every $1 \le i \le \MapNum{u}{v}$, let $S(i) = \left\{(C, u) \in \Map{u}{v}^{-1}(i)\right\} \cup \left\{(C, v) \in \Map{v}{u}^{-1}(i)\right\}$ be the (cohort, individual) pairs appearing in cluster $i$. For $(C,x) \in S(i)$, we take $f(C, x)$ to be $i (\alpha + \epsilon) \D(u,v)$. Since $\D(u,v) < \frac{1}{\alpha \MapNum{u}{v}}$, we can make $\epsilon$ small enough so that all of the scores are in $[0,1]$. 

Now, we show that $f$ is $1$-Lipschitz with respect to $\DS$ restricted to the domain $\mathcal{P}_{u,v}$, for sufficiently small $\epsilon$. Within each cluster, $f$ is constant, so clearly it is $1$-Lipschitz within each cluster. Now, consider (cohort, individual) pairs in different clusters. If $(C, x), (C', y)$ are in different clusters, we know that $\DS((C, x), (C', y)) \ge 1$ based on the fact that $\frac{1}{\alpha} d^{\F}_{u,v}$ is a $\Clustering{\D(u,v)}{\frac{1}{\alpha}}$, so $d^{\F}_{u,v}$ is a $\Clustering{\alpha \D(u,v)}{1}$. Thus, $f$ is $1$-Lipschitz. 

% If $(C, x), (C', y)$ are in different clusters, we know that $\DS((C, x), (C', y)) >  \alpha \D(u,v)$ based on the conditions on $\DS$ in the theorem statement. 

% In fact, since $U$ is finite, there must be a gap between the minimum distance between clusters and $\alpha \D(u,v)$. That is: $\inf_{(C, x), (C', y) \in \mathcal{P}_{u,v} \text{ that are in different clusters }} \DS((C,x), (C', y)) > \alpha \D(u,v)$. This means that $\inf_{(C, x) \in S(i), (C', y)} $
% We can thus make $\epsilon$ sufficiently small so that $f$ is $1$-Lipschitz. 

We apply Lemma \ref{lemma:functioncompletion} to complete $f$ into a function $g: \mathcal{C} \times U \rightarrow [0,1]$ that is $1$-Lipschitz w.r.t. $\DS$. We now show that the mass moving distance between $\UncondDist{\Mech}{g}{u}$ and $\UncondDist{\Mech}{g}{v}$ (resp. $\CondDist{\Mech}{g}{u}$ and $\CondDist{\Mech}{g}{v}$) is larger than $\alpha \D(u,v)$. Assume for sake of contradiction that the mass moving distance is $\le \alpha \D(u,v)$. That would mean that there exist $\tilde{\gamma}_1$ and $\tilde{\gamma}_2$, along with functions $Z_1$ and $Z_2$, that satisfy conditions 1 and 2 in the mass moving distance definition for $(\alpha + \epsilon / 2) \D(u,v)$. By the ``mass does not move far'' condition, we know that for $l = 1,2$, the support of the probability measure $Z_l(i(\alpha + \epsilon)) \in \Delta(\tilde{\gamma}_l)$ must be disjoint from the support of the probability measure $Z_l(j(\alpha + \epsilon)) \in \Delta(\tilde{\gamma}_l)$ for any $0 \le i \neq j \le n_{u,v}$. Thus, $TV(\tilde{\gamma}_1, \tilde{\gamma}_2)$ must be at least $TV(\UncondDist{\Mech}{g}{u}, \UncondDist{\Mech}{g}{v})$ (resp. $TV(\CondDist{\Mech}{g}{u}, \CondDist{\Mech}{g}{v})$). We see that  
$TV(\UncondDist{\Mech}{g}{u}, \UncondDist{\Mech}{g}{v}) = TV(q^1_{u,v}, q^1_{v,u}) + 0.5 |(1 - p(u)) - (1 - p(v))| = (\alpha + \epsilon) \D(u,v)$ (resp. $TV(\CondDist{\Mech}{g}{u}, \CondDist{\Mech}{g}{v}) = TV(q^2_{u,v}, q^2_{v,u}) = (\alpha + \epsilon) \D(u,v)$), which is a contradiction. This proves the desired statement.
\end{proof}

When are our condition in Theorem \ref{thm:manytoonemapping} also ``tight'' for expected score fairness? We prove that our definition is necessary (up to Lipschitz constants) for expected score, again using the structure of $\Clustering{\D(u,v)}{\frac{1}{\alpha}}$s.
\begin{theorem}[Necessity for expected score]
\label{thm:necessarydefexpectedscore}
Let $\F$ be a family of scoring functions, and suppose that $\DS$ has the property that  for all pairs of individuals $u$ and $v$, $\frac{1}{\alpha} \DS_{u,v}$ is a $\Clustering{\D(u,v)}{\frac{1}{\alpha}}$. Let $\left(\Map{u}{v}\right)_{u \neq v \in U}$ be the coarsest mapping for $\frac{1}{\alpha} \DS$. Suppose that $(\alpha+0.5)$-Notion 1 (resp. $\alpha$-Notion 2) is not satisfied for  $\left(\Map{u}{v}\right)_{u \neq v \in U}$. Moreover, suppose that $|p(u) - p(v)| = \D(u,v)$. Then, $\Mech$ is not $\alpha$-robust w.r.t. $\mathcal{F}$ for $\DUncondE$ (resp. $\DCondE$).
\end{theorem}
\begin{proof}[Proof of Theorem \ref{thm:necessarydefexpectedscore}]
Suppose that $\Mech$ does not satisfy $\alpha$-Notion 1 (resp. $\alpha$-Notion 2) for $\left(\Map{u}{v}\right)_{u \neq v \in U}$. (See Definition \ref{def:conditions}.) Then, there exists some pair of individuals $u$ and $v$ such that $TV(q_{u,v}^1, q_{v,u}^1) > \alpha \D(u,v)$ (resp. $TV(q_{u,v}^2, q_{v,u}^2) > \alpha \D(u,v)$). We construct a scoring function $g$ where
$|\mathbb{E}[\UncondDist{\Mech}{g}{u}] - \mathbb{E}[\UncondDist{\Mech}{g}{v}]| > \alpha \D(u,v)$ (resp. $|\mathbb{E}[\CondDist{\Mech}{g}{u}] - \mathbb{E}[\CondDist{\Mech}{g}{v}]| > \alpha \D(u,v)$). 

As before, let $\mathcal{P}_{u,v} = (\mathcal{C}_u \times \left\{u \right\}) \cup (\mathcal{C}_v \times \left\{v \right\})$ be the set of all (cohort, individual) pairs involving $u$ or $v$. We define a scoring function $f: \mathcal{P}_{u,v} \rightarrow \left\{0,1\right\}$. As before, for every $1 \le i \le \MapNum{u}{v}$, let $S(i) = \left\{(C, u) \in \Map{u}{v}^{-1}(i)\right\} \cup \left\{(C, v) \in \Map{v}{u}^{-1}(i)\right\}$ be the (cohort, individual) pairs appearing in cluster $i$. Let's partition $\left\{1, \ldots, \MapNum{u}{v}\right\}$ into two groups $P_1$ and $P_2$ as follows. WLOG, suppose that $p(u) \ge p(v)$. We define $P_1$ such that $q^1_{u,v}(i) \ge q^1_{v,u}(i)$ (resp. $q^2_{u,v}(i) \ge q^2_{v,u}(i)$) and define $P_2$ such that $q^1_{u,v}(i) < q^1_{v,u}(i)$ (resp. $q^2_{u,v}(i) < q^2_{v,u}(i)$).
For $i \in P_1$ and $(C,x) \in S(i)$, we take $f(C, x)$ to be $1$. For $i \in P_2$ and $(C,x) \in S(i)$, we take $f(C, x)$ to be $0$.

Now, we show that $f$ is $1$-Lipschitz with respect to $\DS$ restricted to the domain $\mathcal{P}_{u,v}$, for sufficiently small $\epsilon$. Within each cluster, $f$ is constant, so clearly it is $1$-Lipschitz within each cluster. Now, consider (cohort, individual) pairs in different clusters. If $(C, x), (C', y)$ are in different clusters, we know that $\DS((C, x), (C', y)) = 1$  based on the fact that $\frac{1}{\alpha} d^{\F}_{u,v}$ is a $\Clustering{\D(u,v)}{\frac{1}{\alpha}}$, so $d^{\F}_{u,v}$ is a $\Clustering{\alpha \D(u,v)}{1}$. Thus, $f$ is $1$-Lipschitz.  

We apply Lemma \ref{lemma:functioncompletion} to complete $f$ into a function $g: \mathcal{C} \times U \rightarrow [0,1]$ that is $1$-Lipschitz w.r.t. $\DS$. 
We now show that
$|\mathbb{E}[\UncondDist{\Mech}{g}{u}] - \mathbb{E}[\UncondDist{\Mech}{g}{v}]| > \alpha \D(u,v)$ (resp. $|\mathbb{E}[\CondDist{\Mech}{g}{u}] - \mathbb{E}[\CondDist{\Mech}{g}{v}]| > \alpha \D(u,v)$). We see that $|\mathbb{E}[\UncondDist{\Mech}{g}{u}] - \mathbb{E}[\UncondDist{\Mech}{g}{v}]| = \left|\sum_{i \in P_1} q_{u,v}^1(i) - \sum_{i \in P_1} q_{v,u}^1(i)\right| = TV(q_{u,v}^1, q_{v,u}^1) > \alpha \D(u,v)$. (Similarly, we see that $|\mathbb{E}[\CondDist{\Mech}{g}{u}] - \mathbb{E}[\CondDist{\Mech}{g}{v}]| = \left|\sum_{i \in P_1} q_{u,v}^2(i) - \sum_{i \in P_1} q_{v,u}^2(i)\right| = TV(q_{u,v}^2, q_{v,u}^2) > \alpha \D(u,v)$.)
\end{proof}
\noindent Theorem \ref{thm:necessarydefexpectedscore} is somewhat surprising for the following reason: by Theorem \ref{thm:manytoonemapping}, Notion 1 (resp. Notion 2) in the theorem statement actually gives the stronger notion of mass-moving distance fairness, but we actually show that it is necessary even for the weaker notion of expected score. We can view this result as telling us that for certain classes of post-processing functions, we get the robustness w.r.t mass-moving distance ``for free'' as a consequence of robustness w.r.t expected score.

%% file: sections/existingmechs.tex
% exising mechanisms
Two mechanisms for fair cohort selection were given in \cite{DI2018} based on converting an individually fair classifier for independent classification into a mechanism to select exactly $n$ individuals. The first, ``Permute then Classify'' applies the fair classifier to each element in random order and either (1) stops when $n$ elements are selected or (2) chooses the remaining unclassified elements to get a total of $n$. The second, ``Weighted Sampling'' samples from the set of all cohorts of size $n$ where each cohort is assigned a probability proportional to the sum of the ``weights'' assigned to each element in the cohort by the fair classifier.

In this appendix, we give the formal specifications for each of these mechanisms, show that they satisfy the monotonicity property required in Lemma \ref{lemma:monotone}, and give an extension of weighted sampling to allow individually fair selection from an arbitrary set of cohorts.

\subsection{Permute then Classify}
\begin{mech}[PermuteThenClassify \cite{DI2018}]\label{mech:PTC}
  Given a universe $U$, a cohort size $n \leq |U|$ and an individually fair classifier $C: U \rightarrow \{0,1\}$, first choose a permutation $\pi \sim S_{|U|}$ uniformly at random random from the symmetric group on $|U|$. Initialize an empty cohort $l$. Evaluating the elements of $U$ in the order specified by $\pi$, apply $C$ to each element. If $C(u)=1$ and there are fewer than $n$ elements in the cohort, add $u$ to the cohort. If there are no more than $n-|l|$ elements left to be evaluated (i.e., the only way to select $n$ total is to accept all remaining elements), then add all remaining elements in the permutation to $l$.
\end{mech}

\begin{theorem}[Permute then Classify is individually fair \cite{DI2018} ]\label{theorem:ptc}PermuteThenClassify is a solution to the Cohort Selection Problem for any $C$ that is individually fair when operating on all elements of the universe.
\end{theorem}

To satisfy the requirements of Lemma \ref{lemma:monotone}, it suffices to show that the Permute then Classify mechanism is monotone, i.e. if an individual $u$ is preferred to $v$, then the probability of choosing any cohort with $u$ swapped for $v$ is larger than the probability of selecting the original cohort.

\begin{lemma}[PermuteThenClassify is monotone]
The permute then classify mechanism is monotone, i.e., $\Pr[\mathcal{P}_{C,n,N}= l \cup x] \geq \Pr[\mathcal{P}_{C,n,N}=l \cup y]$ if $\Pr[C(x)=1] \geq \Pr[C(y)=1]$, where $\mathcal{P}_{C,n,N}$ is the permute then classify mechanism instantiated with a randomized classifier $C:U \times r \rightarrow \{0,1\}$ choosing $n$ elements from a set of $N$.
\end{lemma}

\begin{proof}
Recall that Permute then Classify first chooses a permutation uniformly at random from the symmetric group on $[N]$ $\pi \sim S([N])$ for an input of size $N$. It then runs $C$ on each element until either there are $n$ elements "accepted" by $C$, or there only enough elements left in the permutation to make $n$, in which case all remaining elements are selected.

Fix a pair of elements $x$ and $y$. Consider any permutation $\pi$ and any set of elements $l$ such that $|l|=n-1$ and $x \notin l$ and $y \notin l$.
Without loss of generality, suppose that $x$ appears before $y$ in $\pi$. Call the permutation with $x$ and $y$ swapped $\pi'$.

\noindent Call the elements of $l$ that appear before $x$, $l_1$, those that appear in between $x$ and $y$ $l_2$ and those that appear after $y$ $l_3$.

\noindent Given $\pi,$ the probability of choosing $C \cup x = \Pr[l_1 ]* \Pr[C(x) | l_1]* \Pr[l_2 | l_1, C(x)]*\Pr[\bar{C(y)}]*\Pr[l_3 | l_1,l_2,C(x),\bar{C(y)}]$.
Notice that this statement is equivalent with $x$ and $y$ switched under $\pi'$.

\noindent Given $\pi$, the probability of choosing $C \cup y = \Pr[l_1]* \Pr[\bar{C(x)}| l_1]* \Pr[l_2 | l_1 ,\bar{C(x)}]*\Pr[C(y)]*\Pr[l_3 | l_1,l_2,C(y),\bar{C(x)}]$.
As above, this statement is equivalent with $y$ and $x$ switched under $\pi'$.

\noindent Notice that $\Pr[l_3 | l_1,l_2,C(x),\bar{C(y)}]=\Pr[l_3 | l_1,l_2,C(y),\bar{C(x)}]$, etc
as the probability is only dependent on having a sufficient number of slots left.

\noindent Thus, we can relate the probability of $C \cup x$ chosen under $\pi$ or $\pi'$ to the probability of choosing $C \cup y$ :
$$ \Pr[C \cup x| \pi \lor \pi'] - \Pr[C \cup y | \pi\lor\pi'] = \Pr[l_1,l_2,l_3][(1-C(y))C(x) - (1-C(x))C(y)] *2$$
$$ \Pr[C \cup x| \pi \lor \pi'] - \Pr[C \cup y | \pi\lor\pi'] \geq 0$$
Thus, we conclude that Permute then Classify is monotone.

\end{proof}

\subsection{Weighted Sampling}
First we introduce the weighted sampling mechanism, as described in \cite{DI2018}.

\begin{mech}[Weighted Sampling \cite{DI2018}]
  Given an individually fair classifier $C: U \rightarrow [0,1]$, and a cohort size $n$, define the $L$ to be the set of subsets of $U$ of size $n$. Assign each subset $l \in L$ weight $w(l) \leftarrow \sum_{u \in l}\E[C(u)]$. Define a distribution over sets of size $n$, $\mathcal{X} $ such that the weight of $l$ under $\mathcal{X}$ is $\frac{w(l)}{\sum_{l' \in L}w(l')}$. Choose a set according to $\mathcal{X}$ as output.
\end{mech}
%
% \begin{algorithm}[tb]
%    \caption{$\mathsf{WeightedSampling}$ \cite{DI2018}}
%    \label{alg:weightedsampling}
% \begin{algorithmic}
%    \STATE {\bfseries Input:} $n \leftarrow $ the number of elements to select\\
% $C \leftarrow$ a classifier $C: U\times r \rightarrow  \{0,1\}$ \\
% $L \leftarrow$ the set of all subsets of $U$ of size $n$\\
%
% \FOR { $l\in L$}
% \STATE $w(l) \leftarrow \sum_{u \in l}\E[C(u)]$ // set the weight of each set
% \STATE Define $\mathcal{X} \in \Delta(L)$ such that $\forall l \in L$, the weight of $l$ under $\mathcal{X}$ is $\frac{w(l)}{\sum_{l' \in L}w(l')}$\\
% \STATE $M \sim \mathcal{X}$ // Sample a set of size $n$ according to $\mathcal{X}$\\
% \ENDFOR
% \STATE return $M$
%
% \end{algorithmic}
% \end{algorithm}

\begin{theorem}[Weighted sampling is individually fair \cite{DI2018}]\label{thm:fairweightedsampling}For any individually fair classifier $C$ such that the $\Pr_{u \sim U}[C(u)=1] \geq 1/|U|$, weighted sampling is individually fair.
\end{theorem}

Notice that the specification of the weighted sampling mechanism immediately implies that the mechanism is monotone, as $w(C) = w(\{C\backslash\{u\}\}\cup \{v\}) + C(u) - C(v)$.
Despite this monotonicity property, WeightedSampling still runs into issues with Notion 2.
\begin{proposition}
\label{prop:weightedsamplecounterexample}
Suppose that WeightedSampling is run with $\sum_{u \in U} w(u) = 1$. It does not satisfy $\alpha_1$-Notion 2 (Definition \ref{def:conditions}.2) for any constant $\alpha_1$ that is independent of $|U|$, $k$ (the size of the cohort), $\left\{w(u) \right\}_{u \in U}$, $\D$.
\end{proposition}
\begin{proof}
To show this counter-example, we take the (realistic) infinite sequence of $(k, U)$ pairs where $k << U$ and choose weights in terms of these quantities, and show that no such constant $\alpha_1$ independent of $k$ and $U$ exists. Let $S = \sum_{u \in U} w(u)$, which we set to $1$. Suppose that $w(y) = 0.5$ for some $y \in U$. Suppose that $w(u) = 0$ and $w(v) = S \left(\frac{k \log k}{|U|-1} - \frac{k-1}{|U|-1}\right) \frac{|U|-1}{|U| - k}$. A straightforward calculation using the expression for sampling probability in Weighted Sampling and simplifying shows that $p(x) = \frac{w(x)}{S} \frac{|U|-k}{|U| - 1} + \frac{k-1}{|U| - 1}$ for all $x \in X$. Plugging this in, we obtain that  $p(u) = \frac{k-1}{|U| - 1}$ and $p(v) = \frac{k \log k}{|U|-1}$.

We take a particular set of cohorts where the contribution to total variation distance will blow-up.
Let's consider all cohorts $C$ with $y, u$ and $k-2$ elements not including $v$. Let's also consider their corresponding mapped sets using the swapping mapping. 
Observe that the probability that $C = C' \cup \left\{u \right\}$ is chosen is $\frac{w(u) + \sum_{x \in C'} w(x)}{N}$ and the probability that $C' \cup \left\{v \right\}$ is chosen is $\frac{w(v) + \sum_{x \in C'} w(x)}{N}$ where $N = {{|U| - 1} \choose {k-1}} S$. Thus the contribution to the total variation distance of $C$ is:
\begin{align*}
Q &= \left|\frac{w(u) + \sum_{x \in C'} w(x)}{N p(u)} - \frac{w(v) + \sum_{x \in C'} w(x)}{N p(v)}\right| \\
&= \left|\frac{1}{{{|U| - 1} \choose {k-1}}} \left(\frac{w(u) / S}{p(u)} - \frac{w(v)/S}{p(v)} + \frac{\sum_{x \in C'} w(x)}{S} \left(\frac{1}{p(u)} - \frac{1}{p(v)}\right)\right)\right|. 
\end{align*}

Now, observe that \[\frac{w(u) / S}{p(u)} = \frac{w(u) / S}{(w(u) / S)\left[\frac{|U|-k}{|U| - 1}\right] + \frac{k-1}{|U| - 1}} = \frac{|U|-1}{|U| - k} \frac{(w(u) / S) \frac{|U|-k}{|U| - 1}}{(w(u) / S)\left[\frac{|U|-k}{|U| - 1}\right] + \frac{k-1}{|U| - 1}} .\]
This is equal to:
\[ \frac{|U|-1}{|U| - k}\left( 1 - \frac{\frac{k-1}{|U| - 1}}{p(u)}\right).\]
Now, this means that
\[\frac{w(u) / S}{p(u)} - \frac{w(v) / S}{p(v)} =  \frac{k-1}{|U| - k} \left( \frac{1}{p(v)} - \frac{1}{p(u)}\right).  \]
Thus, we know that 
\[Q = \left| \frac{1}{p(u)} - \frac{1}{p(v)}\right| \frac{1}{{{|U| - 1} \choose {k-1}}}\left| \frac{\sum_{x \in C'} w(x)}{S} - \frac{k-1}{|U| - k} \right|. \] Using the fact that $y \in C$, we can rewrite this as: 
\[Q = \left| \frac{1}{p(u)} - \frac{1}{p(v)}\right| \frac{1}{{{|U| - 1} \choose {k-1}}}\left| \frac{w(y) + \sum_{x \neq y \in C'} w(x)}{S} - \frac{k-1}{|U| - k} \right|. \] 

Observe that $\frac{w(y) + \sum_{x \neq y \in C'} w(x)}{S} \ge \frac{0.5}{S} = 0.5$. If  $|U| >> k$, then this is bigger than $\frac{k-1}{|U| - k}$.
Thus, we see that:
\[Q \ge \left| \frac{1}{p(u)} - \frac{1}{p(v)}\right| \frac{1}{{{|U| - 1} \choose {k-1}}}\left| 0.5 - \frac{k-1}{|U| - k} \right|. \] When $|U| >> k$, we  see that:
\[Q \ge \left| \frac{1}{p(u)} - \frac{1}{p(v)}\right| \frac{1}{{{|U| - 1} \choose {k-1}}} 0.4. \]

Now, let's sum over all cohorts $C$ containing $y, u$ and $k-2$ elements not including $v$. Thus, there are ${{|U| - 3} \choose {k-2}}$ cohorts to sum over, so we obtain:
\[\left| \frac{1}{p(u)} - \frac{1}{p(v)}\right| \frac{{{|U| - 3} \choose {k-2}}}{{{|U| - 1} \choose {k-1}}} 0.25 = \left| \frac{1}{p(u)} - \frac{1}{p(v)}\right| 0.4 \frac{(k-1)}{|U|-1} \frac{|U|-k}{|U|-2}. \] When $|U| >> k$, this can be lower bounded by:
\[\left| \frac{1}{p(u)} - \frac{1}{p(v)}\right| 0.25 \frac{(k-1)}{|U|-1}. \]
Using our settings for $p(u)$ and $p(v)$, we see that $\left| \frac{1}{p(u)} - \frac{1}{p(v)}\right| = \frac{|U|-1}{k-1} - \frac{|U|-1}{k \log k}$. When $k$ is sufficiently large, the distance becomes roughly $0.25$ instead of $ \frac{k \log k}{|U|-1}$, so there is no such constant $\alpha_1$ since $\frac{|U| -1}{k \log k}$ is unbounded as $|U| \rightarrow \infty$ when $k << |U|$. 
\end{proof}

\subsection{Adapting weighted sampling for structured cohort sets}\label{appendix:ws}
Permute then Classify and Weighted Sampling behave \textit{almost} as if individuals are classified independently, i.e., the probability of selecting an individual depends only on their independent classification probability and whether there is space left in the cohort.
However, these solutions have a considerable drawback in practice: they are unstructured.
For example, a college admitting a class of 70\% female students or 90\% athletes would cause significant churn in the resources and facilities needed year over year. To best utilize its resources, and perhaps more importantly to expose students to classmates with a variety of backgrounds and interests, the college would naturally want to impose some structure on the classes admitted.

Fortunately, weighted sampling can be adapted to select cohorts with some underlying structure in an individually fair way.
% In practice, however, such unstructured cohort selection mechanisms may not always be useful.
% For example, colleges may want to ensure that the classes of students they admit have a good mix of interests and backgrounds, and may want to tune their mechanism to make sure they get a good mix rather than all one type of student. Indeed, it is entirely possible to construct individually fair cohort selection mechanisms which don't have the properties above. To give an extreme example, one could pick an arbitrary partition of the universe into sets of size $n$ and choose a set uniformly at random. % cilvento: I think this was a copy-paste error
Given a set of ``acceptable'' cohorts $\mathcal{C}$, i.e. cohorts satisfying some property like a diverse set of student interests, weighted sampling (with weights based on a solution to a linear program constraining differences in selection probability for each individual, as in the original linear program in \cite{dwork2012fairness}) can be used to select a single cohort. Roughly speaking, the constraints for solving this linear program concern the number of cohorts in $\mathcal{C}$ in which each individual appears, not their relative qualification or distances within these cohorts.
While we can imagine such a setup being used for good reason, it can also be abused to construct cohorts that justify discrimination in later stages.
Returning to the malicious example from the introduction, notice that a set of ``acceptable cohorts'' could be the set of cohorts which mostly satisfy the property that the most talented person in the cohort is not a minority candidate, giving the veneer of individual fairness to a pipeline explicitly constructed to unfairly discriminate.

%\inote{add formal thm statment and proof here.}

\begin{theorem}\label{theorem:craftedcohorts}
Given a universe $U$ and a distance metric $\D$ and a set of permissible cohorts $\mathcal{C}$, such that the subset of permissible cohorts  $\mathcal{C}^{u}$ containing an individual $u$ and the subset of permissible cohorts  $\mathcal{C}^{v}$ containing an individual $v$ satisfy $\frac{||\mathcal{C}^{u} |-|\mathcal{C}^{v}| |}{|\mathcal{C}|} \leq \D(u, v)$
and there exists a subset of cohorts $\mathcal{C}^P\subseteq \mathcal{C}$ such that no element appears in more than one cohort in $\mathcal{C}^P$ and $\mathcal{C}^P$ forms a partition of $U$,
 then there exists a set of weights for the cohorts in $\mathcal{C}$ such that choosing a single cohort by  sampling proportional to these weights results in individual fairness.
\end{theorem}
\begin{proof}
First, we translate the requirements for individual fairness and constructing the set of weights into a linear program with variables $w_i$ for each cohort in $\mathcal{C}$.
\[\{w_i | i \in [|\mathcal{C}|]\} \text{ s.t. }
\begin{cases}
w_i \geq 0,\\
\sum_{i}w_i=1\\
\sum_{i \in \mathcal{C}^u}w_i - \sum_{i \in \mathcal{C}^v}w_i \leq D(u,v) \forall u,v \in U\\
\sum_{i \in \mathcal{C}^v}w_i - \sum_{i \in \mathcal{C}^u}w_i \leq D(u,v) \forall u,v \in U
\end{cases}\]

To solve the system, we take the following steps:
\begin{enumerate}
  \item Solve the system without the non-negativity constraint.
  \item  Determine the maximum magnitude negative weight, and call the magnitude $w_{*}$.
  \item Add $w_{*}$ to all weights in the original solution.
  \item Take $y = \sum_{i}w_{i} = 1 +|\mathcal{C}|w_{*}$.
  \item Divide all weights by $y$.
\end{enumerate}

Notice that Steps 3-5 ensure that all weights are positive in the total sum of the weights is equal to $1$. Thus it remains to characterize under what conditions the distance constraints are also satisfied after the adjustments in Steps 3-5.
Given the requirement that a subset of $\mathcal{C}$ exactly partition $U$, there always exists a solution to the system which is to place equal weight on each of the cohorts in the partition subset.

Notice that the adjustments do not violate the distance constraints when each element appears an equal number of sets. To handle the more general case, take $y_{u,v} = ||\mathcal{C}^{u} |-|\mathcal{C}^{v}||$, i.e. the difference in the number of cohorts  $u$ and $v$ appear in. Without loss of generality, assume that $u$ participates in more cohorts than $v$. To satisfy the distance constraints we need
\[(y_{u, v}w_{*}+\sum_{i\in \mathcal{C}^{u}}w_{i}-\sum_{i\in \mathcal{C}^{v}}w_{i})\frac{1}{y}\leq \D(u, v)\]
which follows from applying the steps above.

In the worst case, the original solution took on the maximum distance between $u$ and $v$. Substituting $\sum_{i\in \mathcal{C}^{u}}w_{i}-\sum_{i\in \mathcal{C}^{v}}w_{i} = \D(u, v)$,
\[(y_{u, v}w_{*}+\D(u, v))\frac{1}{y}\leq \D(u, v)\]
\[y_{u, v}w_{*}\leq \D(u, v)(y-1)\]
Substituting for the value of $y$:
\[y_{u, v}w_{*}\leq \D(u, v)|\mathcal{C}|w_{*}\]
\[\frac{y_{u, v}}{|\mathcal{C}|}\leq \D(u, v)\]

Thus, so long as the condition requiring that every pair of individuals participate in a similar number of cohorts is satisfied, the theorem statement holds.
\end{proof}

Thus, as long as the difference in number of sets participated in for each pair as a fraction of the total number of sets is less than the distance between the pairs and the permissible cohorts and for partition of the universe, a solution can be found.
Such requirements are reasonably easy to check before attempting to solve the system and in specifying the cohorts.

%% file: sections/mechanisms-detailed.tex
\subsection{A pathological scoring function family.}\label{appendix:pathologicalexample}
Unfortunately, it is not possible to achieve robustness with respect to arbitrary families of intra-cohort individually fair mechanisms: there are pathological classes of scoring functions in which there is no robust cohort selection mechanism. Example \ref{example:impossibility} below illustrates a set of permissible cohorts $\mathcal{C}$ and scoring function $f$ for which there is no robust, individually fair cohort selection mechanism.
\begin{example}
\label{example:impossibility}
Consider a universe $U = \left\{a, b, c\right\}$ of three equivalent individuals, and a fairness metric such that $\D(x,y) = 0$ for all $x, y \in U$. Suppose that $\mathcal{C} = \{a,b\}, \{a,c\}, \{b,c\}$ and $f$ is a scoring function defined so that: $f(\{a,b\}, a) = f(\{a,b\}, b) = 0$, $f(\{a,c\}, a) = f(\{a,c\}, c) = 1$, and $f(\{b,c\}, b) = f(\{b,c\}, c) = 0.5$. 

If $\Mech$ is an individually fair cohort selection mechanism, then $\DistMech(\{a,b\}) = \DistMech(\{a,c\}) = \DistMech(\{b,c\})$. The unconditional expected scores are $\mathbb{E}[\UncondDist{A}{f}{a}] = 1/3$, $\mathbb{E}[\UncondDist{A}{f}{b}] = 1/6$, and $\mathbb{E}[\UncondDist{A}{f}{c}] = 1/2$, and the conditional scores are $\mathbb{E}[\UncondDist{A}{f}{a}] = 1/2$, $\mathbb{E}[\UncondDist{A}{f}{b}] = 1/4$, and $\mathbb{E}[\UncondDist{A}{f}{c}] = 3/4$. Thus, no individually fair cohort selection mechanism $\Mech$ is robust with respect to $f$. 
\end{example}
\noindent The fundamental issue in Example \ref{example:impossibility} is that $f$ is permitted to deviate wildly between cohorts, and as a result $d^{\left\{f\right\}}$ is large on cohort contexts that look intuitively identical. In practice, if the original cohort selection mechanism has some control over the determination of $\mathcal{C}$, such pathological cases may be avoidable.

%We prove the results for Section 4. 

 \subsection{Proofs for Section \ref{sec:mechanisms}}

 We prove Proposition \ref{prop:easiestmech}, restated here. 
 \begin{proposition}[Restatement of Proposition \ref{prop:restatedeasiestmech}]
\label{prop:restatedeasiestmech}
Consider the mapping that, for each pair of individuals $u$ and $v$, places all of the cohort contexts in $(\mathcal{C}_u \times \left\{u\right\}) \cup (\mathcal{C}_v \times \left\{v\right\})$ into the same cluster. If $\Mech$ is individually fair, then $\Mech$ satisfies $0.5$-Notion 1 and $0.5$-Notion 2 w.r.t. this mapping. 
\end{proposition}
\begin{proof}
Pick any pair of individuals $u$ and $v$. Using the mapping described in the proposition statement, we know that $TV(q_{u,v}^1, q_{v,u}^1) = 0.5 |p(u) - p(v) | \le 0.5 \D(u,v)$, and $TV(q_{u,v}^2, q_{v,u}^2) = 0$, as desired. 
\end{proof}
 
% We explore $\F_1$, scoring functions that do not use cohort context. Let $\DS((C,u), (C', v)) = \D(u,v)$ for all $C, C' \in \mathcal{C}$, $u, v \in U$. Then we see that $\F_1 \subseteq \mathcal{F}_{d^S, 1}$. We obtain the following condition using Theorem \ref{thm:manytoonemapping}:
% \begin{proof}[Proof of Proposition \ref{prop:easiestmech}]
% We see that for all $(C_1, x), (C_2, y) \in P(1)$ (as in the definition of a mapping based on $\DS$), it is true that $\DS((C,x), (C', y)) = d(x,y)$. This is $0$ if $x = y$ and $\D(u,v)$ otherwise. Now, we see that for Notion 1, the total variation distance is bounded by $0.5 |p(u) - p(v)| \le 0.5 \D(u,v)$, and for Notion 2, the total variation distance is $0$. 
% \end{proof}
% Thus, Proposition \ref{prop:easiestmech} and Theorem \ref{thm:manytoonemapping} tell us that any individually fair mechanism will be $1$-robust w.r.t $\F_1$ for $\DUncondMMD$. 

   \subsection{Proofs for Section \ref{subsec:F2}}
 We prove Lemma \ref{lemma:monotone}, restated here.   
 The intuition for the link between the monotonicity property and the swapping mapping is that the probability masses on a cohort containing $u$ and a cohort containing $v$ that are paired in the swapping mapping are directionally aligned.
 \begin{lemma}[Restatement of Lemma \ref{lemma:restatedmonotone}]
\label{lemma:restatedmonotone}
Suppose that $\mathcal{C} \subseteq 2^U$ is the set of cohorts of size $k$. If $A$ is monotonic, then $A$ satisfies $0.5$-Notion 1 for the swapping mapping.
\end{lemma}
  \begin{proof}
  Pick any pair of individuals $u$ and $v$. WLOG assume that $p(u) \ge p(v)$. Since $A$ is monotonic, we see that
  
%   \inote{ @Meena, All of the $C \subseteq U$ are supposed to be $C \in \mathcal{C}$, right?}
 \begin{align*}
   TV(q^1_{u,v}, q^1_{v,u}) &=   0.5 \sum_{C \subseteq U, |C| = k-1, u,v \not\in C} |A(C \cup \left\{u\right\}) - A(C \cup \left\{v\right\})| \\
  &= 0.5 \sum_{C \subseteq U, |C| = k-1, u,v \not\in C} (A(C \cup \left\{u\right\}) - A(C \cup \left\{v\right\})) \\
  &= 0.5 \left(\sum_{C \subseteq U, |C| = k-1, u,v \not\in C} A(C \cup \left\{u\right\})\right) - 0.5 \left(\sum_{C \in \mathcal{C}, |C| = k-1, u,v \not\in C} A(C \cup \left\{v\right\})\right) \\
  &= 0.5 \left[\left(\sum_{C \subseteq U, |C| = k-1, u,v \not\in C} A(C \cup \left\{u\right\})\right) +\left(\sum_{C \subseteq U, |C| = k, u,v \in C} A(C)\right)\right] \\
  &- 0.5\left[\left(\sum_{C \subseteq U, |C| = k, u,v \in C} A(C)\right)  +  \left(\sum_{C \subseteq U, |C| = k-1, u,v \not\in C} A(C \cup \left\{v\right\})\right)\right]  \\
    &= 0.5 \left(\sum_{C \subseteq U, u \in C} A(C)\right) - 0.5 \left(\sum_{C \subseteq U, v \in C} A(C)\right) \\
  &= 0.5 p(u) - 0.5 p(v) \le 0.5 \D(u,v).
 \end{align*} 
 \end{proof}

We now analyze the Conditioning Mechanism (Mechanism \ref{mech:modifiedPTC}), restated here: 
\begin{mech}[Conditioning Mechanism]
\label{mech:modifiedPTC:restated}
Given a weight function $w: U \rightarrow [0,1]$, for each $u \in U$, independently draw from $\mathbbm{1}_u \sim \text{Bern}(w(u))$. Denote the set of individuals with $\mathbbm{1}_u=1$ as $S$. If $|S|\geq k$, choose a cohort of $k$ individuals uniformly at random from $S$. Otherwise, repeat.
\end{mech}

We show that under mild conditions, the Conditioning Mechanism is individually fair, robust, and allows for a degree of mechanism expressiveness. % (this is a formal version of Lemma \ref{lemma:informalmodifiedPTC}).
\begin{lemma}
\label{lemma:modifiedPTC}
Let $k$ be the size of the cohorts in $\mathcal{C}$, and assume that $k \ge 2$. Let $\alpha_1$, $\alpha_2$, and $\alpha_3$ be constants defined as follows: $\alpha_1 = 1.6$ when $k\ge 12$ and $\alpha_1 = 13$ when $2 \le k \le 12$, $\alpha_2 = 12$, and $\alpha_3 = 0$ when $k < 54$ and $\alpha_3 = 0.2$ when $k \ge 54$ and $\alpha_3 = 0.485$ when $k \ge 180$. Consider the Conditioning Mechanism with $1/\alpha_1$-individually fair weights satisfying $\sum_{x \in U} w(x) \ge  3k/2$. The mechanism is individually fair, satisfies $\alpha_2$-Notion 2, and concludes in expectation within $\alpha_1$ rounds. Moreover, if $\sum_{x \in U} w(x) = 3k/2$, then $|p(u) - p(v) \ge \alpha_3 |w(u) - w(v)|$ and
$\alpha_3 w(u) \le p(u) \le \alpha_1 w(u)$.
\end{lemma}
%  \inote{Comment: I don't think we really want $\eta_1,\eta_2$ as constants -- these are fairness parameters. We should either change this, or add a sentence of intuition for why we are doing it this way.} \mnote{I don't know how to tune the parameters better.} \inote{I think that's the point though, we don't actually *know* what is better. Someone might be happy with larger or smaller parameters than we are giving. Maybe we want to frame this $\leq$? Basically -- I don't think we have to get the ``best'' tuning, because we don't know the tradeoffs people want. Does that make sense?}
Thus, with a lower bound on the total sum of weights, the Conditioning Mechanism satisfies Notion 2. Moreover, if the sum of weights is tuned exactly to $3k/2$, then the resulting mechanism is also expressive: the difference between $|p(u) - p(v)| \ge 0.485 |w(u) - w(v)|$ for sufficiently large $k$, so dissimilar people have dissimilar probabilities of being selected. Moreover, by setting $w(x) = 0$, the mechanism will make $x$ never appear, and by setting $w(x) = 1$, the mechanism will make $p(x) \ge 0.485$ for sufficiently large $k$.  

Now, we prove Lemma \ref{lemma:modifiedPTC}. We first prove the following helpful Proposition.

\begin{proposition}
\label{prop:helpermodifiedPTC}
The Conditioning Mechanism run with individually fair weights $w(x)$ (i.e. $|w(x) - w(y)| \leq \D(x,y)$) is $\eta_1$-individually fair and satisfies $\eta_2$-Notion 2 for $\eta_1$ and $\eta_2$ defined as follows. For subsets $S \subseteq S' \subseteq U$, let:

\[P^{S'}[S] = \left(\prod_{x \in S} w(x)\right) \left(\prod_{x \not\in S, x \in S'} (1 - w(x))\right). \]
Let:
\[\eta_1  = \left(\sum_{S\subseteq U, |S| \ge k} P^{U}[S]\right)^{-1}. \]
\[\eta_2 = 4 \left(1 + \max\left(\frac{1}{k}, \frac{\sum_{S \subseteq U, u, v \not\in S, |S| = k-2} (P^{U \setminus \left\{u, v\right\}}[S]} {\sum_{S \subseteq U, u, v \not\in S, |S| = k-1} (P^{U \setminus \left\{u, v\right\}}[S]}\right)\right).\]

Moreover, the following equalities are true: 

\[p(u) = \frac{w(u) \sum_{S \subseteq U,u \not\in S, |S| \ge k -1} (P^{U \setminus \left\{u\right\}}[S] \cdot \frac{k}{|S| + 1})}{\sum_{S \subseteq U, |S| \ge k} P^U[S]}.\]
\[|p(u) - p(v)| = \frac{|w(u) - w(v)| \sum_{S \subseteq U, |S| \ge k} P^{U \setminus \left\{u,v\right\}}[S] \frac{k}{|S| + 1}}{\sum_{S \subseteq U, |S| \ge k} P^U[S]}.\]
\end{proposition}

% Observe that $\eta_1$ is the inverse of the probability that the mechanism chooses a set of at least size $k$ in a single round and $\eta_2$ can be thought of as \inote{Meena - this value is not intuitive to me at all. I think we're saying something about the probability that the next of $u$ or $v$ is added to the cohort.}

% .... \inote{Definitely better.} \mnote{We do need $\eta_2$-- hopefully this clarified now.}

\begin{proof}
The proof consists of two main parts. First, we derive the appropriate expressions for $p(u)$ and $p(v)$ to show that the conditioning mechanism is $\eta_1$-individually fair. Second, we prove that the conditioning mechanism satisfies $\eta_2$-Notion 2 (Definition \ref{def:conditions}): we do this by analyzing the total variation distance between the distributions for $u$ and $v$ by breaking into three cases depending on whether $u$ or $v$ or both are in the initial set selected by the mechanism.

\textbf{Part 1.} Observe that $P^{S'}[S]$ 
is the probability that the set $S$ is initially chosen when the Conditioning Mechanism were to be run on the universe $S'$. Although the actual universe is $U$, we introduce this quantity since it turns out to be convenient in the analysis. 
% \mnote{The changes you made here weren't quite correct. I reverted back and added a sentence explanation.}
Observe the probability of having $\ge k$ elements in the set initially chosen by the Conditioning Mechanism is 
$\sum_{S \subseteq U, |S| \ge k} P^U[S] = \eta_1^{-1}$. (Note: for convenience, we will write $\sum_S$ to mean $\sum_{S \subseteq U}$ in subsequent formulae for brevity.). 

First, we compute $p(u)$. Suppose a set $S'$ of $\ge k$ elements is initially drawn by the Conditioning Mechanism. If $u \not\in S'$, then $u$ is not in the cohort. Otherwise, there is a $\frac{k}{|S'|}$ probability that $u$ is in the cohort. 
In this case, let $S = S' \setminus \left\{u\right\}$. We see that this means that: 
\[p(u) = \eta_1 \sum_{u \in S', |S'| \ge k} (P^{U}[S'] \cdot \frac{k}{|S'|}) = \frac{w(u) \sum_{u \not\in S, |S| \ge k -1} (P^{U \setminus \left\{u\right\}}[S] \cdot \frac{k}{|S| + 1})}{\sum_{S, |S| \ge k} P^U[S]}.\]
% \inote{Should subscript on first sum be $u\in S',|S'|\geq k$? or is it supposed to be $|S|\geq k$?}\mnote{Fixed.}

% We see that $w(u) \sum_{u \not\in S, |S| \ge k -1} (P^{U \setminus \left\{u\right\}}[S] \cdot \frac{k}{|S| + 1}) \le w(u) \sum_{u \not\in S, |S| \ge k -1} P^{U \setminus \left\{u\right\}}[S] = \sum_{S, |S| \ge k, u \in S} P^U[S] \le \sum_{S, |S| \ge k} P^U[S]$, so we know that $p(u) \le w(u)$.

We can also write
\[p(u) = \frac{w(u) w(v) \sum_{u, v \not\in S, |S| \ge k -2} (P^{U \setminus \left\{u, v\right\}}[S] \cdot \frac{k}{|S|+2}) + w(u) (1-w(v)) \sum_{u, v \not\in S, |S| \ge k -1} (P^{U \setminus \left\{u, v\right\}}[S] \cdot \frac{k}{|S|+1})}{\sum_{S, |S| \ge k} P^U[S]}.\]
\[p(v) = \frac{w(u) w(v) \sum_{u, v \not\in S, |S| \ge k -2} (P^{U \setminus \left\{u, v\right\}}[S] \cdot \frac{k}{|S|+2}) + w(v) (1-w(u)) \sum_{u, v \not\in S, |S| \ge k -1} (P^{U \setminus \left\{u, v\right\}}[S] \cdot \frac{k}{|S|+1})}{\sum_{S, |S| \ge k} P^U[S]}.\] This implies that
\begin{align*}
    |p(u) - p(v)| &= \frac{|w(u) - w(v)| \sum_{u, v \not\in S, |S| \ge k -1} (P^{U \setminus \left\{u, v\right\}}[S] \cdot \frac{k}{|S|+1})}{\sum_{S, |S| \ge k} P^U[S]} \\
    &\le \frac{|w(u) - w(v)| \sum_{u, v \not\in S, |S| \ge k -1} (P^{U \setminus \left\{u, v\right\}}[S])}{\sum_{S, |S| \ge k} P^U[S]} \\
     &\le \frac{|w(u) - w(v)|}{\sum_{S, |S| \ge k} P^U[S]} \\
    &= \eta_1 |w(u) - w(v)|.
\end{align*}

Thus, the conditioning mechanism is $\eta_1$-individually fair.

\textbf{Part 2.} Now, we compute the TV distance to show the Notion 2 (Definition \ref{def:conditions}.\ref{def:conditions:notion2}) properties. Let $S$ be the set of elements \textit{initially} selected by Conditioning Mechanism. We condition on the event $|S| \geq k$ and compute the total variation distance. There are three relevant cases:
\begin{enumerate}
    \item $u$ and $v$ are both in $S$
    \item $u$ is in $S$ and $v$ is not in $S$
    \item $v$ is in $S$ and $u$ is not in $S$
\end{enumerate}

We define: 
\begin{align*}
    \kappa_1 &= \sum_{u, v \not\in S, |S| \ge k-1} (P^{U \setminus \left\{u, v\right\}}[S] \cdot \frac{k}{|S| + 1}) \\
    \kappa_2 &= \sum_{u, v \not\in S, |S| \ge k-2} (P^{U \setminus \left\{u, v\right\}}[S] \cdot \frac{k}{|S| + 2}).
\end{align*} In this notation, we have that: 
\[p(u) = \eta_1 w(u) (w(v) \kappa_2 + (1-w(v)) \kappa_1)\]
and
\[p(v) = \eta_1 w(v) (w(u) \kappa_2 + (1 - w(u)) \kappa_1.\]

Let $S$ be a subset of size of at least $k$ that is initially chosen by the mechanism, and let $R \subseteq S$ be the size $k$ subset that is finally chosen. We see that the probability $R$ is chosen given $S$ is $\frac{1}{{|S| \choose k}}$, and the probability that $S$ is chosen is $\eta_1 P^U[S]$, where the $\eta_1$ comes from the fact that if the mechanism starts over if it chooses a set of size $< k$. We now consider (1), (2), and (3) separately, and do a triangle inequality between the contributions to the total variation distance of these three cases. For each of these cases, there are three possible states for $R$. Either 
    \begin{enumerate}
        \item[(a)] $R$ contains $u$ and $v$,
        \item[(b)] $R$ contains one of $u$ and $v$, or
        \item[(c)] $R$ contains neither.
    \end{enumerate}
    
% \begin{enumerate}[leftmargin=*]
First, we consider case (1). Here, case (c) contributes nothing to the TV distance.

    Case (a) contributes $\eta_1 \left|\frac{P^U[S] \frac{1}{{|S| \choose k}}}{p(u)} -  \frac{P^U[S] \frac{1}{{|S| \choose k}}}{p(v)}\right|$ by construction. Summing over possible $R$ and $S$ for this case, we see that that the contribution to TV distance is:
    \[\frac{1}{2} \eta_1  \sum_{u, v \in S, |S| \ge k} P^U[S] \frac{1}{{|S| \choose k}} \left(\sum_{u, v \in R \subseteq S, |R| = k} \left|\frac{1}{p(u)} - \frac{1}{p(v)}\right|\right). \]

    Case (b), due to the swapping mapping (Definition \ref{def:swapping}) and symmetry, contributes $\eta_1  \left|\frac{P^U[S] \frac{1}{{|S| \choose k}}}{p(u)} -  \frac{P^U[S] \frac{1}{{|S| \choose k}}}{p(v)}\right|$. For case (b), if $u \in R$, we let $R' = R \setminus \left\{u \right\}$ and if $v \in R$, we let $R'=R \setminus \left\{v\right\}$. Summing over possible $R$ and $S$ for this case, we see that the contribution to TV distance is:
    \[\frac{1}{2}  \eta_1  \sum_{u, v \in S, |S| \ge k} P^U[S] \frac{1}{{|S| \choose k}} \left(\sum_{R' \subseteq S, |R'| = k - 1, u, v \not\in R'} \left|\frac{1}{p(u)} - \frac{1}{p(v)}\right|\right). \]
    
    Thus, the total contribution to the total variation distance for (1) is:
\[\frac{1}{2}  \eta_1  \sum_{u, v \in S, |S| \ge k} P^U[S] \frac{1}{{|S| \choose k}} \left(\sum_{u, v \in R \subseteq S, |R| = k} \left|\frac{1}{p(u)} - \frac{1}{p(v)}\right| + \sum_{R' \subseteq S, |R'| = k - 1, u, v \not\in R'} \left|\frac{1}{p(u)} - \frac{1}{p(v)}\right|\right). \] 
We can write this as: 
\[\frac{1}{2}  \eta_1  \left|\frac{1}{p(u)} - \frac{1}{p(v)}\right| \sum_{u, v \in S, |S| \ge k} P^U[S] \frac{1}{{|S| \choose k}} \left(\sum_{u, v \in R \subseteq S, |R| = k} 1  + \sum_{R' \subseteq S, |R'| = k - 1, u, v \not\in R'} 1 \right),\] which is equal to:
\[\frac{1}{2}  \eta_1  \left|\frac{1}{p(u)} - \frac{1}{p(v)}\right| \sum_{u, v \in S, |S| \ge k} P^U[S] \frac{ {{|S| - 2} \choose {k-2}} + {{|S| - 2} \choose {k-1}}}{{|S| \choose k}} = \frac{1}{2}  \eta_1 \left|\frac{1}{p(u)} - \frac{1}{p(v)}\right| \sum_{u, v \in S, |S| \ge k} P^U[S] \frac{{{|S| - 1} \choose {k-1}}}{{|S| \choose k}},\] which is equal to: 
\[\frac{1}{2}  \eta_1  \left|\frac{1}{p(u)} - \frac{1}{p(v)}\right| \sum_{u, v \in S, |S| \ge k} P^U[S] \frac{k}{|S|} = \frac{1}{2}  \eta_1 w(u)w(v) \kappa_2 \left|\frac{1}{p(u)} - \frac{1}{p(v)}\right|. \] This can be written as:
\[\frac{1}{2}  \left|\frac{w(v) \kappa_2}{w(v) \kappa_2 + (1-w(v)) \kappa_1} - \frac{w(u) \kappa_2}{w(u) \kappa_2 + (1-w(u)) \kappa_1} \right| = \frac{1}{2}  \frac{\kappa_1\kappa_2|w(v) - w(u)|}{(w(v) \kappa_2 + (1-w(v)) \kappa_1)(w(u) \kappa_2 + (1-w(u)) \kappa_1)}\]
    
Now we consider cases (2) and (3), which follow symmetric arguments. Of the three possible states for $R$, (a) is not possible for cases (2) and (3) and (c) contributes nothing to TV distance. Thus, we only need to consider (b). With the swapping mapping, we can match $u \in R \subseteq S$ with $v \in (R \setminus \left\{u\right\}) \cup \left\{v\right\} \subseteq (S \setminus \left\{u\right\}) \cup \left\{v\right\}$. Let $S' = S \setminus \left\{u \right\}$ and $R' = R \setminus \left\{u\right\}$. This contributes $\eta_1 P^{U \setminus \left\{u, v\right\}}[S']  \frac{1}{{{|S'| +1} \choose k}} \left|\frac{w(u)(1-w(v))}{p(u)} - \frac{w(v)(1 - w(u))}{p(v)}\right|$ to the total variation distance. Summing over $S$ and $R$ for this case, the total contribution to the total variation distance is:
\[\frac{1}{2}  \eta_1 \sum_{u, v \not\in S', |S'| \ge k-1} P^{U \setminus \left\{u, v\right\}}[S'] \frac{1}{{{|S'| +1} \choose k}} \left(\sum_{R' \subseteq S', |R'| = k - 1} \left|\frac{w(u)(1-w(v))}{p(u)} - \frac{w(v)(1 - w(u))}{p(v)}\right|.\right). \] This can be written as:
\[\frac{1}{2}  \sum_{u, v \not\in S', |S'| \ge k-1} P^{U \setminus \left\{u, v\right\}}[S'] \frac{1}{{{|S'| +1} \choose k}} \left(\sum_{R' \subseteq S', |R'| = k - 1} \left|\frac{(1 - w(v))}{w(v) \kappa_2 + (1-w(v)) \kappa_1} - \frac{(1 - w(u))}{w(u) \kappa_2 + (1-w(u)) \kappa_1}\right|,\right). \]
which can be simplified
\[ \frac{1}{2}  \left|\frac{(1 - w(v))}{w(v) \kappa_2 + (1-w(v)) \kappa_1} - \frac{(1 - w(u))}{w(u) \kappa_2 + (1-w(u)) \kappa_1}\right| \sum_{u, v \not\in S', |S'| \ge k-1} P^{U \setminus \left\{u, v\right\}}[S'] \frac{1}{{{|S'| +1} \choose k}} \left(\sum_{R' \subseteq S', |R'| = k - 1} 1,\right), \] which is equal to:
\[\frac{1}{2}  \left|\frac{(1 - w(v))}{w(v) \kappa_2 + (1-w(v)) \kappa_1} - \frac{(1 - w(u))}{w(u) \kappa_2 + (1-w(u)) \kappa_1}\right| \sum_{u, v \not\in S', |S'| \ge k-1} P^{U \setminus \left\{u, v\right\}}[S'] \frac{{{|S'|} \choose {k - 1}} }{{{|S'| +1} \choose k}}, \]
%which is equal to: 
\[\frac{1}{2}  \left|\frac{(1 - w(v))}{w(v) \kappa_2 + (1-w(v)) \kappa_1} - \frac{(1 - w(u))}{w(u) \kappa_2 + (1-w(u)) \kappa_1}\right| \sum_{u, v \not\in S, |S| \ge k-1} P^{U \setminus \left\{u, v\right\}}[S] \frac{k}{|S| + 1} \] %This is equal to:
\[\frac{1}{2}  \left|\frac{(1 - w(v))\kappa_1}{w(v) \kappa_2 + (1-w(v)) \kappa_1} - \frac{(1 - w(u))\kappa_1}{w(u) \kappa_2 + (1-w(u)) \kappa_1}\right|, \] %which is equal to:
\[\frac{1}{2}  \frac{\kappa_1\kappa_2|w(v) - w(u)|}{(w(v) \kappa_2 + (1-w(v)) \kappa_1)(w(u) \kappa_2 + (1-w(u)) \kappa_1)}. \]
    
Now, we are ready to put these cases together to compute the overall total variation distances. The total variation distance contribution arising from cases (1), (2), and (3) is bounded by: 
\[\frac{1}{2}  \frac{2 \kappa_1\kappa_2|w(v) - w(u)|}{(w(v) \kappa_2 + (1-w(v)) \kappa_1)(w(u) \kappa_2 + (1-w(u)) \kappa_1)}.\]
We wish to upper bound this by $\eta_2 |w(v) - w(u)|$. Equivalently, we wish to lower bound its reciprocal is lower bounded. Taking the reciprocal and simplifying yield:
\[ \frac{(w(v) \kappa_2 + (1-w(v)) \kappa_1)(w(u) \kappa_2 + (1-w(u)) \kappa_1)}{ \kappa_1\kappa_2|w(v) - w(u)|} \]
\[= \frac{(w(v) w(u) \kappa_2^2 + \kappa_1\kappa_2 (w(u)(1-w(v)) + w(v)(1 - w(u))) + \kappa_1^2 (1 - w(v))(1 - w(u))}{\kappa_1\kappa_2|w(v) - w(u)|}.\]
\[= \frac{(w(v) w(u) \frac{\kappa_2}{\kappa_1} + (w(u)(1-w(v)) + w(v)(1 - w(u))) + \frac{\kappa_1}{\kappa_2}  (1 - w(v))(1 - w(u))}{|w(v) - w(u)|}.\]

Notice that $\max((w(v) w(u), w(u)(1-w(v)), w(v)(1-w(u)), (1-w(u))(1-w(v))) \ge 0.25$. Thus, the numerator is lower bounded by $0.25  \min\left(\frac{\kappa_1}{\kappa_2}, \frac{\kappa_2}{\kappa_1}\right)$. Thus, the whole expression is $\ge \frac{1}{4 |w(v) - w(u)|} \min\left(\frac{\kappa_1}{\kappa_2}, \frac{\kappa_2}{\kappa_1}\right)$. Hence, the total variation distance is upper bounded by $4 |w(v) - w(u)| \max\left(\frac{\kappa_1}{\kappa_2}, \frac{\kappa_2}{\kappa_1}\right)$. Now, it suffices to show that $\max\left(\frac{\kappa_1}{\kappa_2}, \frac{\kappa_2}{\kappa_1}\right) \le \left(1 + \max\left(\frac{1}{k}, \frac{\sum_{S \subseteq U, u, v \not\in S, |S| = k-2} (P^{U \setminus \left\{u, v\right\}}[S]} {\sum_{S \subseteq U, u, v \not\in S, |S| = k-1} (P^{U \setminus \left\{u, v\right\}}[S]}\right)\right)$

Observe that 
\begin{align*}
\kappa_2 &=  \sum_{u, v \not\in S, |S| \ge k-2} (P^{U \setminus \left\{u, v\right\}}[S] \cdot \frac{k}{|S| + 2}) \\
&= \sum_{u, v \not\in S, |S| = k-2} (P^{U \setminus \left\{u, v\right\}}[S] \cdot \frac{k}{|S| + 2}) + \sum_{u, v \not\in S, |S| \ge k-1} (P^{U \setminus \left\{u, v\right\}}[S] \cdot \frac{k}{|S| + 2}) \\
&\le \sum_{u, v \not\in S, |S| = k-2} (P^{U \setminus \left\{u, v\right\}}[S] \cdot \frac{k}{|S| + 2}) + \sum_{u, v \not\in S, |S| \ge k-1} (P^{U \setminus \left\{u, v\right\}}[S] \cdot \frac{k}{|S| + 1}).
\end{align*} Thus, 
\[\frac{\kappa_2}{\kappa_1} \le 1 + \frac{\sum_{u, v \not\in S, |S| = k-2} (P^{U \setminus \left\{u, v\right\}}[S] \cdot \frac{k}{|S| + 2})}{\sum_{u, v \not\in S, |S| \ge k-1} (P^{U \setminus \left\{u, v\right\}}[S] \cdot \frac{k}{|S| + 1})} \le 1 + \frac{\sum_{u, v \not\in S, |S| = k-2} (P^{U \setminus \left\{u, v\right\}}[S] \cdot \frac{k}{|S| + 2})}{\sum_{u, v \not\in S, |S| = k-1} (P^{U \setminus \left\{u, v\right\}}[S] \cdot \frac{k}{|S| + 1})}.\] Also, observe that 
\begin{align*}
\kappa_1 &= \sum_{u, v \not\in S, |S| \ge k-1} (P^{U \setminus \left\{u, v\right\}}[S] \cdot \frac{k}{|S| + 1}) \\
&\le \frac{k+1}{k} \sum_{u, v \not\in S, |S| \ge k-1} (P^{U \setminus \left\{u, v\right\}}[S] \cdot \frac{k}{|S| + 2})\\
&\le \frac{k+1}{k} \sum_{u, v \not\in S, |S| \ge k-2} (P^{U \setminus \left\{u, v\right\}}[S] \cdot \frac{k}{|S| + 2}).
\end{align*} so $\frac{\kappa_1}{\kappa_2} \le \frac{k+1}{k}$. This proves the desired result.

\end{proof}
 
Now, we are ready to prove Lemma \ref{lemma:modifiedPTC}. To show tail bounds, we use the standard multiplicative Chernoff bound, which we recall here for sake of completeness: 
\begin{theorem}
\label{thm:chernoff}
Let $X = \sum_{i=1}^n X_i$, where $X_i = 1$ with probability $p_i$ and $X_i = 0$ with
probability $1 - p_i$, and all $X_i$ are independent. Let $\mu = \mathbb{E}[X] = \sum_{i=1}^n p_i$. Then the following bounds hold:
\begin{enumerate}
    \item \textbf{Upper tail:} $\mathbb{P}[X \ge (1 + \delta)\mu] \le e^{-\frac{\delta^2}{3} \mu}$ for all $0 < \delta < 1$,
    \item \textbf{Lower tail:} $\mathbb{P}[X \le (1 - \delta)\mu] \le e^{-\mu \frac{\delta^2}{2}}$ for all $0 < \delta < 1$. 
\end{enumerate}
(At $\delta = 1/3$, these bounds become $e^{-\frac{1}{27} \mu}$ and $e^{-\frac{1}{18} \mu}$.)
\end{theorem}
We use Theorem \ref{thm:chernoff} to prove Lemma \ref{lemma:modifiedPTC}. 
\begin{proof}[Proof of Lemma \ref{lemma:modifiedPTC}]
 The main ingredient of our proof is Proposition \ref{prop:helpermodifiedPTC}. First, we prove some upper bounds on the values $\eta_1$ and $\eta_2$ in Proposition \ref{prop:helpermodifiedPTC} for the case of $\sum_{x \in U} w(x) \ge 3k/2$. 
 
 \textbf{Bounding $\eta_1$}: We wish to bound $\eta_1 = \left(\mathbb{P}[|S| \ge k]\right)^{-1}$. We use Theorem \ref{thm:chernoff} (a Chernoff bound), since $S$ is a sum of independent indicator random variables with weights according to $w(x)$. Here the mean is $\mu =  \sum_{x \in U} w(x) \ge 3k/2$, and we take $\delta = 1/3$, to obtain that $\mathbb{P}[|S| \le k] \le e^{-\frac{3 k}{2 \cdot 18}} = e^{-\frac{k}{12}}$. Thus, we know that $\mathbb{P}[|S| \ge k] \ge 1 - e^{-\frac{k}{12}}$, so $\eta_1 \le \frac{1}{1 - e^{-\frac{k}{12}}}$. When $k \ge 12$, we can upper bound $\eta_1$ by $1.6$, and when $1 \le k \le 12$, we can upper bound $\eta_1$ by $13$. We define $\alpha_1$ in the theorem statement based on these values.

  \textbf{Bounding $\eta_2$}: We give an upper bound on $\sum_{u, v \not\in S, |S| = k-1} P^{U \setminus \left\{u, v\right\}}[S]$ in terms of $\sum_{u, v \not\in S, |S| = k-2} P^{U \setminus \left\{u, v \right\}}[S] $.  Consider an association between sets of size $k - 1$ and sets of size $k - 2$, where each set of size $k - 1$ is mapped to the $k - 1$ subsets of size $k - 2$. We consider the probability to be equidistributed between the associated sets of size $k - 2$. For a set $|S| = k-1$ and subset $|S'| = k-2$, the associated probability on $S'$ is $P^{U \setminus \left\{u,v\right\}}[S]\frac{1}{k-1} = P^{U \setminus \left\{u,v\right\}}[S']\frac{1}{k-1} \frac{w(x)}{1 - w(x)}$, where $x = S \setminus S'$. Let's assume we perform this process for all $S$ such that $u, v \not\in S$ and $|S| = k-1$, and aggregate the probabilities defined above on sets of size $k - 2$ across all of the different $S$. Then, for a set $S'$ of size $k - 1$ such that $u, v \not\in S'$, the probability is $x \not\in S'$ to obtain $\sum_{x \not\in S', x \neq u,v}P^{U \setminus \left\{u,v\right\}}[S']\frac{1}{k-1} \frac{w(x)}{1 - w(x)} = P^{U \setminus \left\{u,v\right\}}[S]\frac{1}{k-1} \sum_{x \not\in S', x \neq u,v} \frac{w(x)}{1 - w(x)}$. This means that
$$\sum_{u, v \not\in S, |S| = k-1} P^{U \setminus \left\{u, v\right\}}[S] = \sum_{u, v \not\in S, |S'| = k-2} P^{U \setminus \left\{u, v\right\}}[S'] \frac{1}{k-1} \sum_{x \not\in S', x \neq u,v} \frac{w(x)}{1 - w(x)}$$ $$ \ge \sum_{u, v \not\in S, |S'| = k-2} P^{U \setminus \left\{u, v\right\}}[S'] \frac{1}{k-1} \sum_{x \not\in S', x \neq u,v} w(x)$$ Now, observe that $\sum_{x \not\in S', x \neq u,v} w(x) \ge \sum_{x \in U} w(x) - \sum_{x \in S'} w(x) - w(u) - w(v) \ge 3k/2 - k = k/2$. This means that $\sum_{u, v \not\in S, |S| = k-1} P^{U \setminus \left\{u, v\right\}}[S] \ge \sum_{u, v \not\in S, |S'| = k-2} P^{U \setminus \left\{u, v\right\}}[S'] \frac{k}{2(k-1)} \ge 0.5$. Thus, we have that $\frac{\sum_{u, v \not\in S, |S| = k-2} P^{U \setminus \left\{u, v \right\}}[S]}{\sum_{u, v \not\in S, |S| = k-1} P^{U \setminus \left\{u, v \right\}}[S]} \le 2$. Thus, we have that $\eta_2 \le 12$.  We define $\alpha_2$ in the theorem statement based on this value. %\inote{@Meena - it's not clear why we're trying to ``control'' $\eta_2$ rather than show that it is a reasonable expansion of $\eta_1$. i.e., put in $\eta_1$ individual fairness, get out $...\cdot \eta_1$ robustness. It's also not clear why bounding this ratio is what you want to do.} \mnote{Hopefully clarified now?}

 Now, there are four components we must prove: (1) that the conditioning mechanism is $\alpha_1$-individually fair, (2) the mechanism terminates quickly, (3) the mechanism satisfies $\alpha_2$-Notion 2 and (4) when $\sum_{x \in U} w(x) = 3k /2$, then $|p(u) - p(v)| \ge |w(u) - w(v)| C$ and $0.5 (1 - e^{-k/36} - e^{-k/54})w(u) \le p(u) \le \alpha_1 w(u)$.

\textbf{(1) $\alpha_1$-Individual Fairness}: We apply Proposition \ref{prop:helpermodifiedPTC}, and obtain that the mechanism is $\eta_1$-individually fair. We set $\alpha_1$ to be the bounds on $\eta_1$ given above.

\textbf{(2) Bounding the number of rounds in expectation}:
Let $T$ be the expected number of rounds. In (1), we showed that the probability of success is at least $\mathbb{P}[|S| \ge k] \ge 1 - e^{-\frac{k}{12}}$. Thus, we know that the expected number of rounds, $T = \frac{1}{\mathbb{P}[|S| \ge k]} = \eta_1$. We thus know that $T \ge \alpha_1$. 

%\inote{ @Meena - why not just say it terminates in a constant number of rounds with high probability?} \mnote{I just wanted to frame it as a Las Vegas algorithm and give an expected runtime guarantee. I don't feel strongly about this either way.} \inote{OK}

\textbf{(3) $\alpha_2$-Notion 2}: We apply Proposition \ref{prop:helpermodifiedPTC}, and obtain that the mechanism satisfies $\eta_2$-Notion 2. Using the bounds on $\eta_2$ from above yields the desired result.

\textbf{(4) $\sum_{x \in U} w(x)$ tuned to $3k /2$:} We apply Proposition \ref{prop:helpermodifiedPTC}, and obtain that $p(u) = w(u) \sum_{u \not\in S, |S| \ge k -1} (P^{U \setminus \left\{u\right\}}[S] \cdot \frac{k}{|S| + 1})$ and $|p(u) - p(v)| = \frac{|w(u) - w(v)| \sum_{S \subseteq U, |S| \ge k} P^{U \setminus \left\{u,v\right\}}[S] \frac{k}{|S| + 1}}{\sum_{S \subseteq U, |S| \ge k} P^U[S]}$.

First, we bound $p(u)$. Observe that $\sum_{u \not\in S, |S| \ge k -1} (P^{U \setminus \left\{u\right\}}[S] \cdot \frac{k}{|S| + 1}) \le \sum_{u \not\in S, |S| \ge k -1} (P^{U \setminus \left\{u\right\}}[S]) \le 1$, so $p(u) \le \eta_1 w(u)$. 
We now show that when $\sum_{x \in U} w(x) = 3k / 2$, it holds that $p(u) \ge  0.5(1 - e^{-k/36} - e^{-k/54})$. 
We observe that  
\begin{align*}
\sum_{u \not\in S, |S| \ge k -1} (P^{U \setminus \left\{u\right\}}[S]  \frac{k}{|S| + 1}) &\ge 0.5 \sum_{u \not\in S, |S| \ge k -1, |S| \le 2k - 1} (P^{U \setminus \left\{u\right\}}[S]) \\
&\ge 0.5\left(1 - \left(\sum_{u \notin S, |S| \ge k-2} P^{U \setminus \left\{u\right\}}[S]\right) - \left(\sum_{u \notin S, |S| \ge 2k} P^{U \setminus \left\{u\right\}}[S]\right)\right).
\end{align*}
Thus, it suffices to lower bound
\[0.5\left(1 - \left(\sum_{u \notin S, |S| \le k-2} P^{U \setminus \left\{u\right\}}[S]\right) - \left(\sum_{u \notin S, |S| \ge 2k} P^{U \setminus \left\{u\right\}}[S]\right)\right),\] for which we just need to upper bound 
$\sum_{u \notin S, |S| \le k-2} P^{U \setminus \left\{u\right\}}[S]$ and $\sum_{u \notin S, |S| \ge 2k} P^{U \setminus \left\{u\right\}}[S]$. Our main tool is Theorem \ref{thm:chernoff} (a multiplicative Chernoff bound) with $\delta = 1/3$. Here, $\mu = \sum_{x \in U, x \neq u} w(x)$. We use that $k/2 \le 3k/2 - 1 \le \mu \le 3k/2$. We see that 
\[\sum_{u \notin S, |S| \le k-2} P^{U \setminus \left\{u\right\}}[S] \le \sum_{u \notin S, |S| \le k-2/3} P^{U \setminus \left\{u\right\}}[S] \le \sum_{u \notin S, |S| \le 2\mu/3} P^{U \setminus \left\{u\right\}}[S] \le e^{-\frac{k}{36}}.\] Using that $k/2 \le \mu \le 3k/2$, we see that 
\[\sum_{u \notin S, |S| \ge 2k} P^{U \setminus \left\{u\right\}}[S] \le \sum_{u \notin S, |S| \ge 4\mu/3} P^{U \setminus \left\{u\right\}}[S] \le e^{-\frac{k}{54}}.\] Thus, we have that $p(u) \ge \sum_{u \not\in S, |S| \ge k -1} (P^{U \setminus \left\{u\right\}}[S] \cdot \frac{k}{|S| + 1}) \ge 0.5(1 - e^{-k/36} - e^{-k/54}) w(u)$. 

Now, we bound $|p(u) - p(v)|$. We show that When $\sum_{x \in U} w(x) = 3k / 2$, it holds that $|p(u) - p(v)| \ge |w(u) - w(v)| 0.5(1 - e^{-k/36} - e^{-k/54})$. 
We observe that 
\begin{align*}
    \sum_{u, v \not\in S, |S| \ge k -1}  \frac{P^{U \setminus \left\{u, v\right\}}[S] \cdot k}{|S| + 1} &\ge 0.5 \sum_{u,v \not\in S, |S| \ge k -1, |S| \le 2k - 1} (P^{U \setminus \left\{u,v\right\}}[S]) \\
    &\ge 0.5\left(1 - \left(\sum_{u,v\not\in S, |S| \le k-2} P^{U \setminus \left\{u,v\right\}}[S]\right) - \left(\sum_{u,v \not\in S, |S| \ge 2k} P^{U \setminus \left\{u,v\right\}}[S]\right)\right).
\end{align*}
Thus, it suffices to lower bound
\[0.5\left(1 - \left(\sum_{u,v\not\in S, |S| \le k-2} P^{U \setminus \left\{u,v\right\}}[S]\right) - \left(\sum_{u,v \not\in S, |S| \ge 2k} P^{U \setminus \left\{u,v\right\}}[S]\right)\right),\] for which we just need to upper bound 
$\sum_{u,v\not\in S, |S| \le k-2} P^{U \setminus \left\{u,v\right\}}[S]$ and $\sum_{u,v \not\in S, |S| \ge 2k} P^{U \setminus \left\{u,v\right\}}[S]$. Our main tool is Theorem \ref{thm:chernoff} (a multiplicative Chernoff bound) with $\delta = 1/3$. Here, $\mu = \sum_{x \in U, x \neq u, v} w(x)$. We use that $k/2 \le 3k/2 - 2 \le \mu \le 3k/2$. We see that 
\[\sum_{u,v\not\in S, |S| \le k-2} P^{U \setminus \left\{u,v\right\}}[S] \le \sum_{u,v\not\in S, |S| \le k-4/3} P^{U \setminus \left\{u,v\right\}}[S] \le \sum_{u,v\not\in S, |S| \le 2\mu/3} P^{U \setminus \left\{u,v\right\}}[S] \le e^{-\frac{k}{36}}.\] Using that $k/2 \le \mu \le 3k/2$, we see that 
\[\sum_{u,v\not\in S, |S| \ge 2k} P^{U \setminus \left\{u,v\right\}}[S] \le \sum_{u,v\not\in S, |S| \ge 4\mu/3} P^{U \setminus \left\{u,v\right\}}[S] \le e^{-\frac{k}{54}}.\] Thus, we have that $|p(u) - p(v)| \ge |w(u) - w(v)| \sum_{u \not\in S, |S| \ge k -1} (P^{U \setminus \left\{u\right\}}[S] \cdot \frac{k}{|S| + 1}) \ge 0.5(1 - e^{-k/36} - e^{-k/54})$.

We can bound $0.5(1 - e^{-k/36} - e^{-k/54})$. We define $\alpha_3$ in the theorem statement based on such bounds.

\end{proof}

   \subsection{Proofs for Section \ref{subsec:F3}}
 
For convenience, we restate the quality composition mechanism (Mechanism \ref{mech:qualitycomp}):
\begin{mech}[Restatement of the Quality Compositional Mechanisms]\label{mech:qualitycomprestated}

Let $\beta \le 1$ be a constant, and suppose that $\D$ endowed with quality groups $q_1, \ldots, q_n$ is $\beta$-quality-clustered. Suppose also that $\mathcal{C}$ is quality-symmetric. For each $1 \le i \le n$ and each $1 \le x_i \le |q_i|$, let $\Mech_{i, x_i}$ be a $\D^i$-individually fair mechanism selecting $x_i$ individuals in $q_i$. We define the \textbf{quality compositional mechanism} for $\left\{\Mech_{i, x_i}\right\}$ as follows. Let $\mathcal{X}$ be any distribution over $n$-tuples of nonnegative integers $(x_1, \ldots, x_n) \in P(\mathcal{C})$. 

\begin{enumerate}
    \item Draw $(x_1, \ldots, x_n) \sim \mathcal{X}$. 
    \item Independently run $\Mech_{i,x_i}$ for each $1 \le i \le n$, and return the union of the outputs of all of these mechanisms. 
\end{enumerate}
\end{mech}

We prove Lemma \ref{lemma:distquality}, also restated here:
\begin{lemma}[Restatement of Lemma \ref{lemma:distquality}]
\label{lemma:restateddistquality}
Let $\beta \le 0.5$ be a constant, and suppose that $\D$ endowed with quality groups $q_1, \ldots, q_n$ is $\beta$-quality-clustered. Suppose also that $\mathcal{C}$ is quality-symmetric, and let $\mathcal{X}$ be any distribution over $(x_1, \ldots, x_n) \in P(\mathcal{C})$ such that $|\frac{x_i}{|q_i|} - \frac{x_j}{|q_j|}| \le (1 - 2 \beta) D(i,j)$. If $\Mech$ is a quality compositional mechanism, then:
\begin{enumerate}
    \item $\Mech$ is always individually fair.
    \item $\Mech$ always satisfies $0.5$-Notion 1.
    \item $\Mech$ satisfies $0.5$-Notion 2 for $\D$ and $\DS$ if \textbf{either} of the following conditions hold:
    \begin{enumerate}
        \item (One set) $|\text{Supp}(\mathcal{X})| = 1$ (i.e. one ``canonical'' $(x_1, \ldots, x_n)$), or
        \item (0-1 metric) $D(i,j) = 1$ for $1 \le i \neq j \le n$ and $\D^i(u,v) = 0$ for $1 \le i \le n$.
    \end{enumerate}
\end{enumerate}
\end{lemma}
\begin{proof}
First, we handle a single quality profile vector. Then, we show (1) and (2). Lastly, we show (3). 

\textbf{Handling a single quality profile vector.} For $(x_1, \ldots, x_n) \in \text{Supp}(\mathcal{X})$, let $p^{(x_1, \ldots, x_n)}(u)$ be the probability that $u$ is assigned to a cohort if a quality compositional mechanism is run on the distribution $\mathcal{X}'$ with probability $1$ at $(x_1, \ldots, x_n)$. We claim that $|p^{(x_1, \ldots, x_n)}(u) - p^{(x_1, \ldots, x_n)}(v)| \le \D(u,v)$. If $u$ and $v$ are in the same quality group, then individual fairness follows from the individual fairness of $M_{i, c_i}$.

Suppose that $u$ and $v$ are in different quality groups, say $q_i$ and $q_j$. Note that $\sum_{x \in q_i} p^{(x_1, \ldots, x_n)}(x) = x_i$ and $\sum_{x \in q_j} p^{(x_1, \ldots, x_n)}(x) = x_j$. WLOG suppose that $p^{(x_1, \ldots, x_n)}(u) \ge p^{(x_1, \ldots, x_n)}(i)$. 
Let $u'$ be an individual in $q_i$ maximally distant from $u$ and let $v'$ be an individual in $q_j$ maximally distant from $v$. Then, we know that for each $x \in q_i$, it holds that $p^{(x_1, \ldots, x_n)}(x) \ge p^{(x_1, \ldots, x_n)}(u) - \D(u, x) \ge p(u) - \D(u,u')$. Similarly, for each $x \in q_j$, it holds that $p(x) \le p^{(x_1, \ldots, x_n)}(v) + \D(v, x) \le p^{(x_1, \ldots, x_n)}(v) + \D(v,v')$. 

Thus, we know that $x_i = \sum_{x \in q_i} p^{(x_1, \ldots, x_n)}(x) \ge |q_i| (p^{(x_1, \ldots, x_n)}(u) - \D(u,u'))$. This means that 
\[\frac{x_i}{|q_i|} \ge p^{(x_1, \ldots, x_n)}(u) - \D(u,u'),\] and so:
\[p^{(x_1, \ldots, x_n)}(u) \le  \frac{x_i}{|q_i|} + \D(u,u').\]

Similarly, note that $x_j = \sum_{x \in q_j} p^{(x_1, \ldots, x_n)}(x) \le |q_j|(p(v) + \D(v,v'))$. Thus we have $\frac{x_j}{|q_j|} \le p^{(x_1, \ldots, x_n)}(v) + \D(v,v')$, and so: 
\[p^{(x_1, \ldots, x_n)}(v) \ge \frac{x_j}{|q_j|} - \D(v,v').\]

Putting these facts together, we obtain that:
\[p^{(x_1, \ldots, x_n)}(u) - p^{(x_1, \ldots, x_n)}(v)  \le  \frac{x_i}{|q_i|} + \D(u,u') - \left(\frac{x_j}{|q_j|} - \D(v,v')\right) = \frac{x_i}{|q_i|} - \frac{x_j}{|q_j|} + \D(u, u') + \D(v,v').\] We know that $\frac{x_i}{|q_i|} - \frac{x_j}{|q_j|} \le (1 - 2 \beta)$ since this is given in the theorem statement. Since $\D$ is $\beta$-quality-clustered, we know that $\D(u, u') \le \beta D(i,j)$ and $\D(v, v') \le \beta D(i,j)$. Thus, the whole expression is bounded by $\D(u,v)$. 

\textbf{Showing (1) and (2).} Now to show (1) and (2), we essentially combine this with the fact that the quality compositional mechanism borrows features of RandomizeThenClassify in \cite{DI2018}. Let $\gamma$ be the probability measure corresponding to $\mathcal{X}$. Pick a pair of individuals $u$ and $v$. Let the cluster label of the cluster corresponding to $(x_1, \ldots, x_n) \in P(\mathcal{C}_u \cup \mathcal{C}_v)$ be $i((x_1, \ldots, x_n))$. First, we let $C = \sum_{(x_1, \ldots, x_n) \in P(\mathcal{C}_u \cup \mathcal{C}_v)} |\gamma((x_1, \ldots, x_n)) p^{(x_1, \ldots, x_n)}(u) - \gamma((x_1, \ldots, x_n)) p^{(x_1, \ldots, x_n)}(v)|$ and observe that:
\begin{align*}
    C &= \sum_{(x_1, \ldots, x_n) \in P(\mathcal{C}_u \cup \mathcal{C}_v)} |\gamma((x_1, \ldots, x_n)) p^{(x_1, \ldots, x_n)}(u) - \gamma((x_1, \ldots, x_n)) p^{(x_1, \ldots, x_n)}(v)| \\
    &=  \sum_{(x_1, \ldots, x_n) \in P(\mathcal{C}_u \cup \mathcal{C}_v)} \gamma((x_1, \ldots, x_n)) |p^{(x_1, \ldots, x_n)}(u) - p^{(x_1, \ldots, x_n)}(v)| \\
    &\le \sum_{(x_1, \ldots, x_n) \in P(\mathcal{C}_u \cup \mathcal{C}_v)} \gamma((x_1, \ldots, x_n)) \D(u,v) \\
    &\le \D(u,v).
\end{align*}
We use this to show that
\begin{align*}
    TV(q_{u,v}, q_{v,u}) &= 0.5 \sum_{i=1}^{n_{u,v}} |q_{u,v}^2(i) - q_{v,u}^2(i)| \\
    &= \sum_{(x_1, \ldots, x_n) \in P(\mathcal{C}_u \cup \mathcal{C}_v)} |q_{u,v}^2(i((x_1, \ldots, x_n))) - q_{v,u}^2(i((x_1, \ldots, x_n)))| \\
    &=  \sum_{(x_1, \ldots, x_n) \in P(\mathcal{C}_u \cup \mathcal{C}_v)} |\gamma((x_1, \ldots, x_n)) p^{(x_1, \ldots, x_n)}(u) - \gamma((x_1, \ldots, x_n)) p^{(x_1, \ldots, x_n)}(v)| \\
     &= C \\
    &\le \D(u,v),
\end{align*}
proving (2). We similarly see that:
\begin{align*}
    |p(u) - p(v)| &= \left|\sum_{(x_1, \ldots, x_n) \in P(\mathcal{C}_u \cup \mathcal{C}_v)} |\gamma((x_1, \ldots, x_n)) p^{(x_1, \ldots, x_n)}(u) - \gamma((x_1, \ldots, x_n)) p^{(x_1, \ldots, x_n)}(v)\right| \\
    &\le \sum_{(x_1, \ldots, x_n) \in P(\mathcal{C}_u \cup \mathcal{C}_v)} |\gamma((x_1, \ldots, x_n)) p^{(x_1, \ldots, x_n)}(u) - \gamma((x_1, \ldots, x_n)) p^{(x_1, \ldots, x_n)}(v) \\
    &= C \\
    &\le \D(u,v),
\end{align*}
thus proving (1).  

\textbf{Showing (3).} We now show (3). For (3a), pick any pair of individuals $u$ and $v$. Observe that the partition corresponding to $u$ and $v$ in the mapping has a single cluster. Thus, $TV(q_{u,v}^1, q_{v,u}^1) = 0.5 |p(u) - p(v)| \le \D(u,v)$, using the individual fairness of a quality compositional mechanism given by (1). For (3b), pick any pair of individuals $u$ and $v$. If $\D(u,v) = 1$, then there is no condition on $u$ and $v$ in $0.5$-Notion 2, so it trivially holds. If $\D(u,v) < 1$, then we know that $u$ and $v$ are in the same quality group (say $q_i$) and $\D(u,v) = 0$. By (1), we know that the quality compositional mechanism is individually fair so $p(u) = p(v)$. Consider an arbitrary cluster, say cluster $j$. Since the mechanisms $\Mech_{i, x_i}$ are individually fair, and since $\mathcal{C}$ is quality-symmetric, we know that it must true that $q_{u,v}^2(i) = \frac{\sum_{C \in \mathcal{C}_u \mid M_{u,v}(C) = j} \DistMech(C)}{p(u)} = \frac{\sum_{C \in \mathcal{C}_v \mid M_{u,v}(C) = j} \DistMech(C)}{p(v)} = q_{v,u}^2(j)$. Thus, $TV(q_{u,v}^2, q_{v,u}^2) = 0$ as desired. 

\end{proof}

%% file: sections/proofsmodel.tex
We now show that conditional robustness implies unconditional robustness up to a Lipschitz constant.

\begin{proposition}
\label{prop:implicationexpectedscore}
Suppose that $f$ is an individually fair post-processing function and $\Mech$ is an individually fair cohort-selection mechanism. 
\begin{enumerate}
    \item If $\left\{\OriginalDist{\Mech}{f}{u}\right\}_{u \in U}$ is $\alpha$-Lipschitz individually fair w.r.t $\DCondE$, then $\left\{\OriginalDist{\Mech}{f}{u}\right\}_{u \in U}$ is $(\alpha + 1)$-Lipschitz individually fair w.r.t $\DUncondE$.
    \item If $\left\{\OriginalDist{\Mech}{f}{u}\right\}_{u \in U}$ is $\alpha$-Lipschitz individually fair w.r.t $\DCondMMD$, then $\left\{\OriginalDist{\Mech}{f}{u}\right\}_{u \in U}$ is $(\alpha + 1)$-Lipschitz individually fair w.r.t $\DUncondMMD$.
\end{enumerate}
\end{proposition}
\begin{proof}
First, we prove (1). Since $\left\{\OriginalDist{\Mech}{f}{u}\right\}_{u \in U}$ is $\alpha$-Lipschitz individually fair w.r.t $\DCondE$, we know that
\[\left|\sum_{C \in \mathcal{C}, u \in C} \frac{\DistMech(C)}{p(u)} f(C,u) - \sum_{C \in \mathcal{C}, v \in C} \frac{\DistMech(C)}{p(v)} f(C,v) \right| \le \alpha \D(u,v)\]
and
\[|p(u) - p(v)| \le \D(u,v).\]
It suffices to show that
\[\left|\sum_{C \in \mathcal{C}, u \in C} \DistMech(C) f(C,u) - \sum_{C \in \mathcal{C}, v \in C} \DistMech(C) f(C,v) \right| \le (\alpha + 1) \D(u,v).\]
Notice that the first condition is equivalent to
\small{
\begin{align*}
    \left|\sum_{C \in \mathcal{C}, u \in C} \frac{p(v) \DistMech(C)}{p(u) p(v)} f(C,u) - \sum_{C \in \mathcal{C}, v \in C} \frac{\DistMech(C) p(u)}{p(u) p(v)} f(C,v) \right| &\le \alpha \D(u,v)\\
    \left|\sum_{C \in \mathcal{C}, u \in C} p(v) \DistMech(C)f(C,u) - \sum_{C \in \mathcal{C}, v \in C} p(u) \DistMech(C) f(C,v) \right| &\le \alpha p(u) p(v) \D(u,v) \\
    |\sum_{C \in \mathcal{C}, u \in C} p(v) \DistMech(C)f(C,u) - \sum_{C \in \mathcal{C}, v \in C} p(v) \DistMech(C)f(C,v) &+ \sum_{C \in \mathcal{C}, v \in C} p(v) \DistMech(C)f(C,v) - \sum_{C \in \mathcal{C}, v \in C} p(u) \DistMech(C) f(C,v)| \\
    &\le  \alpha p(u) p(v) \D(u,v)
\end{align*}}
\normalsize{}
In the previous line we added and subtracted the term $\sum_{C \in \mathcal{C}, v \in C} p(v) \DistMech(C)f(C,v)$. Now, we use the fact that  $|A| - |B| = |A| - |-B| \le |A + B|$ by the triangle inequality. This implies that 
\small{
\begin{align*}
 \left|\sum_{C \in \mathcal{C}, u \in C} p(v) \DistMech(C)f(C,u) - \sum_{C \in \mathcal{C}, v \in C} p(v) \DistMech(C)f(C,v)\right| &- \left|\sum_{C \in \mathcal{C}, v \in C} p(v) \DistMech(C)f(C,v) - \sum_{C \in \mathcal{C}, v \in C} p(u) \DistMech(C) f(C,v) \right| \\
 &\le \alpha p(u) p(v) \D(u,v) \\
 p(v) \left|\sum_{C \in \mathcal{C}, u \in C} \DistMech(C) F(C,u) - \sum_{C \in \mathcal{C}, v \in C} \DistMech(C) f(C,v)\right| &\le \alpha p(u) p(v) \D(u,v) +  |p(v) - p(u)| \left(\sum_{C \in \mathcal{C}, v \in C} f(C,v) \DistMech(C) \right) \\
 &\le \alpha p(u) p(v) \D(u,v) +  |p(v) - p(u)| \left(\sum_{C \in \mathcal{C}, v \in C} \DistMech(C) \right) \\
  &\le \alpha p(u) p(v) \D(u,v) + p(v) \D(u,v) \\
 \left|\sum_{C \in \mathcal{C}, u \in C} \DistMech(C) F(C,u) - \sum_{C \in \mathcal{C}, v \in C} \DistMech(C) f(C,v)\right| &\le \frac{\alpha p(u) p(v) \D(u,v) +  \D(u,v) p(v)}{p(v)} = (1 + \alpha p(u))\D(u,v) \\
 &\le (\alpha + 1) \D(u,v).
\end{align*}}

\normalsize{}
Now, we prove (2). Recall that we have mass-moving guarantees for the conditional case, i.e. that $MMD(\CondDist{\Mech}{f}{u}, \CondDist{\Mech}{f}{v}) = d'$ for some $d'$. Pick $\epsilon > 0$ and let $d = d' + \epsilon$. By the definition of mass-moving distance, we know that there exist probability measures with finite support $\tilde{\gamma}^C_u$ and $\tilde{\gamma}^C_v$ over $[0,1]$ that achieve the value of $d$ in the mass-moving distance definition. The problem that we now we need to solve boils to handling the extra mass at $0$ in the unconditional distributions. We let $\tilde{\gamma}^N_u(x) = p(u) \tilde{\gamma}^C_u(x)$ for $x \neq 0$ and we let $\tilde{\gamma}^N(0) = (1 - p(u)) + p(u) \tilde{\gamma}^C(0)$. We define $\tilde{\gamma}^N_v(x)$ analogously. Now, we use $\tilde{\gamma}^N_u$ and $\tilde{\gamma}^N_v$ to upper bound $MMD(\UncondDist{\Mech}{f}{u}, \UncondDist{\Mech}{f}{v})$. First, we see that $\tilde{\gamma}^N_u$  and $\tilde{\gamma}^N_v$ have finite support. For condition (2), we let $Z'_u(i) = Z_u(i)$ for $i \neq 0$. If $p(u) = 1$, we let $Z'_u(0) = Z_u(0)$ and otherwise, we let $Z'_u(0)$ have a mass of $\frac{\CondDist{\Mech}{f}{u}(0) + z^0_u(0)}{1 - p(u) + p(u)\CondDist{\Mech}{f}{u}(0)}$ at $0$ and have a mass of $\frac{z^0_u(i)}{1 - p(u) + p(u)\CondDist{\Mech}{f}{u}(0)}$ at $i \neq 0 \in \text{Supp}(\tilde{\gamma}^N_u)$, where $z^0_u$ is the pmf of the distribution $Z_u(0)$. It is straightforward to verify that condition (2) is satisfied.

Thus, we just need to verify condition (1). First, we consider modified measures (not probability measures), where $\tilde{\gamma}^{M}_u(x) = p(u) \tilde{\gamma}^C_u(x)$ for all $x$. We see that the only difference is that there is no extra mass of $1 - p(u)$ on $0$ (analogously for $\tilde{\gamma}^M_v$). Observe that 
\[TV(\tilde{\gamma}^N_u, \tilde{\gamma}^N_v) \le TV(\tilde{\gamma}^M_u, \tilde{\gamma}^M_v) + 0.5 |1 - p(u) - (1 - p(v))| \le  TV(\tilde{\gamma}^M_u, \tilde{\gamma}^M_v) + 0.5 \D(u,v).\] 

We now show that $TV(\tilde{\gamma}^M_u, \tilde{\gamma}^M_v) \le d + 0.5\D(u,v)$. We know that
\begin{align*}
   0.5  \sum |\tilde{\gamma}^C_u(s) - \tilde{\gamma}^C_v(s)| &\le d \\
    0.5  \sum |\frac{\tilde{\gamma}^M_u(s)}{p(u)} - \frac{\tilde{\gamma}^M_v(s)}{p(v)}| &\le d \\
    0.5  \sum |\frac{p(v) \tilde{\gamma}^M_u(s)}{p(u) p(v)} - \frac{p(u) \tilde{\gamma}^M_v(s)}{p(u) p(v)}| &\le d \\
     0.5 \sum |p(v) \tilde{\gamma}^M_u(s) - p(u) \tilde{\gamma}^M_v(s)| &\le d p(u) p(v) 
    \\
    0.5  \sum |p(v) \tilde{\gamma}^M_u(s) - p(v) \tilde{\gamma}^M_v(s) + p(v) \tilde{\gamma}^M_v(s) - p(u) \tilde{\gamma}^M_v(s)| &\le d p(u) p(v).
\end{align*}

Now, we use the fact that $|A| - |B| = |A| - |-B| \le |A + B|$ by the triangle inequality. This implies that 
\begin{align*}
    0.5  \sum (|p(v) \tilde{\gamma}^M_u(s) - p(v) \tilde{\gamma}^M_v(s)| - |p(v) \tilde{\gamma}^M_v(s) - p(u) \tilde{\gamma}^M_v(s)|) &\le  d p(u) p(v) \\
        0.5 \sum (p(v)|\tilde{\gamma}^M_u(s) - \tilde{\gamma}^M_v(s)|- \tilde{\gamma}^M_v(s)|p(v)- p(u)|) &\le d  p(u) p(v) \\
    0.5  \sum p(v)|\tilde{\gamma}^M_u(s) - \tilde{\gamma}^M_v(s)| &\le \alpha_1 p(u) p(v) \D(u,v) +   0.5 \sum \tilde{\gamma}^M_v(s)|p(v)- p(u)| \\
    0.5  p(v) \sum |\tilde{\gamma}^M_u(s) - \tilde{\gamma}^M_v(s)| &\le d p(u) p(v) +  0.5 \D(u,v) \sum \tilde{\gamma}^M_u(s)\\
     0.5  \sum |\tilde{\gamma}^M_u(s) - \tilde{\gamma}^M_v(s)| &\le \frac{d p(u) p(v)  +  0.5 \D(u,v) p(v)}{p(v)} \\
     &= d p(u) + 0.5 \D(u,v)\\ 
     & \le d + 0.5 \D(u,v).
\end{align*}
Since $\left\{\OriginalDist{\Mech}{f}{u}\right\}_{u \in U}$ is $\alpha$-Lipschitz individually fair w.r.t $\DCondMMD$, we know that $d' \le \alpha \D(u,v)$. Thus, we know that for every $\epsilon >0$, we can set $v$ equal to $\D(u,v) + (d' + \epsilon) + \D(u,v) \le (\alpha + 1) \D(u,v) + \epsilon$. This gives the desired statement.  
\end{proof}

Next, we show that mass moving distance is at least as strong as expected score (up to Lipschitz constants). %\inote{It's not clear to me that this is actually what the lemma says -- we're just saying that difference in expected score is always upper bounded by a constant times MMD. I think what we meant to show was that MMD -> Expected score and then show an example where expected score is satisfied but MMD is not. @Meena thoughts?} \mnote{This wording used to be ambiguous -- good catch. Does it look okay now?} \inote{Yes}
\begin{proposition}
\label{prop:mmdimpliesexpectedscore}
Consider distributions $\mathcal{X}_1, \mathcal{X}_2 \in \Delta([0,1])$. Then, $|\mathbb{E}[\mathcal{X}_1] - \mathbb{E}[\mathcal{X}_2]| \le 3 MMD(\mathcal{X}_1, \mathcal{X}_2)$.
\end{proposition}
\begin{proof}
We know that
\[|\mathbb{E}[\mathcal{X}_1] - \mathbb{E}[\mathcal{X}_2]| \le |\mathbb{E}[\mathcal{X}_1] - \mathbb{E}[\tilde{\mathcal{X}}_1]| +|\mathbb{E}[\tilde{\mathcal{X}}_1] - \mathbb{E}[\tilde{X}_2]| + |\mathbb{E}[\mathcal{X}_2] - \mathbb{E}[\tilde{\mathcal{X}}_2]|. \]
Let $d' = MMD(\mathcal{X}_1, \mathcal{X}_2)$. For $\epsilon > 0$, let $d = d' + \epsilon$. Then, by the definition of mass-moving distance, we know that there exist probability measures (can be viewed as distributions) $\tilde{\mathcal{X}}_1$ and $\tilde{\mathcal{X}}_2)$ that achieve $d$. By condition (2) in the mass-moving distance definition, we know that $TV(\tilde{\mathcal{X}}_1, \tilde{\mathcal{X}}_2) \le d$. This means that the $\ell_1$ distance is at most $2 d$. Since scores are in $[0,1]$, this implies that $|\mathbb{E}[\tilde{\mathcal{D}}_1] - \mathbb{E}[\tilde{\mathcal{D}}_2]| \le 2 d$. For the first and last terms, we use condition (3) in the definition of mass-moving distance. We see that $|\mathbb{E}[\mathcal{X}_i] - \mathbb{E}[\tilde{\mathcal{X}}_i]| \le 0.5 d$, and this means that $|\mathbb{E}[\mathcal{X}_1] - \mathbb{E}[\mathcal{X}_2]| \le 3(d' + \epsilon)$. Taking the limit as $\epsilon \rightarrow 0$, this gives the desired answer. 
\end{proof}

%% file: arxiv.bbl
\begin{thebibliography}{10}

\bibitem{BowerKNSVV17}
Amanda Bower, Sarah~N. Kitchen, Laura Niss, Martin~J. Strauss, Alexander
  Vargas, and Suresh Venkatasubramanian.
\newblock Fair pipelines.
\newblock {\em CoRR}, abs/1707.00391, 2017.

\bibitem{celis2019toward}
L.~Elisa Celis, Anay Mehrotra, and Nisheeth~K. Vishnoi.
\newblock Toward controlling discrimination in online ad auctions.
\newblock In {\em Proceedings of the 36th International Conference on Machine
  Learning, {ICML} 2019, 9-15 June 2019, Long Beach, California, {USA}}, pages
  4456--4465, 2019.

\bibitem{chouldechova2017fair}
Alexandra Chouldechova.
\newblock Fair prediction with disparate impact: {A} study of bias in
  recidivism prediction instruments.
\newblock {\em Big Data}, 5(2):153--163, 2017.

\bibitem{datta2015automated}
Amit Datta, Michael~Carl Tschantz, and Anupam Datta.
\newblock Automated experiments on ad privacy settings.
\newblock {\em Proceedings on Privacy Enhancing Technologies}, 2015(1):92--112,
  2015.

\bibitem{dwork2012fairness}
Cynthia Dwork, Moritz Hardt, Toniann Pitassi, Omer Reingold, and Richard~S.
  Zemel.
\newblock Fairness through awareness.
\newblock In {\em Innovations in Theoretical Computer Science 2012, Cambridge,
  MA, USA, January 8-10, 2012}, pages 214--226, 2012.

\bibitem{DI2018}
Cynthia Dwork and Christina Ilvento.
\newblock Fairness under composition.
\newblock In {\em 10th Innovations in Theoretical Computer Science Conference,
  {ITCS} 2019, January 10-12, 2019, San Diego, California, {USA}}, pages
  33:1--33:20, 2019.

\bibitem{evidencerankings2019}
Cynthia Dwork, Michael Kim, Omer Reingold, Guy Rothblum, and Gal Yona.
\newblock Learning from outcomes: Evidence-consistent rankings.
\newblock In {\em 60th Annual IEEE Symposium on Foundations of Computer Science
  November 9-12, 2019, Baltimore, Maryland}, pages 106--125, 2019.

\bibitem{FNSV17}
Danielle Ensign, Sorelle~A. Friedler, Scott Neville, Carlos Scheidegger, and
  Suresh Venkatasubramanian.
\newblock Runaway feedback loops in predictive policing.
\newblock In {\em Conference on Fairness, Accountability and Transparency,
  {FAT} 2018, 23-24 February 2018, New York, NY, {USA}}, pages 160--171, 2018.

\bibitem{gillen2018online}
Stephen Gillen, Christopher Jung, Michael~J. Kearns, and Aaron Roth.
\newblock Online learning with an unknown fairness metric.
\newblock In {\em Advances in Neural Information Processing Systems 31: Annual
  Conference on Neural Information Processing Systems 2018, NeurIPS 2018, 3-8
  December 2018, Montr{\'{e}}al, Canada}, pages 2605--2614, 2018.

\bibitem{hardt2016equality}
Moritz Hardt, Eric Price, Nati Srebro, et~al.
\newblock Equality of opportunity in supervised learning.
\newblock In {\em Advances in Neural Information Processing Systems}, pages
  3315--3323, 2016.

\bibitem{hebert2017calibration}
{\'{U}}rsula H{\'{e}}bert{-}Johnson, Michael~P. Kim, Omer Reingold, and Guy~N.
  Rothblum.
\newblock Multicalibration: Calibration for the (computationally-identifiable)
  masses.
\newblock In Jennifer~G. Dy and Andreas Krause, editors, {\em Proceedings of
  the 35th International Conference on Machine Learning, {ICML} 2018,
  Stockholmsm{\"{a}}ssan, Stockholm, Sweden, July 10-15, 2018}, volume~80 of
  {\em Proceedings of Machine Learning Research}, pages 1944--1953. {PMLR},
  2018.

\bibitem{HuC17}
Lily Hu and Yiling Chen.
\newblock Fairness at equilibrium in the labor market.
\newblock {\em CoRR}, abs/1707.01590, 2017.

\bibitem{CIJ2019}
Christina Ilvento, Meena Jagadeesan, and Shuchi Chawla.
\newblock Multi-category fairness in sponsored search auctions.
\newblock In {\em FAT* '20: Conference on Fairness, Accountability, and
  Transparency, Barcelona, Spain, January 27-30, 2020}, pages 348--358, 2020.

\bibitem{jung2019eliciting}
Christopher Jung, Michael~J. Kearns, Seth Neel, Aaron Roth, Logan Stapleton,
  and Zhiwei~Steven Wu.
\newblock Eliciting and enforcing subjective individual fairness.
\newblock {\em CoRR}, abs/1905.10660, 2019.

\bibitem{kearns2017gerrymandering}
Michael~J. Kearns, Seth Neel, Aaron Roth, and Zhiwei~Steven Wu.
\newblock Preventing fairness gerrymandering: Auditing and learning for
  subgroup fairness.
\newblock In {\em Proceedings of the 35th International Conference on Machine
  Learning, {ICML} 2018, Stockholmsm{\"{a}}ssan, Stockholm, Sweden, July 10-15,
  2018}, pages 2569--2577, 2018.

\bibitem{kilbertus2017avoiding}
Niki Kilbertus, Mateo Rojas{-}Carulla, Giambattista Parascandolo, Moritz Hardt,
  Dominik Janzing, and Bernhard Sch{\"{o}}lkopf.
\newblock Avoiding discrimination through causal reasoning.
\newblock In {\em Advances in Neural Information Processing Systems 30: Annual
  Conference on Neural Information Processing Systems 2017, 4-9 December 2017,
  Long Beach, CA, {USA}}, pages 656--666, 2017.

\bibitem{kim2018fairness}
Michael~P. Kim, Omer Reingold, and Guy~N. Rothblum.
\newblock Fairness through computationally-bounded awareness.
\newblock In {\em Advances in Neural Information Processing Systems 31: Annual
  Conference on Neural Information Processing Systems 2018, NeurIPS 2018, 3-8
  December 2018, Montr{\'{e}}al, Canada}, pages 4847--4857, 2018.

\bibitem{DBLP:journals/corr/KleinbergMR16}
Jon~M. Kleinberg, Sendhil Mullainathan, and Manish Raghavan.
\newblock Inherent trade-offs in the fair determination of risk scores.
\newblock In {\em 8th Innovations in Theoretical Computer Science Conference,
  {ITCS} 2017, January 9-11, 2017, Berkeley, CA, {USA}}, pages 43:1--43:23,
  2017.

\bibitem{lambrecht2016algorithmic}
Anja Lambrecht and Catherine Tucker.
\newblock Algorithmic bias? {A}n empirical study of apparent gender-based
  discrimination in the display of {STEM} career ads.
\newblock {\em Management Science}, 65(7):2966--2981, 2019.

\bibitem{liu2018delayed}
Lydia~T. Liu, Sarah Dean, Esther Rolf, Max Simchowitz, and Moritz Hardt.
\newblock Delayed impact of fair machine learning.
\newblock In {\em Proceedings of the Twenty-Eighth International Joint
  Conference on Artificial Intelligence, {IJCAI} 2019, Macao, China, August
  10-16, 2019}, pages 6196--6200, 2019.

\bibitem{lum2016predict}
Kristian Lum and William Isaac.
\newblock To predict and serve?
\newblock {\em Significance}, 13(5):14--19, 2016.

\bibitem{madras2018learning}
David Madras, Elliot Creager, Toniann Pitassi, and Richard~S. Zemel.
\newblock Learning adversarially fair and transferable representations.
\newblock In {\em Proceedings of the 35th International Conference on Machine
  Learning, {ICML} 2018, Stockholmsm{\"{a}}ssan, Stockholm, Sweden, July 10-15,
  2018}, pages 3381--3390, 2018.

\bibitem{mishra2007}
Nina Mishra, Robert Schreiber, Isabelle Stanton, and Robert~E. Tarjan.
\newblock Clustering social networks.
\newblock In {\em Algorithms and Models for the Web-Graph}, pages 56--67,
  Berlin, Heidelberg, 2007. Springer Berlin Heidelberg.

\bibitem{ritov2017conditional}
Ya'acov Ritov, Yuekai Sun, and Ruofei Zhao.
\newblock On conditional parity as a notion of non-discrimination in machine
  learning.
\newblock {\em arXiv preprint arXiv:1706.08519}, 2017.

\bibitem{rothblum2018probably}
Gal Yona and Guy~N. Rothblum.
\newblock Probably approximately metric-fair learning.
\newblock In {\em Proceedings of the 35th International Conference on Machine
  Learning, {ICML} 2018, Stockholmsm{\"{a}}ssan, Stockholm, Sweden, July 10-15,
  2018}, pages 5666--5674, 2018.

\bibitem{zemel2013learning}
Richard~S. Zemel, Yu~Wu, Kevin Swersky, Toniann Pitassi, and Cynthia Dwork.
\newblock Learning fair representations.
\newblock In {\em Proceedings of the 30th International Conference on Machine
  Learning, {ICML} 2013, Atlanta, GA, USA, 16-21 June 2013}, pages 325--333,
  2013.

\end{thebibliography}
